\documentclass[showkeys]{openjournal}
\usepackage{amsmath, amssymb}
\usepackage{subfigure}
\usepackage{graphicx}
\usepackage{float} 
\usepackage[colorlinks=true, linkcolor=black, allcolors=black]{hyperref}
\usepackage[utf8]{inputenc}

\newcommand{\coloneqq}{\mathrel{\mathop:}=}
\newcommand{\coloneq}{\mathrel{\mathop:}=}

\begin{document}

\title{Deep Neural Networks Hunting Ultra-Light Dark Matter}

\author{
Pavel Kůs$^{1,2}$,
Diana López Nacir$^{3,4}$,
Federico R.\ Urban$^1$
}

\address{
$^1$ CEICO, FZU--Institute of Physics of the Czech Academy of Sciences, Na Slovance 2, 182 00 Praha 8, Czechia \\
$^2$ Charles University, Faculty of Mathematics and Physics, Institute of Theoretical Physics, V Holešovičkách 2, 180 00 Praha 8, Czechia \\
$^3$ Universidad de Buenos Aires, Facultad de Ciencias Exactas y Naturales, Departamento de Física, Buenos Aires, Argentina \\
$^4$ CONICET - Universidad de Buenos Aires, Instituto de Física de Buenos Aires (IFIBA), Buenos Aires, Argentina
}

\begin{abstract}  
Ultra-light dark matter (ULDM) is a compelling candidate for cosmological dark matter. If ULDM interacts with ordinary matter, it can induce measurable, characteristic signals in pulsar-timing data because it causes the orbits of pulsars in binary systems to osculate. In this work, we investigate the potential of machine learning (ML) techniques to detect such ULDM signals. To this end, we construct three types of neural networks: an autoencoder, a binary classifier, and a multiclass classifier. We apply these methods to four theoretically well-motivated ULDM models: a linearly coupled scalar field, a quadratically coupled scalar field, a vector field and a tensor field. We show that the sensitivity achieved using ML methods is comparable to that of a semi-analytical Bayesian approach, which to date has only been applied to the linear scalar case. The ML approach is readily applicable to all four ULDM models and, in the case of the multiclass classifier, can distinguish between them. Our results, derived from simulated data, lay the foundation for future applications to real pulsar-timing observations.
\end{abstract}


\maketitle

\tableofcontents

\section{Introduction}
\label{sec:intro}

Dark matter constitutes about 27\% of the energy density of the Universe and is about five times more abundant than baryonic matter, yet its fundamental nature remains elusive. A promising solution to this problem is a class of dark matter candidates known as \textit{ultra-light dark matter} (ULDM). This type of dark matter consists of light bosons with masses in the range $10^{-23} \lesssim m \lesssim 1$\,eV. If these particles account for the entirety of the dark matter, their number density must be enormous, allowing the dark matter to be described by a coherent classical field oscillating at a frequency determined by the particle's mass. When the mass approaches the lower end of the interval, the de Broglie wavelength of ULDM stretches across astrophysical scales of order of~pc to~kpc. Consequently, ULDM exhibits wave-like behaviour on galactic scales, leading to a rich phenomenology while still reproducing the behaviour of cold dark matter on cosmological scales. The dynamics of non-relativistic ULDM is governed by the Schrödinger–Poisson system of equations; numerical simulations predict that ULDM halos form with a central solitonic core, while the outer regions exhibit a granular structure due to wave interference effects~\citep{Schive:2014}.\footnote{For studies on soliton merger simulations for spin-1 and spin-2 ULDM see~\citet{Amin:2022pzv,Lopez-Sanchez:2025osk}.} For a comprehensive review of ULDM properties and tests we refer to~\citet{Ferreira:2021}.

\begin{figure}[htbp]
    \centering
    \includegraphics[width=0.5\linewidth]{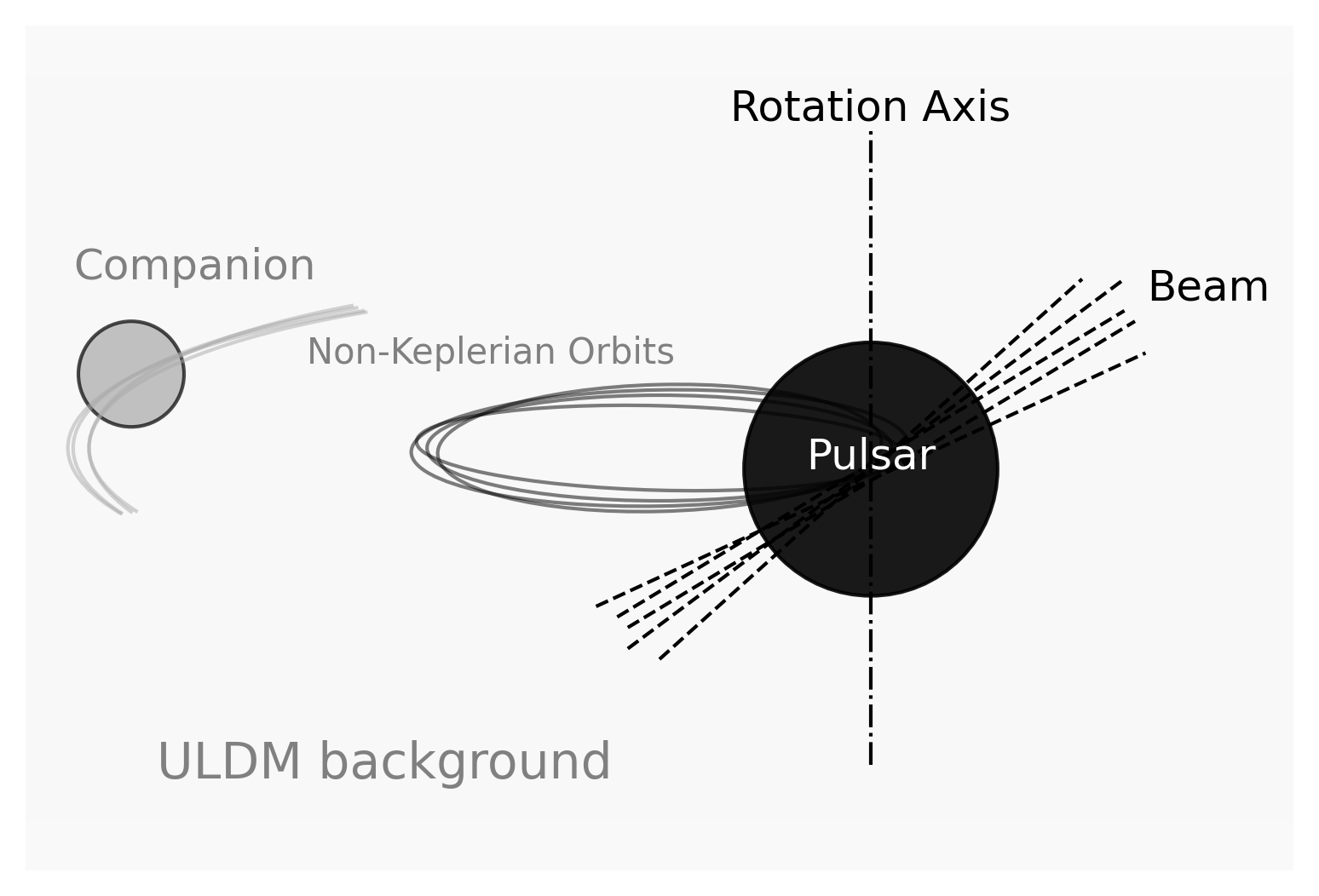}
    \caption{Illustration of a binary pulsar inside a dark matter halo that follows a non-Keplerian orbit caused by interaction with ULDM.}
    \label{fig:binary_pulsar_picture}
\end{figure}

Binary pulsars---binary star systems where one star is a pulsar---have proven to be promising ULDM detectors. Studies in~\citet{Blas:2016ddr, Blas:2019hxz, LopezNacir:2018epg, Armaleo:2020yml, Kus:2024} have shown that both direct and gravitational interactions between ULDM and the components of a binary pulsar lead to perturbations in the dynamics of the binary---see Fig.~\ref{fig:binary_pulsar_picture} for an artistic illustration---which leave their imprints on pulsar timing data. If this effect is not incorporated into the timing model describing the times of arrival of the pulses on Earth, it results in a discrepancy between the theoretically predicted values and the actual measurements. This discrepancy manifests as \textit{time residuals} with a characteristic pattern---a unique fingerprint of each ULDM model---which we refer to as a \textit{signal}.

Although the gravitational imprint on pulsar timing data appears too weak to be detectable~\citep{Blas:2019hxz}, a signal from direct ULDM-matter interaction could be observed. If such a signal is not detected, it would place stringent constraints on the ULDM parameter space. Initial studies were conducted for models with spins~0, 1 and~2, within the mass range of $10^{-23}$\,to $10^{-18}$\,eV, beyond which the binary pulsar's backreaction to dark matter can not be ignored~\citep{Blas:2019hxz}. Recently, a new two-step Bayesian method to determine the sensitivity of binary systems to the linear spin~0 ULDM interaction constant $\alpha$ (which characterises a direct universal interaction between binary pulsars and ULDM) was given in~\citet{Kus:2024}, inspired by a similar method developed for gravitational wave (GW) signals in pulsar timing~\citep{Moore:2014eua}. The sensitivity is described by the curve $\alpha(m)$, above which the ULDM parameter space is excluded by signal non-detection. The Bayesian method significantly extends previous studies in that it applies to the whole mass range and not only when the ULDM and the binary system are in resonance, that is, when $m = Q \omega_b$ where $\omega_b$ is the orbital angular frequency of the binary system and $Q$ is the integer resonance number. Despite the success of this Bayesian method, it requires handling and solving complex algebraic equations that make its generalisation to higher spins a challenging task. 

In order to surmount these difficulties, we look again at GW research, which also seeks to detect coherent signals in very noisy data. Recently, researchers have been exploring the use of machine learning (ML) tools for GW signal detection. These methods leverage deep neural networks (DNNs) with various architectures, such as convolutional neural networks (CNNs) and residual neural networks, see for instance~\citet{Baltus:2021, Baltus:2022} and references therein. Another architecture that is naturally suited for coherent signal detection is the autoencoder, a specialised type of DNN that can be used as an anomaly detector~\citep{Erdmann:2021}.

Inspired by advancements in the application of ML algorithms for GW searches, we aim to adapt these tools to address two key objectives.

\begin{enumerate}
	\item Our primary goal is to determine the sensitivity of ML techniques to ULDM signals for the models discussed in~\citet{Blas:2016ddr, Blas:2019hxz, LopezNacir:2018epg, Armaleo:2020yml}. This sensitivity sets limits beyond which signal non-detection excludes a given region of the ULDM parameter space. To achieve this, we employ two independent approaches: an autoencoder, which is an unsupervised ML technique, and a binary classifier, which instead is a supervised ML technique. In order to assess the power of the ML method, once we obtain the sensitivity across the full mass range of $10^{-23}$\,eV to $10^{-18}$\,eV for all ULDM spins, we compare them to the Bayesian sensitivity limits for the spin-0 case presented in~\citet{Kus:2024}.
	\item Our second goal is to develop a multiclass classifier capable of distinguishing between five types of inputs: pure noise and four different ULDM signal types embedded in the noise. We demonstrate that, given a sufficiently strong ULDM signal, this classification can be performed with high accuracy.
\end{enumerate}

Our aim is to provide \textit{a proof of concept} that ML tools can be used to detect and classify ULDM signals of various nature and, in the absence of a detected signal, to impose constraints on coupling constants. Optimising DNN hyperparameters for peak performance and exploring the applicability of alternative architectures are not central to this work and are left for future investigation. We focus on a single binary pulsar with low eccentricity, although~\citet{Kus:2024} show that more eccentric systems exhibit strong, rich resonances, and combining multiple binary pulsars enhances sensitivity. Finally, the methods used here might also be applied in other scenarios---for example, detecting ULDM signals in pulsar timing from an isolated pulsar~\citep{Smarra:2024} or investigating ULDM-induced modifications to the gravitational waveform from inspiralling binaries~\citep{Kim:2024,Chase:2025wwj}.

The main text is divided into four sections. Section~\ref{sec:signal} covers the modelling of scalar, vector and tensor dark matter signals, along with the noise. Section~\ref{sec:ml} introduces the three different ML architectures, the autoencoder, the convolutionary binary and the multiclass classifiers, describes how we obtain the sensitivities to ULDM and compares them with the results from~\citet{Kus:2024}. We discuss the main features of our results in Section~\ref{sec:discussion}. Lastly, in Section~\ref{sec:end} we provide a summary and an outlook for future work. The codes we used to build and train our DNNs as well as to generate our results can be found at \href{https://github.com/pavelkuspro/ULDM-BinaryPulsars-DNN.git}{\textcolor{blue}{https://github.com/pavelkuspro/ULDM-BinaryPulsars-DNN.git}}.

\section{Signal and noise modelling}
\label{sec:signal}

\subsection{Overview of pulsar timing}

\subsubsection{Timing model and ULDM}
Pulsars are astronomical clocks with remarkable stability. This stability is crucial, as it allows us to accurately predict the times of arrival, as clocked on Earth, of pulses emitted from pulsars, even many years into the future. To achieve this, we must model the propagation of the pulse from the moment it is emitted until it is captured by the detector, building what is known as a \textit{timing model}. Exotic physics, such as phenomena related to modified gravity or dark matter, can also influence pulse emission and propagation, making pulsars effective probes of these phenomena---for a comprehensive overview of the wide range of pulsars' applications, we refer to~\citet{Lorimer:2008se}. If the timing model is inaccurate, it results in a discrepancy between the actual and predicted pulse arrival times, leading to \textit{time residuals}. In addition to the stochastic component, these residuals contain a signal from the incorrect or incomplete modelling of the system, which manifests as a characteristic coherent pattern.

In this work, we explore a scenario where ULDM leads to an additional signal in the time residuals which was not accounted for in the timing model.\footnote{In principle time residuals can be computed using pulsar timing software such as TEMPO2~\citep{tempo2006} and PINT~\citep{pint2021}, with data from pulsar-timing arrays such as NANOGrav~\citep{NANOGrav:2023hde}, EPTA~\citep{epta2023}, PPTA~\citep{ppta2023} and InPTA~\citep{inpta2022}. However, in this study we rely exclusively on mock data to test ML techniques.} These time residuals are simulated for each ULDM model and ML techniques are then used to search for the imprints of the ULDM.

\subsubsection{Time residuals}
Let \( T \) be the pulsar’s proper time and \( \Theta \) the pulsar’s phase at time \( T \). They are related by the following expression:
\begin{align}
\Theta = \Theta_0 + \nu T + \frac{1}{2} \dot{\nu} T^2 + \frac{1}{6} \ddot{\nu} T^3\,,
\end{align}
where \( \Theta_0 \) is the initial phase and \( \nu \), \( \dot{\nu} \) and \( \ddot{\nu} \) represent the pulsar's spin frequency, its first time derivative and its second time derivative, respectively. The proper time \( T \) can be expressed in terms of the \textit{infinite-frequency barycentric arrival time} \( t \), a coordinate time referenced to the solar system barycenter, which is used to measure pulse arrival times---further details are available in~\citet{Kus:2024} and~\citet{Teukolsky1976}. Importantly, \( \Theta \) is a function of \( t \), the pulsar’s parameters and, if in a binary, the orbital parameters:
\begin{align}
\Theta = \Theta(t, \Theta_0, \nu, \ldots)\,.
\end{align}
Even after a measurement we do not know the exact values of these parameters, only their best-fit estimates which we denote as \( \Theta_0^{(1)}, \nu^{(1)} \), etc. The time residuals \( R(t) \) are derived from the phase residuals:
\begin{align}
- \nu^{(1)} R(t) \coloneq \Theta(t, \Theta_0, \nu, \dots) - \Theta(t, \Theta_0^{(1)}, \nu^{(1)}, \dots)\,.
\end{align}
Here \( \Theta(t, \Theta_0, \nu, \dots) \) is the true phase of the pulsar at coordinate time \( t \), while \( \Theta(t, \Theta_0^{(1)}, \nu^{(1)}, \dots) \) represents the estimated phase based on the best-fit parameters. Assuming that we have a model \( \Theta \) and that the estimated parameters are close to the true values, we can express \( R(t) \) in terms of the small differences \( \delta \Theta_0 \coloneq \Theta_0 - \Theta_0^{(1)} \), etc.:
\begin{align}
R(t) = - \frac{1}{\nu^{(1)}} \frac{\partial \Theta}{\partial \Theta_0}(t, \Theta_0^{(1)}, \dots) \, \delta \Theta_0 + \dots\,. 
\end{align}
These time residuals arise from an inaccurate fit of the model to the data, with additional noise modelled as Gaussian, see Eq.~(3.2) in~\citet{Kus:2024} for further details.

The \(a\)-th time residual, $R_a \coloneq R(t_a)$, corresponding to the \(a\)-time of arrival $t_a$ of a series of $N$ data points, has both stochastic and deterministic components:
\begin{align}
R_a = R_a^{\mathrm{sto}} + R_a^{\mathrm{det}}\,.
\end{align}
The stochastic component of the timing residuals arises from various sources, including receiver noise, clock noise, intrinsic timing noise (irregularities of the pulsar beam rotation), fluctuations in the refractive index of the interstellar medium and potentially the gravitational-wave background~\citep{Haasteren:2009}. In this paper, we focus exclusively on intrinsic timing noise, which follows a Gaussian normal distribution:
\begin{align}
    R_a^{\mathrm{sto}} &\sim \mathcal{N}\left(0, \mathbf{C}\right), \\
    P(\vec{R}^{\mathrm{sto}}) &= \frac{1}{\sqrt{(2\pi)^\mathrm{N} \det \mathbf{C}}} \exp\left( -\frac{1}{2} \sum_{a,b=1}^\mathrm{N} R_a^{\mathrm{sto}} \left(\mathbf{C}^{-1}\right)_{ab} R_b^{\mathrm{sto}} \right),
\end{align}
where \(P\) denotes the probability distribution of the time-series noise vector, $\vec{R}^{\mathrm{sto}} \coloneq (R_1^{\mathrm{sto}}, \dots, R_N^{\mathrm{sto}})$. 

The form of the covariance matrix \( \mathbf{C} \) depends on the choice of the power spectral density \( S \). We consider two types: flat (white) and Lorentzian (red). For white noise, \( S \) is constant, \( S = \epsilon^2 / f_s \), where \( f_s \) is the sampling frequency and the covariance matrix is \( \mathbf{C}_{ab} = \epsilon^2 \delta_{ab} \). In the Lorentzian case, the power spectral density is \(
S = \epsilon^2 \left\{f_0 \left[ 1 + \left( f/f_0 \right)^2 \right] \right\}^{-1} \) and the covariance matrix is \( \mathbf{C}_{ab} = \epsilon^2 \exp\left(- 2 \pi f_0 |t_a - t_b| \right) \), where \( f_0 \) is the characteristic reddening frequency, chosen as \( f_0 = 1 \, \mathrm{yr}^{-1} \) as a typical value~\citep{Haasteren:2009}.

\begin{figure}[htbp]
    \centering
    \includegraphics[width=0.48\linewidth]{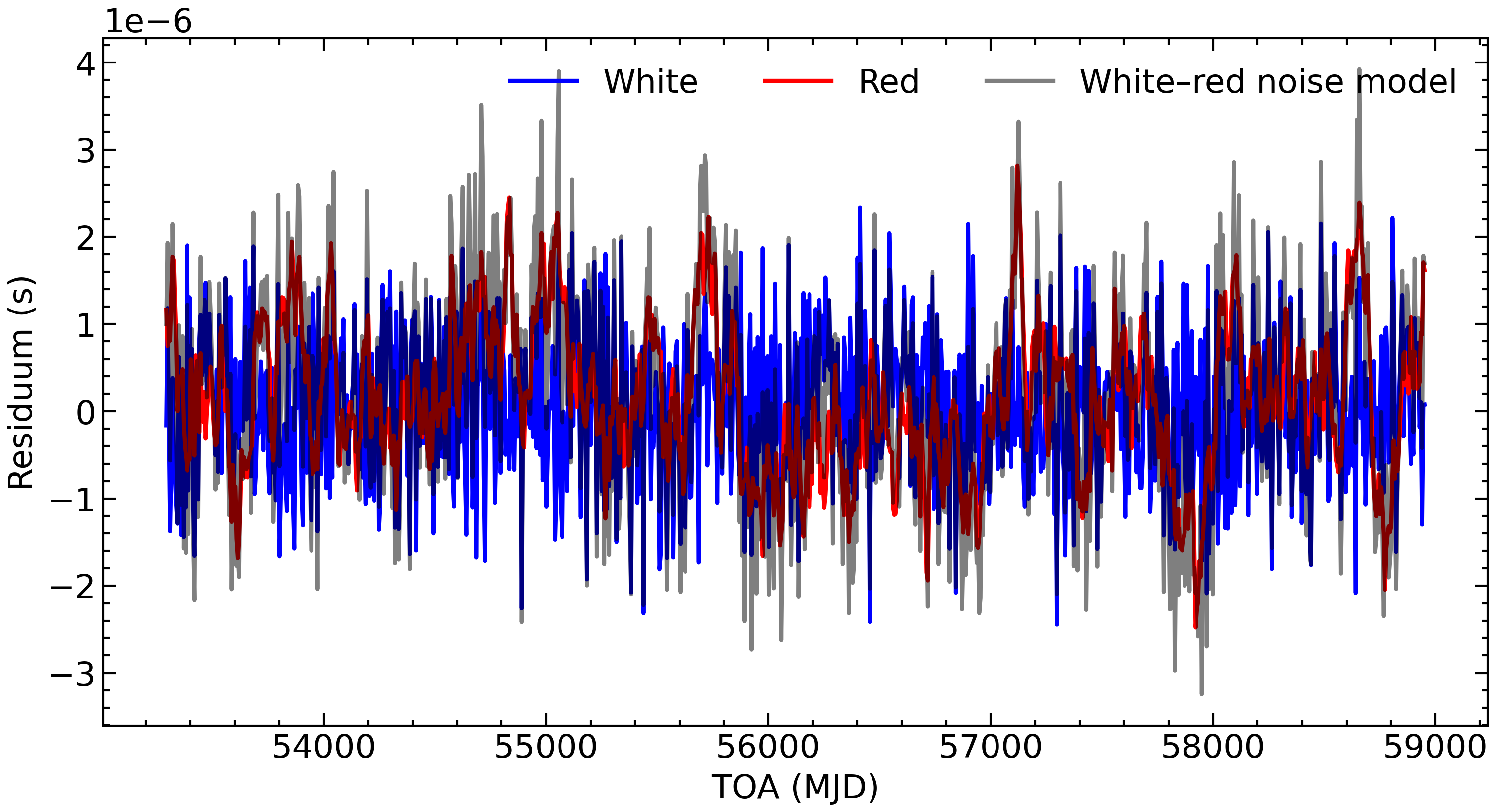}
    \hfill
    \includegraphics[width=0.48\linewidth]{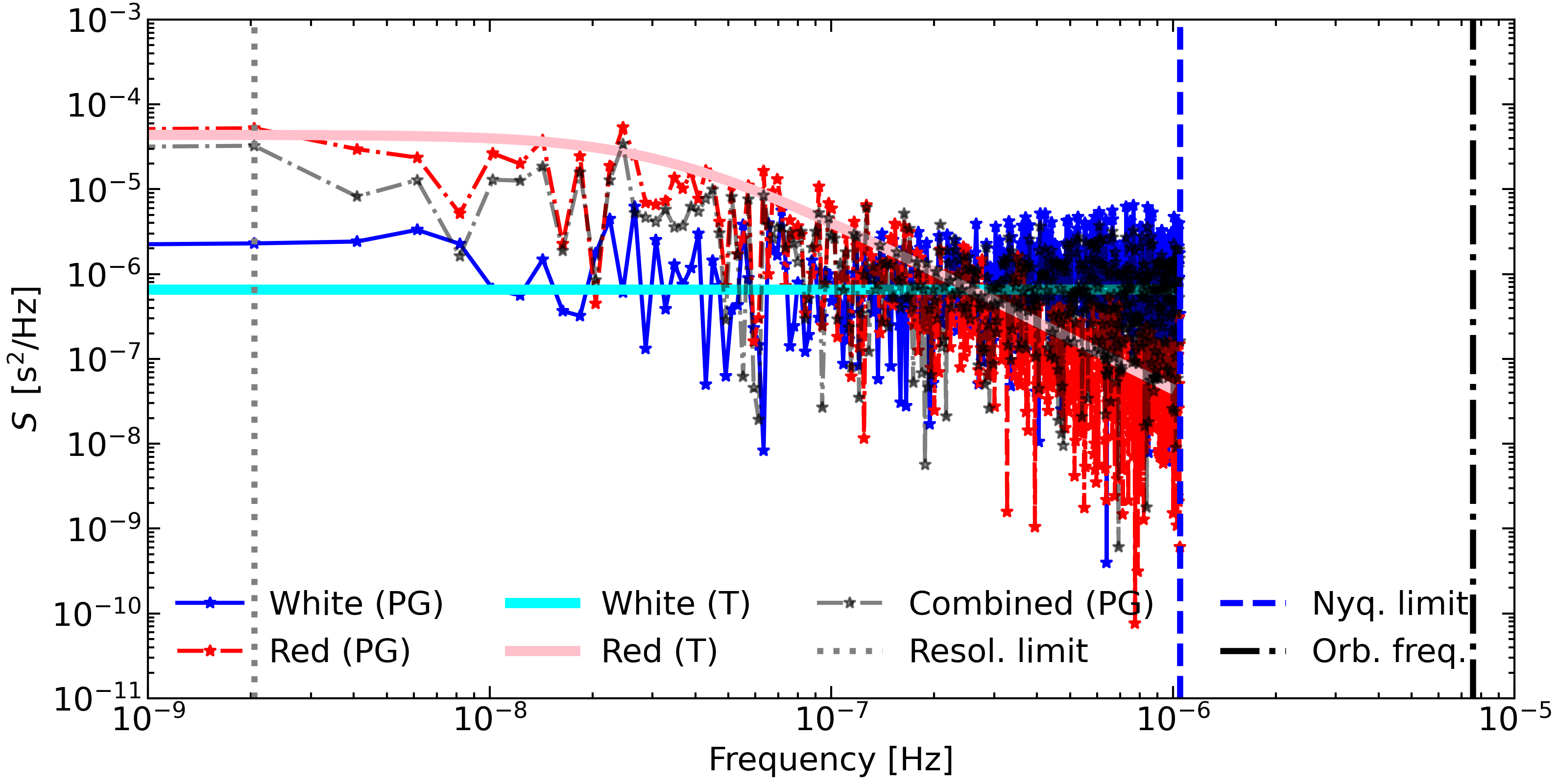}
    \caption{\textit{Left panel:} Examples of time residuals for white noise, red noise and an equally weighted combination of white and red noise. These apply to the low-eccentricity pulsar PSR~J1909-3744, with \( N = 1024 \) uniformly distributed data points. \textit{Right panel:} Power spectral densities for the same simulated data. PG stands for periodogram, T stands for theory prediction.}
    \label{fig:noise_example}
\end{figure}

In practice, both white and red noise contribute to pulsar timing noise. For mock data generation, we assume either pure white noise or an equally weighted sum of white and red noise, ensuring that the standard deviation remains \(\epsilon\). The data are generated uniformly over the observational campaign, consisting of \( N = 1024 \) data points. Examples of noise realisations and their corresponding power spectral density functions \( S(f) \) are shown in Fig.~\ref{fig:noise_example}, specific to the low-eccentricity pulsar PSR~J1909-3744.

We assume that the deterministic signal has a constant contribution, a linear contribution in time and a quadratic contribution in time---the purpose of these terms is to model potentially unincorporated or inaccurately modelled physics:
\begin{equation}
R_a^{\mathrm{det}} = K_0 + K_1 (t_a - T_\mathrm{asc}) + K_2 (t_a - T_\mathrm{asc})^2 + R_a^{\mathrm{binary}}\,,
\end{equation}
where \(T_\mathrm{asc}\) represents the time when the pulsar passes the ascending node, $R_a^{\mathrm{binary}}$ represents the contribution to the time residuals from variations of the orbital parameters and $K_0$, $K_1$ and $K_2$ are constants with unspecified values at this moment; these values will be determined later during mock data generation by sampling from random distributions.

\subsubsection{PSR~J1909-3744}

We will apply ML methods to mock pulsar timing data from the specific pulsar PSR~J1909-3744~\citep{NANOGrav:2023hde}, whose parameters are provided in Table~\ref{tab:pulsar}. To account for time delays in the binary system, we use the ELL1 binary model~\citep{Lange:2001rn}, which is specifically designed for systems with low eccentricities. Given that most binary systems have nearly circular orbits with eccentricities \(e \approx 0\), the ELL1 model is broadly applicable to a large number of pulsars. Let us note that the value $N = 1024$ and the requirement for uniform sampling are fictitious but serve well for illustrative purposes. Generalisation beyond these constraints is straightforward with our method.  

\begin{table}[htbp]
\centering
\caption{Parameters for PSR~J1909-3744~\citep{NANOGrav:2023hde}.}
\label{tab:pulsar}
\vspace{2mm}
\small 
\begin{tabular}{|c|c|l|}
\hline
\textbf{Symbol} & \textbf{Value} & \textbf{Description} \\
\hline
$x$ & 1.9 s & Projected semi-major axis \\
$\omega$ & $156^\circ$ & Argument of periapsis \\
$\omega_b$ & 4.1 day$^{-1}$ & Angular orbital frequency \\
$\omega_b$ & $3.1 \times 10^{-20}$ eV & Angular orbital frequency (converted to energy units) \\
$\epsilon$ & $1.18 \times 10^{-6}$ s & Noise level \\
$T_{\mathrm{obs}}$ & 15.5 years & Length of the observational campaign \\
$T_{\mathrm{asc}}$ & 56121 MJD & Time of ascending node \\
$T_{\mathrm{min}}$ & 53292 MJD & Start of observational campaign \\
$T_{\mathrm{max}}$ & 58951 MJD & End of observational campaign \\
$\eta$ & $4.5 \times 10^{-8}$ & Laplace-Lagrange parameter ($e\sin \omega$) \\
$\kappa$ & $-9.9 \times 10^{-8}$ & Laplace-Lagrange parameter ($e\cos \omega$) \\
$N$ & 1024 & Number of mock data points \\
\hline
\end{tabular}
\end{table}


\subsubsection{Binary system description}

\begin{figure}
    \centering
    \includegraphics[width=0.5\linewidth]{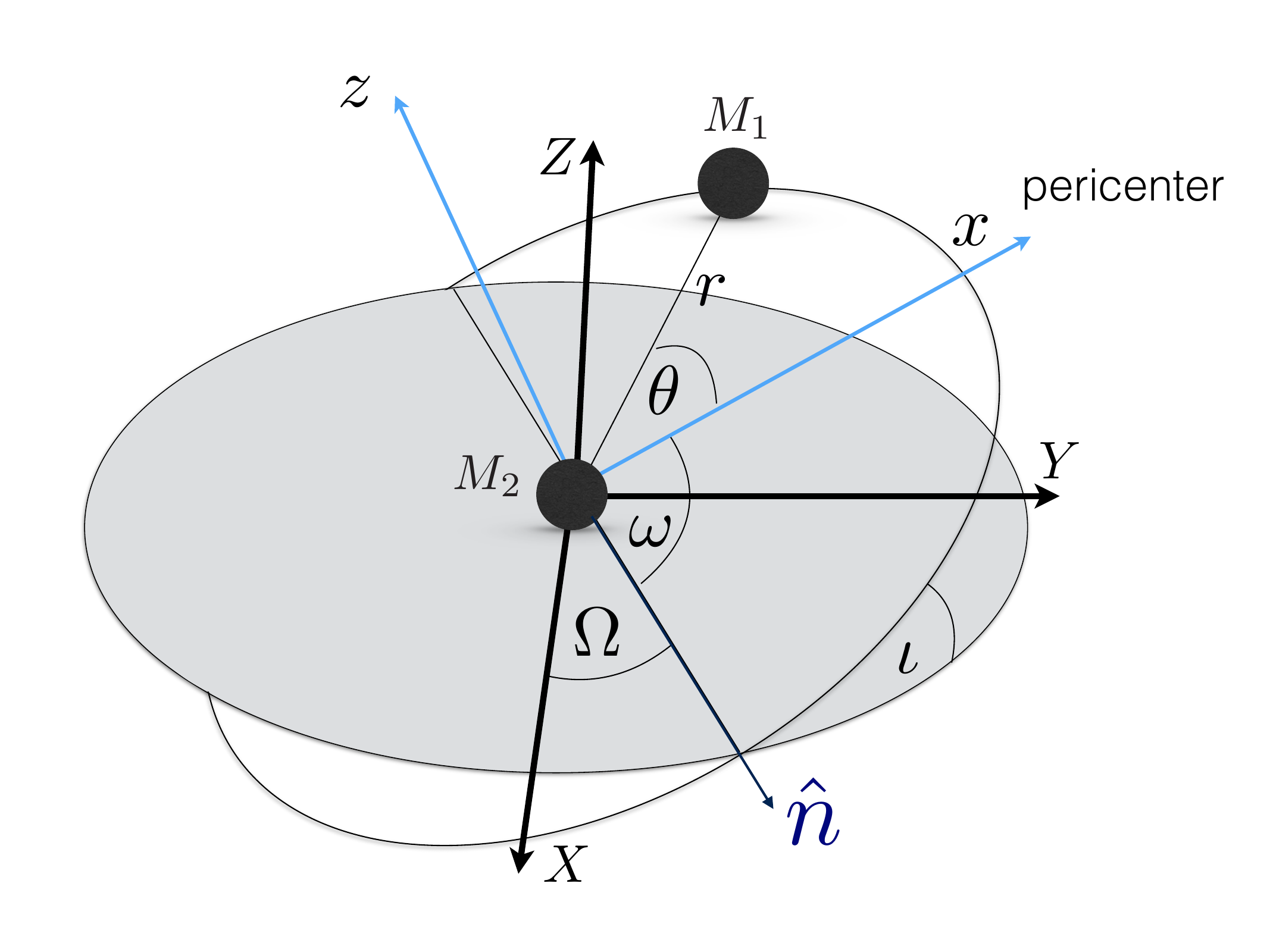}
    \caption{Description of Keplerian orbits in terms of the orbital elements viewed in the fundamental reference frame \((X, Y, Z)\). The Cartesian orbital frame \((x, y, z)\) and the polar one \((r, \theta, z)\) are also shown (centred on \(M_2\) for convenience).}
    \label{fig:orbit}
\end{figure}

In an idealised binary system stars follow elliptical orbits described by six orbital parameters. Two of these parameters define the shape and size of the ellipse: the semi-major axis \( a \) and the eccentricity \( e \). Two others specify the orientation of the orbital plane: the inclination \( \iota \) and the longitude of the ascending node \( \Omega \). The remaining parameters are the argument for periapsis \( \omega \) and the time of passage of periapsis \( T_0 \). Altogether, the system is characterised by the set \( \{a, e, \omega, \iota, \Omega, T_0\} \). A diagram illustrating the orbital plane and its key parameters is shown in Fig.~\ref{fig:orbit}.  Further details on Keplerian mechanics can be found, for instance, in~\citet{Danby:1970}.

If a pulsar is part of a binary system, the propagation of light within the system must be reflected in the timing model---it includes a \textit{binary model}. For non-relativistic, elliptic systems, the timing model is given by the BT model~\citep{Teukolsky1976}. However, for systems with low eccentricities (\( e \to 0 \)), it is often more practical to use an alternative set of parameters: \( \{a, \kappa, \eta, \iota, \Omega, T_{\mathrm{asc}}\} \), where \( \kappa \) and \( \eta \) are the Laplace-Lagrange parameters, $\kappa \coloneqq e \cos \omega\,,\eta \coloneqq e \sin \omega$ and \( T_{\mathrm{asc}} \) is the time of passage through the ascending node. This set is employed in the binary ELL1 model, which is specifically designed for systems with nearly circular orbits~\citep{Lange:2001rn}.

In practice no real two-body system follows a perfectly Keplerian orbit due to effects such as gravitational wave emission, perturbations from other celestial bodies, or potential interactions between ULDM and the binary components. Such orbital evolution can be studied within the post-Keplerian framework, also known as the \emph{osculating orbits formalism}. In this approach, the binary dynamics are still described by six orbital parameters, but these parameters are now time-dependent---for example, \(a = a(t)\), and so on. The two stars therefore move along instantaneous Keplerian orbits that evolve over time.

\subsubsection{Binary model and time residuals}

Although tracking the variation of all orbital parameters would lead to the highest sensitivity to ULDM, in order to illustrate our method it is sufficient to track three of them: the semi-major axis \( a \) and the Laplace-Lagrange parameters \( \eta \) and \( \kappa \).\footnote{Whilst we focus on three parameters in this work, in~\citet{Kus:2024} it was shown that low-eccentricity systems are most sensitive to variations in $\Psi' = \int_{T_{\mathrm{asc}}}^t \omega_b \, dt'$ for masses in the range $10^{-23}$~eV to $\sim 2 \times 10^{-20}$~eV. A full discussion of orbital parameter variations across all models is left to future work.}

The ELL1 model is not directly sensitive to the variation \(\delta a\), but to \(\delta x\), where \(x \coloneqq a_1 \sin \iota\) is the projected semi-major axis of the pulsar orbit \(a_1\) and \(\iota\) is the inclination of the orbit. The time residuals coming from \(\delta x\), \(\delta \eta\) and \(\delta \kappa\) are as follows (see Eq.~(F.2) in~\citet{Kus:2024}):
\begin{equation}
R_a^{\mathrm{binary}} = \sin (\Psi'_a) \delta x_a - x\frac{\cos(2\Psi'_a) + 3}{2} \delta \eta_a +  x\frac{\sin(2\Psi'_a)}{2} \delta \kappa_a\,,
\end{equation}
where \(\Psi'_a \coloneqq \omega_b (t_a - T_\mathrm{asc})\) and \(\delta x_a = \delta x(t_a)\) is a time-dependent function---further details are provided below. In what follows we omit the contribution of $\delta \iota$ and focus solely on the contribution from $a_1 \coloneqq a M_2 / (M_1 + M_2)$, where $M_{1,2}$ represents the pulsar and companion mass, respectively. We have
\begin{align}
    \delta x^{(a_1)} \coloneqq (\delta a_1) \sin \iota\,;
\end{align}
to reduce clutter we will denote \(\delta x^{(a_1)}\) simply as \(\delta x\) in what follows, unless specified otherwise. Apart from the contribution from dark matter, the variations in $x, \eta$ and $ \kappa$ have a constant component due to measurement errors and a component that is linear in time due to potential secular effects:
\begin{subequations} \label{eq:variation_general}
\begin{align}
    \delta x_a &= \delta x_0 + A_x (t_a - T_\mathrm{asc}) + \delta x_a^{\mathrm{DM}} \,, \\
    \delta \eta_a &= \delta \eta_0 + A_\eta (t_a - T_\mathrm{asc}) + \delta \eta_a^{\mathrm{DM}} \,, \\
    \delta \kappa_a &= \delta \kappa_0 + A_\kappa (t_a - T_\mathrm{asc}) + \delta \kappa_a^{\mathrm{DM}} \,.
\end{align}
\end{subequations}
During synthetic data generation, the $\delta x_0, \delta \eta_0, \delta \kappa_0, A_x, A_\eta$ and $A_\kappa$ parameters are treated as random variables, as well as $K_0, K_1$ and $K_2$, see below.

Lastly, in order to derive the formulas for the ULDM-induced variations, such as \(\delta x_a^{\mathrm{DM}}\), we employ the post-Keplerian formalism. When an external force \(\vec{F} = F_r \hat{r} + F_{\theta} \hat{\theta} + F_z \hat{z}\) acts on the binary system, it introduces a perturbation on the binary parameters as follows~\citep{Danby:1970}:
\begin{align}
    \label{eq:variation_a}
   \dot{a} &= \frac{2}{\omega_b} \left( \frac{F_r e}{\sqrt{1-e^2}} \sin \theta + \frac{F_{\theta}}{r} a \sqrt{1-e^2} \right)\,,\\
   \label{eq:variation_e}
    \dot{e} &= \frac{\sqrt{1-e^2}}{a \omega_b} \left[ (\cos \theta + \cos E') F_\theta + \sin \theta F_{r} \right]\,,\\
   \label{eq:variation_omega}
    \dot{\omega} &= \frac{\sqrt{1-e^2}}{a e \omega_b} \left[ \left(1 + \frac{r}{a(1-e^2)} \right) \sin \theta F_\theta - \cos \theta F_r  \right] + \left[ 2 \sin^2\left(\frac{\iota}{2}\right) - 1 \right] \frac{r \sin(\theta + \omega)}{a^2 \omega_b \sqrt{1-e^2}} F_z \,,
\end{align}
where \( F_r \), \( F_\theta \) and \( F_z \) are force components specific to a given ULDM model~\citep{Blas:2019hxz, LopezNacir:2018epg, Armaleo:2020yml}, \( r \) is the radial distance between the two bodies, \(\theta\) is the true anomaly and \( E' \) is the eccentric anomaly, satisfying the Kepler equation \( E' - e \sin E' = \Psi' - \omega \). We obtain $\dot{\eta}$ and $\dot{\kappa}$ from $\dot{e}$ and $\dot{\omega}$ using the relations $\dot{\eta} = \dot{e}\sin \omega + e\dot{\omega} \cos \omega$ and 
$\dot{\kappa} = \dot{e} \cos \omega - e \dot{\omega} \sin \omega$. The equations for $\dot{a}\,,\,\, \dot{\eta}$ and $\dot{\kappa}$ for each ULDM model are listed in the following section.

\subsection{ULDM signal modelling}

\subsubsection{ULDM coherent patch}
The de~Broglie wavelength of ULDM particles is given by
\[
\lambda_{\mathrm{dB}} \sim 1.3 \times 10^{12} \, \mathrm{km} \times \left( \frac{10^{-3}}{V_0} \right) \left( \frac{10^{-18} \, \mathrm{eV}}{m} \right),
\]  
where \( V_0 \simeq 10^{-3} \) is the velocity dispersion of ULDM in the Milky Way halo~\citep{Blas:2016ddr}. Since the occupation number must be extremely large in order to account for the observed dark matter density, a classical field theory description becomes applicable. The characteristic scale over which the field can be considered homogeneous is approximately \( \lambda_{\mathrm{dB}} /2 \), and the field oscillations remain coherent over a timescale given by
\begin{equation}
    t_{\mathrm{coh}} \sim 65~\mathrm{years} \left( \frac{10^{-3}}{V_0} \right)^2 \left( \frac{10^{-18}~\mathrm{eV}}{m} \right) \,.
\end{equation}

For the mass scales of interest, ULDM exhibits a pronounced wave nature on astrophysical scales, which manifests as interference patterns. Consequently, the local ULDM amplitude is modulated by an interference factor, \( \varrho \geq 0 \), treated as a random variable that follows the Rayleigh distribution, \( P(\varrho) = 2\varrho \exp(-\varrho^2) \)~\citep{Foster:2017hbq}. The ULDM configuration local to each binary system also differs in the local phase $\Upsilon \in [0, 2\pi)$ (see below).

\begin{figure}[htbp]
    \centering
    \includegraphics[width=0.6\linewidth]{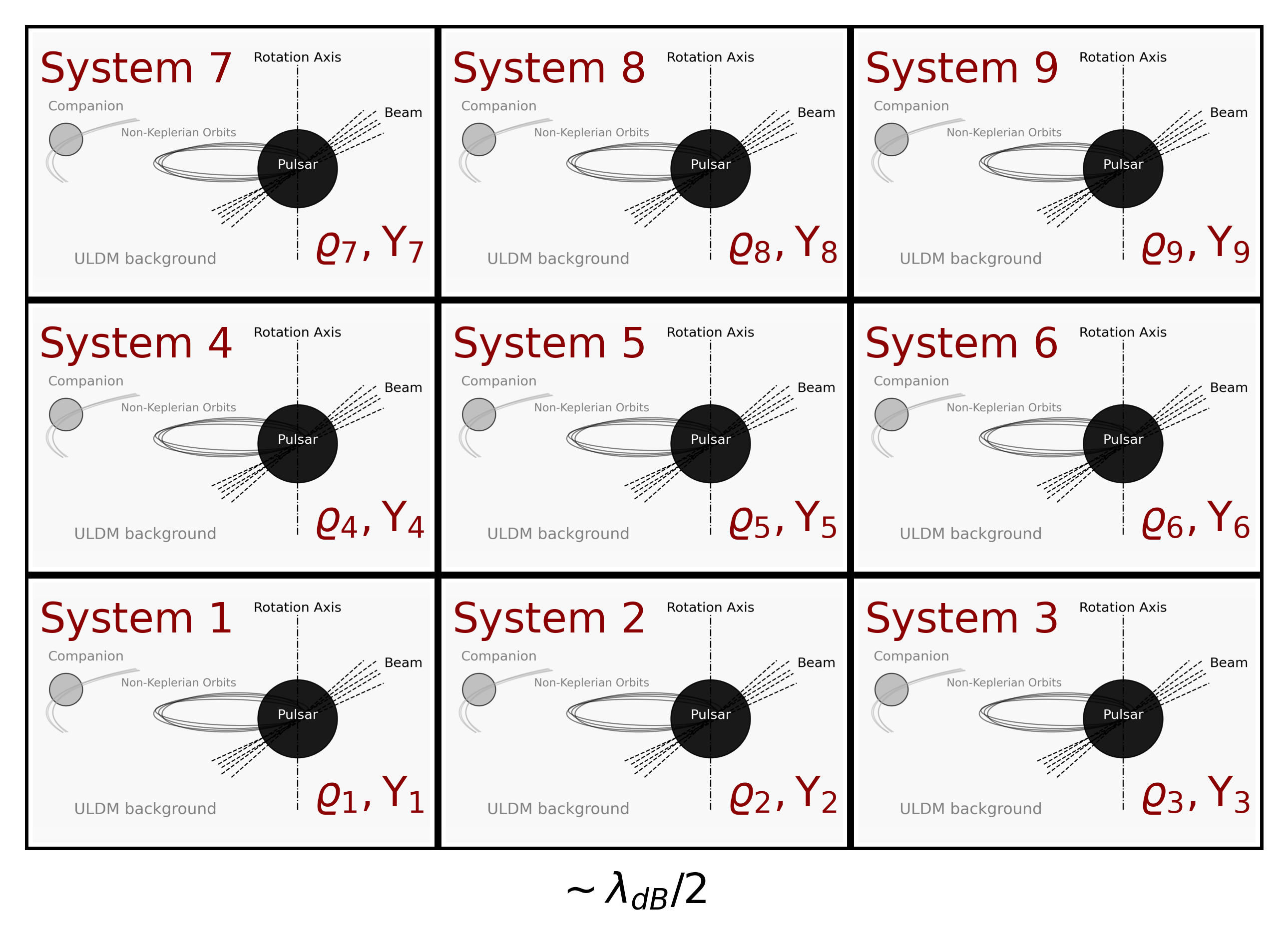}
    \caption{Binary systems are located in distinct ULDM patches, each approximately \( \lambda_{\mathrm{dB}} / 2 \) in size.}
    \label{fig:many_systems}
\end{figure}

In the case of binary pulsars, the separation between the two components is much smaller than the de Broglie wavelength, which allows the ULDM configuration to be treated as identical for both components. However, the typical separation between binary pulsars exceeds the de Broglie wavelength, resulting in distinct ULDM configurations for different binary systems, as depicted in Fig.~\ref{fig:many_systems}. We stress that only the subrange \(10^{-23} \text{ to } 10^{-18}\, \mathrm{eV}\) is accessible to binary pulsars, whereas beyond this value we would need to take into account the backreaction of the binaries on the ULDM (see Eq.~(13) in~\citet{Blas:2019hxz}).

ULDM is not a single theory but rather a class of models. We consider four viable ULDM models: (pseudo)scalar fields with linear or quadratic direct couplings to ordinary matter via dilatonic portal~\citep{Marsh:2016, Thibault:2010}; a vector field coupled to the baryonic (or baryonic-leptonic) current~\citep{Nelson:2011, Knapen:2017}; and a tensor field motivated by e.g.\ the bimetric theory of gravity~\citep{Schmidt:2016,Aoki:2017cnz,Marzola:2018}. In the following, we describe the influence of each field on the three orbital parameters of the binary system discussed above: $a$, $\eta$ and $\kappa$, following the results of~\citet{Blas:2016ddr, Blas:2019hxz, LopezNacir:2018epg, Armaleo:2020yml}.

\subsubsection{Scalar field $\Phi$ (Linear dilatonic coupling)}

This ULDM model is described by a classical scalar field \(\Phi\) given by
\begin{align}
    \Phi(t) = \Phi_0 \varrho \cos(mt + \Upsilon)\,,
\end{align}
where \(\Upsilon \in [0, 2\pi)\) represents the ULDM local phase, assumed to be a random variable with a uniform distribution. The amplitude of the field is defined as \(\Phi_0 \coloneqq \sqrt{2 \rho_{\mathrm{DM}}}/m\) and it is modulated by the interference factor \(\varrho\). In this study we adopt a conservative value of the dark matter density normalisation: \(\rho_{\mathrm{DM}} = 0.3\, \mathrm{GeV/cm^3}\)~\citep{Weber:2010}.

Following~\citet{Blas:2016ddr, Blas:2019hxz}, we consider the masses of the binary components to depend on the scalar field \(\Phi\). Expanding the mass as a power series in \(\Phi\), we retain only the leading-order term:
\begin{align}
\label{eq:spin0_lin_coupling}
    M^{\alpha}_A(\Phi) = \bar{M}_A \left( 1 + \alpha \Phi \right)\,,
\end{align}
where \(\alpha\) denotes the (effective) coupling constant. This interaction arises from the fundamental coupling between the scalar field and Standard Model fields, but it can also receive contributions from the binding energy due to self-gravity. While \(\alpha\) could, in principle, vary between the two components because of the latter effect, we assume a universal coupling for both bodies in this work. For a discussion on possible non-universality, we refer to~\citet{Blas:2019hxz}.

The mass of each binary component (\(A = 1,2\)) depends on time through the oscillating dark matter field \(\Phi\), leading to a perturbation in the binary's dynamics. The Lagrangian describing this system is
\begin{align}
    L = \frac{1}{2}M_1(\Phi) \dot{\boldsymbol{r}}_1^2 + \frac{1}{2}M_2(\Phi) \dot{\boldsymbol{r}}_2^2 + \frac{GM_1(\Phi) M_2(\Phi)}{|\boldsymbol{r}_1 - \boldsymbol{r}_2|}\,,
\end{align}
where \(G\) is the gravitational constant and \(\boldsymbol{r}_1, \boldsymbol{r}_2\) are the positions of the two binary components.

The dependence on \(\Phi(t)\) introduces an additional force that perturbs the Keplerian motion, beyond the gravitational force. As a result, \(a(t)\), \(\eta(t)\) and \(\kappa(t)\) are no longer constant and evolve over time~\citep{Blas:2019hxz}:
\begin{subequations} \label{eq:time_evolution_spin0_alpha}
\begin{align}
    \frac{\dot{a}}{a} = &- 2 \alpha \dot{\Phi} \,, \\
    \dot{\eta} = &+\alpha \Phi_0 \varrho \left( \frac{\omega_b}{2} + m \right) \cos\left[ (m-\omega_b)(t-T_{\mathrm{asc}}) + \Upsilon_{\mathrm{asc}} \right]\, \nonumber \\
    &+ \alpha \Phi_0 \varrho \left( \frac{\omega_b}{2} - m \right) \cos\left[ (m+\omega_b)(t-T_{\mathrm{asc}}) + \Upsilon_{\mathrm{asc}} \right]\,, \\
    \dot{\kappa} = &+ \alpha \Phi_0 \varrho \left( \frac{\omega_b}{2} + m \right) \sin\left[ (m-\omega_b)(t-T_{\mathrm{asc}}) + \Upsilon_{\mathrm{asc}} \right]\, \nonumber \\
    &- \alpha \Phi_0 \varrho \left( \frac{\omega_b}{2} - m \right) \sin\left[ (m+\omega_b)(t-T_{\mathrm{asc}}) + \Upsilon_{\mathrm{asc}} \right]\,,
\end{align}
\end{subequations}
where $\Upsilon_{\mathrm{asc}} \coloneqq \Upsilon + m T_{\mathrm{asc}}$. After integration from $T_{\mathrm{asc}}$ to $t$ we get:
\begin{subequations} \label{eq:integrated_time_evolution_spin0_alpha}
\begin{align}
    \delta a^{\mathrm{S0L}} = &- 2 \alpha a \Phi_0 \varrho \left[ \cos(k \Psi' + \Upsilon_{\mathrm{asc}}) -  \cos\Upsilon_{\mathrm{asc}} \right]\,,\\ 
    \delta \eta^{\mathrm{S0L}} = &+\alpha \Phi_0 \varrho \left( \frac{1}{2} + k \right) \frac{\sin\left[(k-1)\Psi' + \Upsilon_\mathrm{asc}\right] - \sin \Upsilon_\mathrm{asc}}{k-1} \nonumber \\
    &+ \alpha \Phi_0 \varrho \left( \frac{1}{2} - k \right) \frac{\sin\left[(k+1)\Psi' + \Upsilon_\mathrm{asc}\right] - \sin \Upsilon_\mathrm{asc}}{k+1}\,,\\
    \delta \kappa^{\mathrm{S0L}} = &+\alpha \Phi_0 \varrho \left( \frac{1}{2} + k \right) \frac{\cos \Upsilon_\mathrm{asc} - \cos\left[(k-1)\Psi' + \Upsilon_\mathrm{asc}\right]}{k-1}\, \nonumber \\
    &- \alpha \Phi_0 \varrho \left( \frac{1}{2} - k \right)\frac{\cos \Upsilon_\mathrm{asc} - \cos\left[(k+1)\Psi' + \Upsilon_\mathrm{asc}\right]}{k+1}\,,
\end{align}
\end{subequations}
where we define $k \coloneqq m / \omega_b$ and the $\mathrm{S0L}$ superscripts mean that these variations refer to the linear spin-0 model. In order to obtain the variation of \(x\) we define the pulsar's semi-major axis:
\begin{align}
    a_1 \coloneqq \frac{M_2}{M_1 + M_2} a\,.
\end{align}
Although the component masses are time-dependent, their ratio in front of \(a\) is time-independent, i.e.~$M_2 / (M_1 + M_2) = \bar{M}_2 / (\bar{M}_1 + \bar{M}_2) = \text{const}$. Therefore, the variation of \(a_1\) depends only on the variation of \(a\). This leads to the result
\begin{align}
    \delta x^{\mathrm{S0L}} = (\delta a_1^{\mathrm{S0L}}) \sin \iota = \frac{\bar{M}_2}{\bar{M}_1 + \bar{M}_2} (\delta a^{\mathrm{S0L}}) \sin \iota = - 2 \alpha x \Phi_0 \varrho \left[ \cos(k \Psi' + \Upsilon_{\mathrm{asc}}) -  \cos\Upsilon_{\mathrm{asc}} \right]\,.
\end{align}
It is convenient to normalise $\alpha \Phi_0$ as
\begin{align*}
    \alpha \Phi_0 &= \frac{\alpha}{\mathrm{GeV}^{-1}} \times 3.92 \times 10^{-12} \left(\frac{m}{\mathrm{eV}} \right)^{-1} \times \sqrt{\frac{\rho_{\mathrm{DM}}}{\mathrm{GeV/cm^3}}}\,.
\end{align*}

\subsubsection{Scalar field $\Phi$ (Quadratic dilatonic coupling)}

The scalar field theory may possess an underlying symmetry which forbids odd powers in \(\Phi\); in that case the leading component of the Taylor series becomes the quadratic term in \(\Phi\):
\begin{align}
\label{eq:spin0_quad_coupling}
    M^{\beta}_A(\Phi) = \bar{M}_A \left( 1 + \beta \frac{\Phi^2}{2} \right)\,.
\end{align}
The variation of the orbital parameters in this case can be obtained from the expressions for the linear coupling case by substituting \(\alpha \rightarrow \beta\), \(\Phi \rightarrow \Phi^2 / 2\)~\citep{Blas:2019hxz}, leading to:
\begin{subequations} \label{eq:time_evolution_spin0_beta}
\begin{align}
    \frac{\dot{a}}{a} = &- \beta \left(\Phi^2\right)^. \,, \\
    \dot{\eta} = &+ \frac{\beta \omega_b \Phi_0^2 \varrho^2}{4} \cos \left[ \omega_b \left( t - T_{\mathrm{asc}} \right) \right] \nonumber \\
    &+ \frac{\beta \Phi_0^2 \varrho^2}{2} \left( \frac{\omega_b}{4} + m \right) \cos \left[ (2m-\omega_b)(t-T_{\mathrm{asc}}) + 2\Upsilon_{\mathrm{asc}}\right] \nonumber \\
    &+ \frac{\beta \Phi_0^2 \varrho^2}{2} \left( \frac{\omega_b}{4} - m \right) \cos \left[ (2m+\omega_b)(t-T_{\mathrm{asc}}) + 2\Upsilon_{\mathrm{asc}}\right]\,, \\
    \dot{\kappa} = &- \frac{\beta \omega_b \Phi_0^2 \varrho^2}{4} \sin \left[ \omega_b \left( t - T_{\mathrm{asc}} \right) \right] \nonumber \\
    &+ \frac{\beta \Phi_0^2 \varrho^2}{2} \left( \frac{\omega_b}{4} + m \right) \sin \left[ (2m-\omega_b)(t-T_{\mathrm{asc}}) + 2\Upsilon_{\mathrm{asc}}\right] \nonumber \\
    &- \frac{\beta \Phi_0^2 \varrho^2}{2} \left( \frac{\omega_b}{4} - m \right) \sin \left[ (2m+\omega_b)(t-T_{\mathrm{asc}}) + 2\Upsilon_{\mathrm{asc}}\right]\,.
\end{align}
\end{subequations} 
After integration from $T_{\mathrm{asc}}$ to $t$ we get:
\begin{subequations} \label{eq:integrated_time_evolution_spin0_beta}
\begin{align}
    \delta a^{\mathrm{S0Q}} =  &- \beta a \Phi_0^2 \varrho^2 \left[ \cos^2(k \Psi' + \Upsilon_{\mathrm{asc}}) - \cos^2\Upsilon_{\mathrm{asc}} \right]\,, \label{eq:delta_a_spin0_beta} \\
    \delta \eta^{\mathrm{S0Q}} = &+ \frac{\beta \Phi_0^2 \varrho^2}{4} \sin \Psi' \nonumber\\
    &+ \frac{\beta \Phi_0^2 \varrho^2}{2}\left( k +  \frac{1}{4} \right) \frac{\sin\left[(2k-1)\Psi' + 2\Upsilon_\mathrm{asc}\right] - \sin (2\Upsilon_\mathrm{asc})}{2k-1} \nonumber \\
    &+ \frac{\beta \Phi_0^2 \varrho^2}{2} \left( \frac{1}{4} - k \right) \frac{\sin\left[(2k+1)\Psi' + 2\Upsilon_\mathrm{asc}\right] - \sin (2\Upsilon_\mathrm{asc})}{2k+1}\,,\\
    \delta \kappa^{\mathrm{S0Q}} =&+ \frac{\beta \Phi_0^2 \varrho^2}{4} \left( \cos \Psi' - 1 \right) \nonumber\\
    &+ \frac{\beta \Phi_0^2 \varrho^2}{2}\left( k +  \frac{1}{4} \right) \frac{\cos (2\Upsilon_\mathrm{asc}) - \cos\left[(2k-1)\Psi' + 2\Upsilon_\mathrm{asc}\right]}{2k-1} \nonumber \\
    &- \frac{\beta \Phi_0^2 \varrho^2}{2} \left( \frac{1}{4} - k \right) \frac{\cos (2\Upsilon_\mathrm{asc}) - \cos\left[(2k+1)\Psi' + 2\Upsilon_\mathrm{asc}\right]}{2k+1}\,,
\end{align}
\end{subequations} 
where the $\mathrm{S0Q}$ superscript means that these variations refer to the quadratic spin-0 model. From Eq.~(\ref{eq:delta_a_spin0_beta}) we obtain
\begin{align}
     \delta x^{\mathrm{S0Q}} &=  - \beta x \Phi_0^2 \varrho^2 \left[ \cos^2(k \Psi' + \Upsilon_{\mathrm{asc}}) - \cos^2\Upsilon_{\mathrm{asc}} \right]\,.
\end{align}
The expression $\beta \Phi_0^2$ can be normalised as
\begin{align*}
    \beta \Phi_0^{2} &= \frac{\beta}{\mathrm{GeV}^{-2}} \times 1.54 \times 10^{-23} \left(\frac{m}{\mathrm{eV}} \right)^{-2} \times \frac{\rho_{\mathrm{DM}}}{\mathrm{GeV/cm^3}}\,. 
\end{align*}

\subsubsection{Vector field $\boldsymbol{A}$}

The vector ULDM field is
\begin{align}
    \boldsymbol{A} = -\frac{\sqrt{2\rho_{\mathrm{DM}}}}{m} \varrho \cos(m t + \Upsilon) (\sin \vartheta \cos \varphi, \sin \vartheta \sin \varphi, \cos \vartheta)\,,
\end{align}
and is characterised by four parameters: the amplitude \(\varrho\), the phase \(\Upsilon\) and two spherical angles \((\varphi, \vartheta)\). The dynamics of the binary system is described by the standard non-relativistic Lagrangian, with two extra terms describing the interaction with the ULDM:
\begin{align}
    L = \frac{1}{2}M_1 \dot{\boldsymbol{r}}_1^2 + \frac{1}{2}M_2 \dot{\boldsymbol{r}}_2^2 + \frac{GM_1 M_2}{|\boldsymbol{r}_1 - \boldsymbol{r}_2|} + q_1 \dot{\boldsymbol{r}}_1 \cdot \boldsymbol{A} + q_2 \dot{\boldsymbol{r}}_2 \cdot \boldsymbol{A}\,.
\end{align}
Each component carries a charge
\begin{align*}
    q_A = g N_A = g c_A \frac{M_A}{m_n}\,,
\end{align*}
where \(g\) represents the fundamental coupling constant between the vector field and baryonic current, \(N_A\) is the number of baryons and \(c_A\) is a phenomenological parameter that depends primarily on the baryonic-to-gravitational mass ratio and the baryon content of the star. We denote the difference in \(c_A\) between the binary's components as \(\Delta c \coloneqq c_1 - c_2\). Finally, \(m_n\) is the neutron mass.

The direct interaction induces a perturbing force~\citep{LopezNacir:2018epg} that leads to these evolution equations:
\begin{subequations} \label{eq:time_evolution_spin1}
\begin{align}
    \dot{a} =&-\frac{1}{\omega_b}  \frac{g \Delta c}{m_n} \sqrt{2 \rho_{\mathrm{DM}}} \varrho \sin \vartheta \left[ \cos\left( (k+1) \Psi' + \Upsilon_{\mathrm{asc}} - \varphi^1_{\mathrm{asc}} \right) - \cos\left( (k-1) \Psi' + \Upsilon_{\mathrm{asc}} + \varphi^1_{\mathrm{asc}} \right) \right]\,, \\
    \dot{\eta} =&-\frac{3}{4 a \omega_b}  \frac{g \Delta c}{m_n} \sqrt{2 \varrho_{\mathrm{DM}}} \varrho \sin \vartheta \left\{ \sin \left[ (m-2\omega_b)(t - T_\mathrm{asc}) + \Upsilon_\mathrm{asc} - \varphi_\mathrm{asc}^2 \right] \right. \nonumber \\
    &\,\,\,\,\,\,\,\,\,\,\,\,\,\,\,\,\,\,\,\,\,\,\,\,\,\,\,\,\,\,\,\,\,\,\,\,\,\,\,\,\,\,\,\,\,\,\,\,\,\,\,\,\,\,\,\,\,\,\,\,\,\,\,\,\,\,\,\,\,\,\,\,+\sin \left[ (m+2\omega_b)(t - T_\mathrm{asc}) + \Upsilon_\mathrm{asc} + \varphi_\mathrm{asc}^2 \right]  \nonumber\\
    &\,\,\,\,\,\,\,\,\,\,\,\,\,\,\,\,\,\,\,\,\,\,\,\,\,\,\,\,\,\,\,\,\,\,\,\,\,\,\,\,\,\,\,\,\,\,\,\,\,\,\,\,\,\,\,\,\,\,\,\,\,\,\,\,\,\,\,\,\,\,\,\,+\frac{2}{3}\cos \varphi \sin \left[m (t - T_\mathrm{asc}) + \Upsilon_\mathrm{asc}\right] \left. \right\} \sin \omega \nonumber \\
    &-\frac{3}{4 a \omega_b}  \frac{g \Delta c}{m_n} \sqrt{2 \varrho_{\mathrm{DM}}} \varrho \sin \vartheta \left\{ \cos \left[ (m-2\omega_b)(t - T_\mathrm{asc}) + \Upsilon_\mathrm{asc} - \varphi_\mathrm{asc}^2 \right] \right. \nonumber \\
    &\,\,\,\,\,\,\,\,\,\,\,\,\,\,\,\,\,\,\,\,\,\,\,\,\,\,\,\,\,\,\,\,\,\,\,\,\,\,\,\,\,\,\,\,\,\,\,\,\,\,\,\,\,\,\,\,\,\,\,\,\,\,\,\,\,\,\,\,\,\,\,\,+\cos \left[ (m+2\omega_b)(t - T_\mathrm{asc}) + \Upsilon_\mathrm{asc} + \varphi_\mathrm{asc}^2 \right]  \nonumber\\
    &\,\,\,\,\,\,\,\,\,\,\,\,\,\,\,\,\,\,\,\,\,\,\,\,\,\,\,\,\,\,\,\,\,\,\,\,\,\,\,\,\,\,\,\,\,\,\,\,\,\,\,\,\,\,\,\,\,\,\,\,\,\,\,\,\,\,\,\,\,\,\,\,+\frac{2}{3}\sin \varphi \sin \left[m (t - T_\mathrm{asc}) + \Upsilon_\mathrm{asc}\right] \left. \right\} \cos \omega\,, \\
 \dot{\kappa} =&-\frac{3}{4 a \omega_b}  \frac{g \Delta c}{m_n} \sqrt{2 \varrho_{\mathrm{DM}}} \varrho \sin \vartheta \left\{ \sin \left[ (m-2\omega_b)(t - T_\mathrm{asc}) + \Upsilon_\mathrm{asc} - \varphi_\mathrm{asc}^2 \right] \right. \nonumber \\
    &\,\,\,\,\,\,\,\,\,\,\,\,\,\,\,\,\,\,\,\,\,\,\,\,\,\,\,\,\,\,\,\,\,\,\,\,\,\,\,\,\,\,\,\,\,\,\,\,\,\,\,\,\,\,\,\,\,\,\,\,\,\,\,\,\,\,\,\,\,\,\,\,+\sin \left[ (m+2\omega_b)(t - T_\mathrm{asc}) + \Upsilon_\mathrm{asc} + \varphi_\mathrm{asc}^2 \right]  \nonumber\\
    &\,\,\,\,\,\,\,\,\,\,\,\,\,\,\,\,\,\,\,\,\,\,\,\,\,\,\,\,\,\,\,\,\,\,\,\,\,\,\,\,\,\,\,\,\,\,\,\,\,\,\,\,\,\,\,\,\,\,\,\,\,\,\,\,\,\,\,\,\,\,\,\,+\frac{2}{3}\cos \varphi \sin \left[m (t - T_\mathrm{asc}) + \Upsilon_\mathrm{asc}\right] \left. \right\} \cos \omega \nonumber \\
    &+\frac{3}{4 a \omega_b}  \frac{g \Delta c}{m_n} \sqrt{2 \varrho_{\mathrm{DM}}} \varrho \sin \vartheta \left\{ \cos \left[ (m-2\omega_b)(t - T_\mathrm{asc}) + \Upsilon_\mathrm{asc} - \varphi_\mathrm{asc}^2 \right] \right. \nonumber \\
    &\,\,\,\,\,\,\,\,\,\,\,\,\,\,\,\,\,\,\,\,\,\,\,\,\,\,\,\,\,\,\,\,\,\,\,\,\,\,\,\,\,\,\,\,\,\,\,\,\,\,\,\,\,\,\,\,\,\,\,\,\,\,\,\,\,\,\,\,\,\,\,\,+\cos \left[ (m+2\omega_b)(t - T_\mathrm{asc}) + \Upsilon_\mathrm{asc} + \varphi_\mathrm{asc}^2 \right]  \nonumber\\
    &\,\,\,\,\,\,\,\,\,\,\,\,\,\,\,\,\,\,\,\,\,\,\,\,\,\,\,\,\,\,\,\,\,\,\,\,\,\,\,\,\,\,\,\,\,\,\,\,\,\,\,\,\,\,\,\,\,\,\,\,\,\,\,\,\,\,\,\,\,\,\,\,+\frac{2}{3}\sin \varphi \sin \left[m (t - T_\mathrm{asc}) + \Upsilon_\mathrm{asc}\right] \left. \right\} \sin \omega\,,
\end{align}
\end{subequations} 
where we define:
\begin{equation*}
    \Psi' \coloneqq \omega_b(t-T_{\mathrm{asc}})\,,~~\Upsilon_{\mathrm{asc}} \coloneqq \Upsilon + m T_{\mathrm{asc}}\,,~~\varphi_{\mathrm{asc}}^1 \coloneqq \varphi - \omega\,,~~\varphi_{\mathrm{asc}}^2 \coloneqq \varphi - 2\omega\,.
\end{equation*}
After integration  from $T_{\mathrm{asc}}$ to $t$ we get:
\begin{subequations} 
\begin{align*}
    \delta a^{\mathrm{S1}} &= \frac{1}{\omega_b^2}  \frac{g \Delta c}{m_n} \sqrt{2 \varrho_{\mathrm{DM}}} \varrho \sin \vartheta \left[S^a\cos \Upsilon_{\mathrm{asc}} + C^a\sin \Upsilon_{\mathrm{asc}} \right]\,,\\
    \delta \eta^{\mathrm{S1}} &= -\frac{3}{4 a \omega_b^2}  \frac{g \Delta c}{m_n}
    \sqrt{2 \varrho_{\mathrm{DM}}} \varrho \sin \vartheta \left[ \left(C_c^{\eta,\kappa} \cos \Upsilon_{\mathrm{asc}} + S_c^{\eta,\kappa} \sin \Upsilon_{\mathrm{asc}} \right) \sin \omega + \left(C_s^{\eta,\kappa} \cos \Upsilon_{\mathrm{asc}} + S_s^{\eta,\kappa} \sin \Upsilon_{\mathrm{asc}} \right) \cos \omega \right] \,, \\
    \delta \kappa^{\mathrm{S1}} &= -\frac{3}{4 a \omega_b^2}  \frac{g \Delta c}{m_n} \sqrt{2 \varrho_{\mathrm{DM}}} \varrho \sin \vartheta \left[ \left(C_c^{\eta,\kappa} \cos \Upsilon_{\mathrm{asc}} + S_c^{\eta,\kappa} \sin \Upsilon_{\mathrm{asc}} \right) \cos \omega - \left(C_s^{\eta,\kappa} \cos \Upsilon_{\mathrm{asc}} + S_s^{\eta,\kappa} \sin \Upsilon_{\mathrm{asc}} \right) \sin \omega \right] \,,
\end{align*}
\end{subequations} 
where the $\mathrm{S1}$ superscript means that these variations refer to the spin-1 model and we define:
\begin{align*}
    S^a &\coloneqq -\frac{\sin\left[(k+1)\Psi'- \varphi^1_{\mathrm{asc}}\right]+\sin \varphi^1_{\mathrm{asc}}}{k+1} + \frac{\sin\left[(k-1)\Psi'+ \varphi^1_{\mathrm{asc}}\right]-\sin \varphi^1_{\mathrm{asc}}}{k-1}\,,\\
    C^a &\coloneqq -\frac{\cos\left[(k+1)\Psi'- \varphi^1_{\mathrm{asc}}\right]-\cos \varphi^1_{\mathrm{asc}}}{k+1} + \frac{\cos\left[(k-1)\Psi'+ \varphi^1_{\mathrm{asc}}\right]-\cos \varphi^1_{\mathrm{asc}}}{k-1}\,,\\
    C_c^{\eta,\kappa} &\coloneqq A + \frac{2}{3}\cos \varphi \frac{1-\cos (k\Psi')}{k}\,,\,\,\,S_c^{\eta,\kappa} = B + \frac{2}{3}\cos \varphi \frac{\sin(k\Psi')}{k}\,,\\
     C_s^{\eta,\kappa} &\coloneqq B + \frac{2}{3}\sin \varphi \frac{1-\cos (k\Psi')}{k}\,,\,\,\,S_s^{\eta,\kappa} = -A + \frac{2}{3}\sin \varphi \frac{\sin(k\Psi')}{k}\,,\\
     A &\coloneqq \frac{\cos \varphi^2_{\mathrm{asc}} - \cos\left((k-2)\Psi' - \varphi^2_{\mathrm{asc}}\right)}{k-2} + \frac{\cos \varphi^2_{\mathrm{asc}} - \cos\left((k+2)\Psi' + \varphi^2_{\mathrm{asc}}\right)}{k+2}\,,\\
     B &\coloneqq -\frac{\sin(\varphi^2_{\mathrm{asc}} - (k-2)\Psi') - \sin \varphi^2_{\mathrm{asc}} }{k-2} + \frac{\sin((k+2)\Psi' + \varphi^2_{\mathrm{asc}} ) - \sin \varphi^2_{\mathrm{asc}} }{k+2}\,.
\end{align*}
The variation in $x$ thus becomes
\begin{equation}
    \delta x^{\mathrm{S1}} = (\delta a_1^{\mathrm{S1}}) \sin \iota = \frac{M_2 \sin \iota}{M_1 + M_2} \delta a^{\mathrm{S1}} \coloneqq p \delta a^{\mathrm{S1}} \,, 
\end{equation}
where \(p \coloneqq \frac{M_2}{M_1 + M_2} \sin \iota \) is a numerical constant whose value, for PSR~J1909-3744, is $p \approx 0.1$~\citep{Jacoby:2003, Liu:2020}. If we denote $\lambda \coloneqq g \Delta c p$, then the variations take the form:
\begin{align}
    \delta x^{\mathrm{S1}} &= \lambda \frac{\sqrt{2 \rho_\mathrm{DM}}}{m_n \omega_b^2} \varrho \sin \vartheta \left[S^a\cos \Upsilon_{\mathrm{asc}} + C^a\sin \Upsilon_{\mathrm{asc}} \right]\,, \\
    x \delta \eta^{\mathrm{S1}} &= -\frac{3}{4}\lambda \frac{\sqrt{2 \rho_\mathrm{DM}}}{m_n \omega_b^2} \varrho \sin \vartheta \left[ \left(C_c^{\eta,\kappa} \cos \Upsilon_{\mathrm{asc}} + S_c^{\eta,\kappa} \sin \Upsilon_{\mathrm{asc}} \right) \sin \omega + \left(C_s^{\eta,\kappa} \cos \Upsilon_{\mathrm{asc}} + S_s^{\eta,\kappa} \sin \Upsilon_{\mathrm{asc}} \right) \cos \omega \right]\,, \\
     x \delta \kappa^{\mathrm{S1}} &= -\frac{3}{4}\lambda \frac{\sqrt{2 \rho_\mathrm{DM}}}{m_n \omega_b^2} \varrho \sin \vartheta \left[ \left(C_c^{\eta,\kappa} \cos \Upsilon_{\mathrm{asc}} + S_c^{\eta,\kappa} \sin \Upsilon_{\mathrm{asc}} \right) \cos \omega - \left(C_s^{\eta,\kappa} \cos \Upsilon_{\mathrm{asc}} + S_s^{\eta,\kappa} \sin \Upsilon_{\mathrm{asc}} \right) \sin \omega \right]\,.
\end{align}
It is also useful to normalise  $\frac{\sqrt{2 \rho_\mathrm{DM}}}{m_n \omega_b^2}$ as
\begin{align*}
    \frac{\sqrt{2 \rho_\mathrm{DM}}}{m_n \omega_b^2} = 6339 \sqrt{\frac{\rho_{\mathrm{DM}}}{\frac{\mathrm{GeV}}{\mathrm{cm}^3}}} \left(\frac{\omega_b}{\mathrm{s}^{-1}}\right)^{-2}\,\,\mathrm{s} \,. 
\end{align*}

\subsubsection{Tensor field $M_{ij}$}

The equations of motion for the massive tensor field enforce \( M_{00} \approx V_0 M_{0i} \approx V_0^2 M_{ij} \ll M_{ij} \), such that the dominant components of \( M_{\mu\nu} \) are the spatial ones~\citep{Armaleo:2020yml}. The solutions to their equations of motion are
\begin{align}
    M_{ij} = \frac{\sqrt{2\rho_{\mathrm{DM}}}}{m} \varrho \cos(m t + \Upsilon) \epsilon_{ij}\,,
\end{align}
where $\epsilon_{ij}$ is the angular quadrupole matrix, which has unit norm, zero trace and is symmetric:
\begin{align}
    \epsilon_{ij} = \frac{1}{\sqrt{2}}\begin{pmatrix}
\epsilon_T \cos \chi - \frac{\epsilon_s}{\sqrt{3}} & \epsilon_T \sin \chi & \epsilon_V \cos \eta\\
\epsilon_T \sin \chi & -\epsilon_T \cos \chi - \frac{\epsilon_s}{\sqrt{3}} & \epsilon_V \sin \eta \\
\epsilon_V \cos \eta & \epsilon_V \sin \eta & \frac{2\epsilon_s}{\sqrt{3}}
\end{pmatrix}\,.
\end{align}
The three real parameters $\epsilon_S, \epsilon_V$ and $\epsilon_T$ satisfy $\epsilon_S^2 + \epsilon_V^2 + \epsilon_T^2 = 1$ and $\chi$ and $\eta$ are angular variables.

The interaction of the spin-2 field with ordinary matter is given by the interaction action~\citep{Schmidt:2016}
\begin{align}
    S_{\mathrm{int}} = \Lambda \int \mathrm{d}^4x \sqrt{-g} M_{\mu\nu} T^{\mu\nu}\,,
\end{align}
where $T_{\mu\nu}$ is the energy-momentum tensor of two point-like non-relativistic objects:
\begin{align}
    T^{\mu\nu} \simeq M_1 u_1^\mu u_1^\nu \delta(\boldsymbol{x}-\boldsymbol{x}_1) + M_2 u_2^\mu u_2^\nu \delta(\boldsymbol{x}-\boldsymbol{x}_2)\,.
\end{align}
The perturbing force causes the variation of semi-major axis as
\begin{align}
   \frac{1}{4 \Lambda \sqrt{\rho_{\mathrm{DM}}} \varrho } \frac{\dot{a}}{a} &= \frac{\epsilon_T}{2} \left( \frac{\omega_b}{m} - 1 \right) \sin \left[ ( m - 2 \omega_b) (t - T_{\mathrm{asc}}) + \Upsilon_{\mathrm{asc}} + \chi_{\mathrm{asc}}^2 \right] \nonumber \\
   & -\frac{\epsilon_T}{2} \left( \frac{\omega_b}{m} + 1\right) \sin \left[ (m + 2 \omega_b) (t - T_{\mathrm{asc}}) + \Upsilon_{\mathrm{asc}} - \chi_{\mathrm{asc}}^2 \right] \nonumber \\
   & -\frac{\epsilon_S}{\sqrt{3}} \sin \left[  m (t - T_{\mathrm{asc}}) + \Upsilon_{\mathrm{asc}} \right]\,,
\end{align}
where we define $\chi_{\mathrm{asc}}^j \coloneqq \chi + j \omega$~\citep{Armaleo:2020yml}. After the integration we get
\begin{align}
\label{deltaa_spin2}
    \frac{\omega_b}{4 \Lambda \sqrt{\rho_{\mathrm{DM}}} \varrho} \frac{\delta a^{\mathrm{S2}}}{a}  = & +\frac{\epsilon_T}{2}\left( \frac{1}{k} - 1 \right) \frac{\cos(\Upsilon_{\mathrm{asc}} + \chi_{\mathrm{asc}}^2) - \cos\left[(k-2)\Psi' + \Upsilon_{\mathrm{asc}} + \chi_{\mathrm{asc}}^2\right]}{k-2} \nonumber \nonumber \\
    & -\frac{\epsilon_T}{2}\left( \frac{1}{k} + 1 \right) \frac{\cos(\Upsilon_{\mathrm{asc}} - \chi_{\mathrm{asc}}^2) - \cos\left[(k+2)\Psi' + \Upsilon_{\mathrm{asc}} - \chi_{\mathrm{asc}}^2\right]}{k+2} \nonumber \\
    & -\frac{\epsilon_S}{\sqrt{3}} \frac{\cos \Upsilon_{\mathrm{asc}} - \cos\left( k \Psi' + \Upsilon_{\mathrm{asc}} \right)}{k}\,,
\end{align}
where the $\mathrm{S2}$ superscript means that these variations refer to the spin-2 model. The equations for the Laplace-Lagrange parameters are $\dot{\eta} = \dot{e}\sin \omega + e\dot{\omega} \cos \omega$ and 
$\dot{\kappa} = \dot{e} \cos \omega - e \dot{\omega} \sin \omega$, where $\dot{e}$ and $e \dot{\omega}$ are given by Eq.~(\ref{eq:dote_spin2}) and (\ref{eq:edotomega_spin2}), thus:
\begin{align}
\label{eq:dote_spin2}
    \frac{\dot{e}}{2 \Lambda \sqrt{\rho_{\mathrm{DM}}} \varrho } = & -\frac{3}{4} \frac{\omega_b}{m} \epsilon_T \sin\left[(m+\omega_b)(t-T_{\mathrm{asc}}) + \Upsilon_{\mathrm{asc}} - \chi_{\mathrm{asc}}^1\right] \nonumber \\
    & +\frac{3}{4} \frac{\omega_b}{m} \epsilon_T \sin\left[(m-\omega_b)(t-T_{\mathrm{asc}}) + \Upsilon_{\mathrm{asc}} + \chi_{\mathrm{asc}}^1\right] \nonumber \\
    & - \left[ \frac{1}{\sqrt{3}} \left( \frac{1}{2}\frac{\omega_b}{m} + 1 \right) \epsilon_S + \frac{1}{4}\epsilon_T \right] \sin\left[(m+\omega_b)(t-T_{\mathrm{asc}}) + \Upsilon_{\mathrm{asc}} - \omega\right] \nonumber \\
    & + \left[ \frac{1}{\sqrt{3}} \left( \frac{1}{2}\frac{\omega_b}{m} - 1 \right) \epsilon_S - \frac{1}{4}\epsilon_T \right] \sin\left[(m-\omega_b)(t-T_{\mathrm{asc}}) + \Upsilon_{\mathrm{asc}} + \omega\right] \nonumber \\
     & - \frac{1}{2}\epsilon_T  \sin\left[(m+2\omega_b)(t-T_{\mathrm{asc}}) + \Upsilon_{\mathrm{asc}} - \chi_{\mathrm{asc}}^2\right] \nonumber \\
     & - \frac{1}{2}\epsilon_T  \sin\left[(m-2\omega_b)(t-T_{\mathrm{asc}}) + \Upsilon_{\mathrm{asc}} + \chi_{\mathrm{asc}}^2\right] \nonumber \\
     & - \frac{1}{4} \frac{\omega_b}{m} \epsilon_T  \sin\left[(m+3\omega_b)(t-T_{\mathrm{asc}}) + \Upsilon_{\mathrm{asc}} - \chi_{\mathrm{asc}}^3\right] \nonumber \\
     & + \frac{1}{4} \frac{\omega_b}{m} \epsilon_T  \sin\left[(m-3\omega_b)(t-T_{\mathrm{asc}}) + \Upsilon_{\mathrm{asc}} + \chi_{\mathrm{asc}}^3\right] \nonumber \\
     & - \frac{1}{4} \epsilon_T  \sin\left[(m+3\omega_b)(t-T_{\mathrm{asc}}) + \Upsilon_{\mathrm{asc}} - 3\omega\right] \nonumber \\
      & - \frac{1}{4} \epsilon_T  \sin\left[(m-3\omega_b)(t-T_{\mathrm{asc}}) + \Upsilon_{\mathrm{asc}} + 3\omega\right]\,, 
\end{align}
\begin{align}
\label{eq:edotomega_spin2}
 \frac{e\dot{\omega}}{2 \Lambda \sqrt{\rho_{\mathrm{DM}}} \varrho } = & -\epsilon_T\left(\frac{\omega_b}{m} + 1 \right) \cos\left[(m+\omega_b)(t-T_{\mathrm{asc}}) + \Upsilon_{\mathrm{asc}} - \chi_{\mathrm{asc}}^1\right] \nonumber \nonumber \\
 & -\epsilon_T\left(\frac{\omega_b}{m} - 1  \right) \cos\left[(m-\omega_b)(t-T_{\mathrm{asc}}) + \Upsilon_{\mathrm{asc}} + \chi_{\mathrm{asc}}^1\right] \nonumber \\
 & + \frac{1}{\sqrt{3}} \epsilon_S \left(\frac{1}{2}\frac{\omega_b}{m} + 1\right) \cos\left[ (m+\omega_b)(t-T_{\mathrm{asc}}) + \Upsilon_{\mathrm{asc}} - \omega \right] \nonumber \\
 & + \frac{1}{\sqrt{3}} \epsilon_S \left(\frac{1}{2}\frac{\omega_b}{m} - 1\right) \cos\left[ (m-\omega_b)(t-T_{\mathrm{asc}}) + \Upsilon_{\mathrm{asc}} + \omega \right]\,. 
\end{align}
\begin{align}
\label{eq:e_spin2}
    \frac{\omega_b \delta e^{\mathrm{S2}}} {2 \Lambda \sqrt{\rho_{\mathrm{DM}}} \varrho } = & -\frac{3}{4} \frac{1}{k} \epsilon_T \frac{\cos(\Upsilon_{\mathrm{asc}} - \chi_{\mathrm{asc}}^1) - \cos\left[(k+1)\Psi' + \Upsilon_{\mathrm{asc}} - \chi_{\mathrm{asc}}^1\right]}{k+1} \nonumber \\
    & +\frac{3}{4} \frac{1}{k} \epsilon_T \frac{\cos(\Upsilon_{\mathrm{asc}} + \chi_{\mathrm{asc}}^1) - \cos\left[(k-1)\Psi' + \Upsilon_{\mathrm{asc}} + \chi_{\mathrm{asc}}^1\right]}{k-1} \nonumber \\
    & - \left[ \frac{1}{\sqrt{3}} \left( \frac{1}{2}\frac{1}{k} + 1 \right) \epsilon_S + \frac{1}{4}\epsilon_T \right] \frac{\cos\left(\Upsilon_{\mathrm{asc}} - \omega\right) - \cos\left[(k+1)\Psi' + \Upsilon_{\mathrm{asc}} - \omega\right]}{k+1} \nonumber \\
    & + \left[ \frac{1}{\sqrt{3}} \left( \frac{1}{2}\frac{1}{k} - 1 \right) \epsilon_S - \frac{1}{4}\epsilon_T \right] \frac{\cos\left(\Upsilon_{\mathrm{asc}} + \omega\right) - \cos\left[(k-1)\Psi' + \Upsilon_{\mathrm{asc}} + \omega\right]}{k-1} \nonumber \\
     & - \frac{1}{2}\epsilon_T  \frac{\cos\left( \Upsilon_{\mathrm{asc}} - \chi_{\mathrm{asc}}^2\right) - \cos\left[(k+2)\Psi' + \Upsilon_{\mathrm{asc}} - \chi_{\mathrm{asc}}^2\right]}{k+2} \nonumber \\
     & - \frac{1}{2}\epsilon_T  \frac{\cos\left( \Upsilon_{\mathrm{asc}} + \chi_{\mathrm{asc}}^2\right) - \cos\left[(k-2)\Psi' + \Upsilon_{\mathrm{asc}} + \chi_{\mathrm{asc}}^2\right]}{k-2} \nonumber \\
     & - \frac{1}{4}\epsilon_T \frac{1}{k} \frac{\cos\left( \Upsilon_{\mathrm{asc}} - \chi_{\mathrm{asc}}^3\right) - \cos\left[(k+3)\Psi' + \Upsilon_{\mathrm{asc}} - \chi_{\mathrm{asc}}^3\right]}{k+3} \nonumber \\
     & + \frac{1}{4}\epsilon_T \frac{1}{k} \frac{\cos\left( \Upsilon_{\mathrm{asc}} + \chi_{\mathrm{asc}}^3\right) - \cos\left[(k-3)\Psi' + \Upsilon_{\mathrm{asc}} + \chi_{\mathrm{asc}}^3\right]}{k-3} \nonumber \\
      & - \frac{1}{4}\epsilon_T \frac{\cos\left( \Upsilon_{\mathrm{asc}} - 3\omega\right) - \cos\left[(k+3)\Psi' + \Upsilon_{\mathrm{asc}} - 3\omega\right]}{k+3} \nonumber \\
      & - \frac{1}{4}\epsilon_T \frac{\cos\left( \Upsilon_{\mathrm{asc}} + 3\omega\right) - \cos\left[(k-3)\Psi' + \Upsilon_{\mathrm{asc}} + 3\omega\right]}{k-3} \,.
\end{align}
\begin{align}
\label{eq:eomega_spin2}
 \frac{\omega_b e \delta\omega^{\mathrm{S2}}}{2 \Lambda \sqrt{\rho_{\mathrm{DM}}} \varrho } = & -\epsilon_T\left(\frac{1}{k} + 1 \right) \frac{ \sin\left[(k+1) \Psi' + \Upsilon_{\mathrm{asc}} - \chi_{\mathrm{asc}}^1\right] - \sin(\Upsilon_{\mathrm{asc}} - \chi_{\mathrm{asc}}^1)  }{ k + 1 } \nonumber \\
 & -\epsilon_T\left(\frac{1}{k} - 1 \right) \frac{ \sin\left[(k-1) \Psi' + \Upsilon_{\mathrm{asc}} + \chi_{\mathrm{asc}}^1\right] - \sin(\Upsilon_{\mathrm{asc}} + \chi_{\mathrm{asc}}^1)  }{ k - 1 } \nonumber \\
 & + \frac{1}{\sqrt{3}} \epsilon_S \left(\frac{1}{2}\frac{1}{k} + 1\right) \frac{ \sin\left[ (k+1)\Psi' + \Upsilon_{\mathrm{asc}} - \omega \right] - \sin(\Upsilon_{\mathrm{asc}} - \omega) }{ k + 1} \nonumber \\
 & + \frac{1}{\sqrt{3}} \epsilon_S \left(\frac{1}{2}\frac{1}{k} - 1\right) \frac{ \sin\left[ (k-1)\Psi' + \Upsilon_{\mathrm{asc}} + \omega \right] - \sin(\Upsilon_{\mathrm{asc}} + \omega) }{ k - 1}\,. 
\end{align}
The integration of the Laplace-Lagrange parameters gives
$$
\delta \eta^{\mathrm{S2}} = (\delta e^{\mathrm{S2}}) \sin \omega + (e \delta \omega^{\mathrm{S2}}) \cos \omega\,,\,\, \delta \kappa^{\mathrm{S2}} = (\delta e^{\mathrm{S2}}) \cos \omega - (e \delta \omega^{\mathrm{S2}}) \sin \omega\,,
$$
where $\delta e^{\mathrm{S2}}$ and $e \delta \omega^{\mathrm{S2}}$ are given by Eq.~(\ref{eq:e_spin2}) and (\ref{eq:eomega_spin2}), respectively. We again consider only the contribution of $\delta a^{\mathrm{S2}}$ when computing $\delta x^{\mathrm{S2}}$, i.e.~$\delta x^{\mathrm{S2}} / x = \delta a^{\mathrm{S2}} / a$. It is useful to denote the right hand sides of Eqs.~(\ref{deltaa_spin2}), (\ref{eq:e_spin2}) and~(\ref{eq:eomega_spin2}) as $T_k^a$, $T_k^e$ and $T_k^\omega$, respectively. Then the variations caused by ULDM become:
\begin{align}
    \delta x^{\mathrm{S2}} &= 4 \frac{\Lambda \sqrt{2 \rho_\mathrm{DM}}}{\omega_b} x \varrho T_k^a\,, \\
    \delta e^{\mathrm{S2}} &= 2 \frac{\Lambda \sqrt{2 \rho_\mathrm{DM}}}{\omega_b}  \varrho T_k^e\,, \\
    e \delta \omega^{\mathrm{S2}} &= 2 \frac{\Lambda \sqrt{2 \rho_\mathrm{DM}}}{\omega_b}  \varrho T_k^\omega\,, 
\end{align}
where we can normalise the amplitude as
\begin{align*}
    \frac{\Lambda \sqrt{2 \rho_\mathrm{DM}}}{\omega_b} = 8.77\times10^{-16} \lambda \sqrt{\frac{\rho_{\mathrm{DM}}}{\frac{\mathrm{GeV}}{\mathrm{cm}^3}}} \left(\frac{\omega_b}{\mathrm{s}^{-1}}\right)^{-1}\,, 
\end{align*}
and we have introduced a new dimensionless coupling constant $\lambda$, $\Lambda \coloneqq \lambda / (2 M_P)$, where $M_P \approx 2.4\times10^{18}$ $\mathrm{GeV}$ is the reduced Planck mass. 

\section{Machine learning}
\label{sec:ml}

\subsection{Deep neural networks}

Machine learning (ML) is a subfield of artificial intelligence that focuses on building systems capable of recognising and learning patterns from data. Unlike traditional programming, where explicit instructions are provided, ML algorithms aim to uncover data structures autonomously from the data provided: \textit{the machine is learning}. A key feature of ML is its ability to generalise beyond the data it was trained on. Once trained and tested on previously unseen data, the model can be used to make predictions or decisions on new datasets.

Neural networks are a type of specialised machine learning algorithms inspired by the structure of the human brain. The fundamental unit of a neural network is the artificial neuron, which connects to other neurons through edges, analogous to synapses between biological neurons. These neurons are organised into layers: input layer receives data, hidden layers process it and output layer generates results. Data transfer between layers involves linear transformations followed by activation functions, which introduce non-linearities to capture complex patterns. Neural network architectures can vary widely, often including specialised layers, such as convolutional layers, for specific tasks. When a neural network has more than two hidden layers, it is referred to as a deep neural network (DNN)~\citep{Erdmann:2021}.

Neural networks have found applications across various domains, including natural language processing, computer vision and speech recognition. Recently, researchers have begun exploring these techniques to detect GW signals and estimate the parameters of their astrophysical sources (e.g.~\citet{Baltus:2021, Baltus:2022}). This development is particularly exciting because the search for ULDM in pulsar timing data shares similarities with GW searches, hence similar techniques can be employed.


Our goal is to enhance the application of ML algorithms in the search for ULDM signals within time-series data. To achieve this, we explore three independent ML-based classifiers designed to detect ULDM signals in noisy time series: an \textit{anomaly detector}, a \textit{binary classifier} and a \textit{multiclass classifier}. Our approach leverages two specialised neural network architectures---an \textit{autoencoder} and a \textit{convolutional neural network} (CNN). We aim to demonstrate that these networks are capable to effectively detect ULDM signals and classify them.

\paragraph{Anomaly detector } It is based on the autoencoder architecture. It is trained on simulated pulsar timing residuals without ULDM signals, referred to as the \textit{normal dataset}, and is capable to identify irregularities in a new dataset containing ULDM signals, referred to as the \textit{anomalous dataset}. We use the autoencoder to estimate the minimum signal strength required for a time series to be classified as anomalous. This is an example of unsupervised learning.

\paragraph{Convolutional Classifier } We employ a CNN that functions as either a binary classifier or a multiclass classifier. The binary classifier can distinguish between pure noise and a single type of ULDM signal mixed with noise, or between pure noise and any type of ULDM signals combined with noise. However, it cannot differentiate between individual ULDM signals. In contrast, the multiclass classifier can identify various ULDM models while distinguishing them from pure noise. In all settings, the neural network can be used to estimate the threshold value of the coupling constants required for detection and even differentiate between various spins and coupling types.\\

In the following sections we delineate concrete DNNs and analyse their performance, compare their sensitivity to ULDM and to the Bayesian sensitivity estimates given in~\citet{Kus:2024}.

\subsection{ULDM parameter-space sampling}

ML is a data-driven approach. For DNNs to successfully learn data structures, a large amount of data must be generated. Regardless of which DNN we use, the process of data generation is always the same. Here, we describe how we sample from the phase space of parameters associated with ULDM.

In all models, two parameters are always present: the phase $\Upsilon$ and the interference factor $\varrho$. The phase $\Upsilon$ is sampled uniformly from the interval $[0, 2\pi)$, while $\varrho$ follows a Rayleigh distribution $P(\varrho) = 2 \varrho \exp(-\varrho^2)$, where $\varrho \in [0,+\infty)$. This is all we need for ULDM models with spin-0.

\begin{figure}[htbp]
	\centering
	\rotatebox{0}{\includegraphics[width=0.5\linewidth]{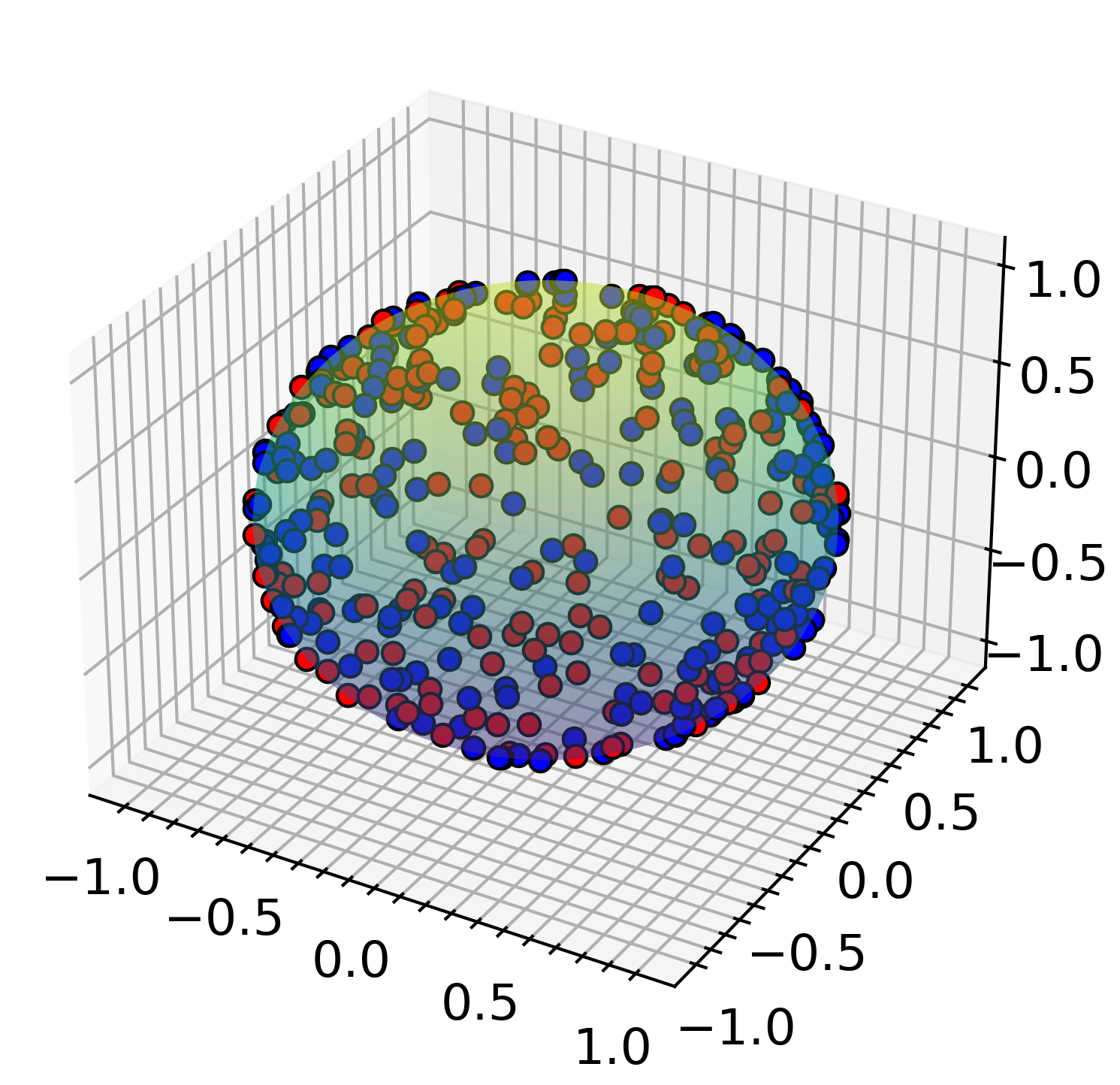}}
	\caption{Distribution on the sphere of 200 red (blue) points when using the Gaussian (direct) sampling approach.}
	\label{fig:sphere_plot}
\end{figure}

For spin-1 ULDM, two additional angles, \(\phi\) and \(\theta\), describe the orientation of the vector field. We aim to generate data in such a way as to ensure uniform coverage of the sphere, whose points are parameterised by these two angles. We can employ one of two approaches:
\begin{enumerate}
    \item Gaussian sampling approach: we randomly sample three values, \(X\), \(Y\) and \(Z\), from a Gaussian distribution \({\cal N}(0, 1)\). The radius is given by \(R^2 = X^2 + Y^2 + Z^2\) and the coordinates are normalised as \(X \rightarrow X/R\), \(Y \rightarrow Y/R\), \(Z \rightarrow Z/R\). Finally, the spherical angles are derived as \(\phi = \arctan(Y/X)\) and \(\theta = \arccos Z\).
    \item Direct sampling approach: we sample \(\phi\) uniformly from \([0, 2\pi)\) and a helper variable \(U\) uniformly from \([-1, 1]\). The polar angle \(\theta\) is then obtained as \(\theta = \arccos U\) which ensures uniform coverage of the sphere.
\end{enumerate}
Both processes generate points uniformly distributed over the sphere, as shown in Fig.~\ref{fig:sphere_plot}.

For the spin-2 ULDM field, the configuration is parametrised by three real numbers, $\epsilon_S$, $\epsilon_V$ and $\epsilon_T$, which satisfy the normalisation condition, along with two angular variables $(\chi, \eta)$. Although this gives five parameters in total, only four are independent, meaning the full parameter space corresponds to the sphere $S^4$. Alternatively, we can describe this space using four angular coordinates, $(\zeta, \beta, \xi, \psi)$. In terms of these angular coordinates, the embedding coordinates on $S^4$ are given by
\begin{align}
    X_1 &= \sin \zeta \cos \beta \sin \psi\,, & X_4 &= \sin \zeta \cos \beta \cos \psi\,, \\
    X_2 &= \sin \zeta \sin \beta \sin \xi\,, & X_5 &= \sin \zeta \sin \beta \cos \xi\,, \\
    X_3 &= \cos \zeta\,.
\end{align}
The map between the two parametrisations reads~\citep{Maggiore:2007, Armaleo:2020yml}
\begin{align}
    X_1 &= \epsilon_T \sin \chi\,, & X_4 &= \epsilon_T \cos \chi\,, \\
    X_2 &= \epsilon_V \sin \eta\,, & X_5 &= \epsilon_V \cos \eta\,, \\
    X_3 &= \epsilon_S\,.
\end{align}
To uniformly sample points on $S^4$ we draw five independent random variables $x_1, \dots, x_5$ from a standard normal distribution ${\cal N}(0,1)$. We then normalise the resulting vector by computing the radius \(R^2 = X_1^2 + X_2^2 + X_3^2 + X_4^2 + X_5^2 \) and normalise \(X_p \rightarrow x_p/R\) for \(p = 1, \dots, 5\). From these normalised coordinates we can extract the parameters $\epsilon_S$, $\epsilon_V$, $\epsilon_T$, $\chi$ and $\eta$.

\subsection{ULDM signal strength, \(S\)}

To quantify the strength of the ULDM signal we introduce a dimensionless quantity \(S\), defined as
\begin{align}
	S \coloneqq \frac{\sqrt{\left(\vec{R}^{\mathrm{binary\,DM}}\right)^2}}{\epsilon}\,,
\end{align}
where the \(O(\lambda)\) time-series vector
$$
R_a^{\mathrm{binary\,DM}} = \sin (\Psi'_a) \delta x_a^{\mathrm{DM}} - x\frac{\cos(2\Psi'_a) + 3}{2} \delta \eta_a^{\mathrm{DM}} +  x\frac{\sin(2\Psi'_a)}{2} \delta \kappa_a^{\mathrm{DM}}\,,
$$
represents the contribution of the ULDM signal (here the DM superscript refers to one of the four ULDM models we consider: $\mathrm{S0L}$, $\mathrm{S0Q}$, $\mathrm{S1}$ and $\mathrm{S2}$) to the pulsar time residuals and \(\epsilon\) is the amplitude of the intrinsic timing noise of the pulsar, introduced to make \(S\) dimensionless. We define the squared magnitude of the time-series vector $\vec{R}^{\mathrm{binary,DM}}$ as $\left(\vec{R}^{\mathrm{binary,DM}}\right)^2 \coloneqq \sum_{a = 1}^{N} \left(R_a^{\mathrm{binary,DM}}\right)^2$.

\subsection{Estimating \( \lambda \): Machine learning vs Bayesian approach}

Since \(\vec{R}^{\mathrm{binary\,DM}} = \lambda \vec{R}^{\mathrm{binary\,DM}}(\lambda = 1)\), we can estimate the sensitivity limit to $\lambda$ assuming a critical value of $S$ for detection $S_c$:
\begin{align}
	\label{eq:ML_lambda_estimation_formula}
	|\lambda_c|_{\mathrm{ML}} = \frac{S_c~\epsilon}{\sqrt{\left(\vec{R}^{\mathrm{binary\,DM}}(\lambda = 1)\right)^2}}\,.
\end{align}
The goal is to use the ML methods to determine the value of $S_c$ needed ti detect the ULDM signal.

For the spin-0 linear coupling model the critical value \(\lambda_c\) can be directly compared to the Bayesian sensitivity estimate presented in~\citet{Kus:2024}, which employs the two-step approach detailed therein and takes into account white (but not red) Gaussian noise and nuisance effects. However, this method has been developed specifically for this spin and coupling case, and extending it to other models poses a significant technical challenges because of the larger number of parameters needed to describe spin-1 and spin-2 ULDM. Instead, here we adopt a simplified one-step approach, described in the Appendix of~\citet{Kus:2024}, for which we assume delta-function priors. This method provides an estimate of the Bayesian sensitivity limit for arbitrary spin and coupling scenarios. Moreover, although it neglects nuisance effects and assumes only white Gaussian noise, it achieves an accuracy comparable to the two-step approach:
\begin{align}
	\label{eq:Bayes_lambda_estimation_formula}
	|\lambda_c|_{\mathrm{Bayes}} = \frac{\sqrt{\ln\mathcal{B}}~\epsilon}{\sqrt{\left(\vec{R}^{\mathrm{binary\,DM}}(\lambda = 1)\right)^2}}\,,
\end{align}
where \(\mathcal{B}\) is the Bayes factor, chosen to be 1000, yielding \(\sqrt{\ln\mathcal{B}} \simeq 2.63\). Both equations (\ref{eq:ML_lambda_estimation_formula}) and (\ref{eq:Bayes_lambda_estimation_formula}) have the same form, differing only in the multiplicative factor: \(S_c\) for the ML approach and \(\sqrt{\ln\mathcal{B}}\) for the Bayesian approach. The two approaches can be therefore directly contrasted by comparing these factors.

\subsection{Nuisance effects}
\label{subsec:nuisance_effects}

\begin{figure}[htbp]
	\centering
	\rotatebox{0}{\includegraphics[width = 0.7\linewidth]{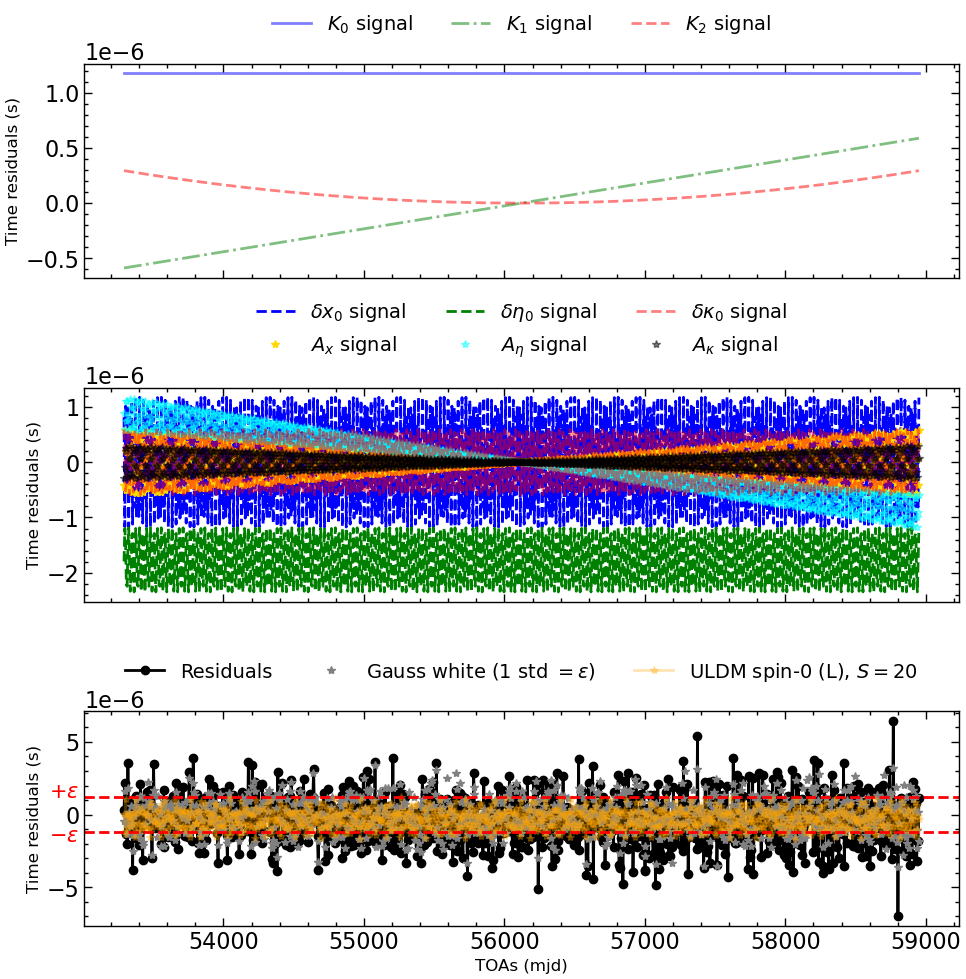}}
	\caption{{\itshape Top panel:} Example time series with maximal nuisance parameters \(K_0\) (blue solid line), \(K_1\) (green dot-dashed) and \(K_2\) (red dashed). {\itshape Middle panel:} Example time series with maximal nuisance parameters \(\delta x_0\) (blue dashed like), \(\delta \eta_0\) (green dashed), \(\delta \kappa_0\) (red dashed), \(A_x\) (yellow stars), \(A_\eta\) (teal) and \(A_\kappa\) (black). {\itshape Bottom panel:} Example time series for injected spin-0 linear ULDM signal at \(S=20\) (yellow) with Gaussian white noise (grey); the residuals are displayed in black.}
	\label{fig:time_residual_example}
\end{figure}

Time residuals may include contributions from nuisance effects. In this work, the nuisance parameters are sampled uniformly for each time series realisation from the following ranges:
\begin{align*}
	K_0\,,\,\delta x_0 & \in [-\epsilon,\epsilon]\,;\,\,\,\, K_1\,,\,A_x \in [-\epsilon,\epsilon] \times \frac{1}{T_{\mathrm{max}} - T_{\mathrm{min}}}\,;\,\,\,\, K_2 \in [-\epsilon,\epsilon] \times \frac{1}{ (T_{\mathrm{max}} - T_{\mathrm{min}} )^2}\,;   \\
	\delta \eta_0\,,\,\delta \kappa_0 & \in [-\epsilon,\epsilon] \times \frac{1}{x}\,;\,\,\,\, A_\eta\,,\,A_\kappa \in [-\epsilon,\epsilon] \times \frac{1}{x} \times \frac{1}{T_{\mathrm{max}} - T_{\mathrm{min}}}\,.
\end{align*}
Fig.~\ref{fig:time_residual_example} illustrates the contributions of each signal type and how they combine to form complex time residuals. Specifically, the top and middle panels show the contributions of individual nuisance effects when their parameters are set to their maximum values, i.e.~\( K_0 = \epsilon \) and \( K_1 = \epsilon / (T_{\mathrm{max}} - T_{\mathrm{min}}) \). The bottom panel displays the resulting time residuals obtained by summing the signals from the top and middle panels (nuisance effects), a spin-0 linear coupling ULDM signal with strength \( S = 20 \) and added white Gaussian noise \( {\cal N}(0, \epsilon) \).

\subsection{Noise modelling}

We investigate the ability of ML methods to detect ULDM signals in three distinct noise environments labelled A, B and C. Noise of type A consists purely of white Gaussian noise. Type B includes both white Gaussian noise and all nuisance parameter contributions discussed previously. Finally, type C builds upon type B by adding an additional red noise component. An overview of the noise configurations is presented in Table~\ref{tab:noise_models}.

\begin{table}[htbp]
\centering
\begin{tabular}{|c||c|c|c|c|c|c|c|c|c|c|}
\hline
\textbf{Noise type} & \textbf{Gauss} & $K_0$ & $K_1$ & $K_2$ & $\delta x_0$ & $\delta \eta_0$ & $\delta \kappa_0$ & $A_x$ & $A_{\eta}$ & $A_{\kappa}$ \\
\hline
A & White & $\times$ & $\times$ & $\times$ & $\times$ & $\times$ & $\times$ & $\times$ & $\times$ & $\times$ \\
\hline
B & White & \checkmark & \checkmark & \checkmark & \checkmark & \checkmark & \checkmark & \checkmark & \checkmark & \checkmark \\
\hline
C & White and red & \checkmark & \checkmark & \checkmark & \checkmark & \checkmark & \checkmark & \checkmark & \checkmark & \checkmark \\
\hline
\end{tabular}
\caption{Parameters in the three noise models.}
\label{tab:noise_models}
\end{table}

\subsection{Computational Environment}

All computations were performed in a Jupyter notebook (Python 3.10) using \texttt{TensorFlow}. The codes ran on CPU only (Intel Core i5-7200U at 2.50\,GHz, 2~cores, 4~threads) with 8~GB RAM. No GPU acceleration was used. Training times ranged from a few seconds to several dozen minutes, depending on the dataset size, learning strategy and specific task. 

\subsection{Autoencoder}

\begin{figure}[htbp]
    \centering
    \includegraphics[height = 5cm]{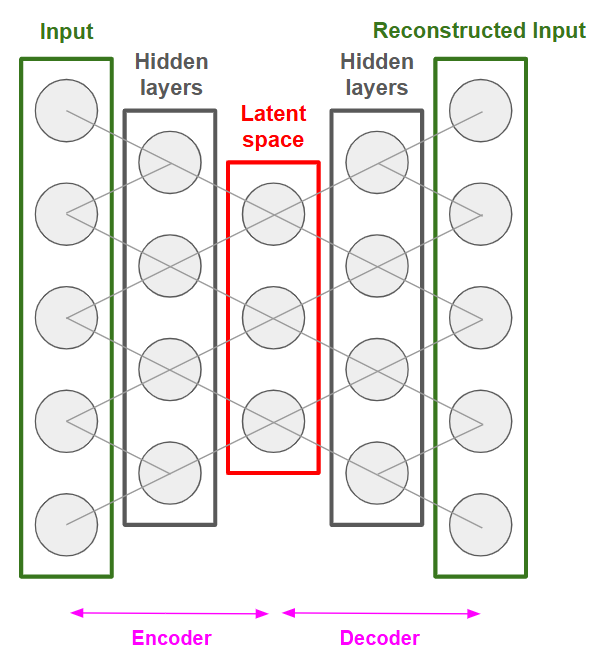}
    \hspace{0.5cm}%
    \includegraphics[height = 5cm]{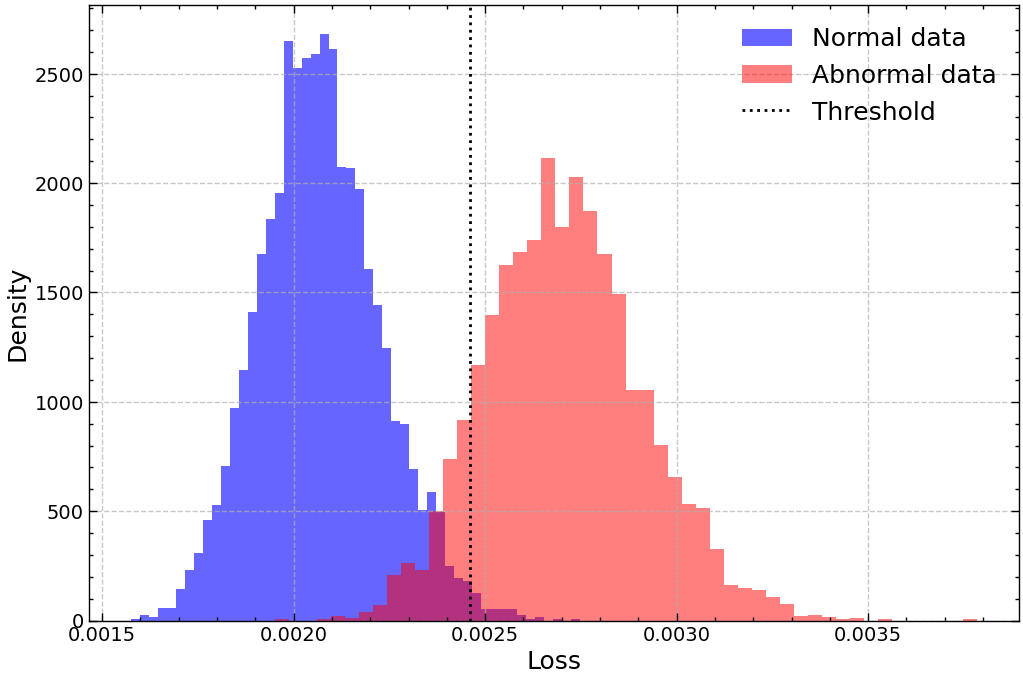}
    \caption{{\itshape Left:} Schematic illustration of an autoencoder. The encoder compresses the input into a lower-dimensional representation, which is then reconstructed by the decoder to match the original input. {\itshape Right:} Distribution of reconstruction errors for a normal dataset---i.e.~a dataset with the same statistical properties as the one used during training (blue histogram) and an abnormal dataset containing novel patterns (ULDM signals), which results in significantly higher reconstruction errors (red histogram).}
    \label{fig:autoencoder_scheme_and_reconstruction}
\end{figure}

An autoencoder is a neural network designed for unsupervised learning tasks, such as dimensionality reduction, feature extraction, noise reduction, image restoration and anomaly detection.\footnote{For a hands-on introduction we refer to the official \texttt{TensorFlow} tutorial: \url{https://www.tensorflow.org/tutorials/generative/autoencoder}} It compresses input data into a lower-dimensional representation, known as the latent space, representing extracted features, and then reconstructs the original data from this compressed form, as illustrated in Fig.~\ref{fig:autoencoder_scheme_and_reconstruction}, left panel. The objective is to generate an output that closely matches the input by minimising the reconstruction error, achieved by learning the underlying patterns in the data. For more details about the autoencoder we recommend~\citet{Erdmann:2021}.

When used as an anomaly detector, the autoencoder is trained exclusively on \textit{normal data}, learning to minimise reconstruction error for this concrete data type. In our case, the normal data comprises a time series of Gaussian noise combined with nuisance signals but without any ULDM signals, i.e.~\(S = 0\).

The autoencoder struggles to reconstruct inputs with unfamiliar patterns, such as ULDM signals present in a new dataset. This results in a significant increase in reconstruction error, as illustrated in Fig.~\ref{fig:autoencoder_scheme_and_reconstruction}, right panel. By monitoring these errors, the autoencoder can effectively detect anomalies. A time series is flagged as an anomaly if its reconstruction error exceeds a predefined threshold. We refer to the dataset containing anomalous time series as \textit{abnormal}.

\subsubsection{Architecture}

In this work, we utilise a convolutional autoencoder (CAE, or simply ``autoencoder'' in what follows), a type of autoencoder that incorporates convolutional layers into its architecture. The specific structure of our CAE is detailed in Table~\ref{tab:CAE_structure}. This design was chosen after limited hyperparameter tuning, prioritising simplicity and fast learning. While alternative architectures may yield better reconstruction accuracy, our model strikes an optimal balance for our purposes.

\begin{table}[htbp]
\centering
\begin{tabular}{|l|c|}
\hline
\textbf{Layers} & \textbf{Output shape} \\
\hline
\hline
\# Encoder &  \\
\hline
\hline
InputLayer(input\_shape=(1024, 1))  & (1024, 1) \\
\hline
Conv1D(32, kernel\_size=5, padding="same") & (1024, 32) \\
\hline
LeakyReLU(0.1) & (1024, 32) \\
\hline
MaxPooling1D(pool\_size=2, padding="same") & (512, 32) \\
\hline
Conv1D(16, kernel\_size=5, padding="same") & (512, 16) \\
\hline
LeakyReLU(0.1) & (512, 16) \\
\hline
MaxPooling1D(pool\_size=2, padding="same") & (256, 16) \\
\hline
Conv1D(8, kernel\_size=3, padding="same") & (256, 8) \\
\hline
LeakyReLU(0.1) & (256, 8) \\
\hline
MaxPooling1D(pool\_size=2, padding="same") & (128, 8) \\
\hline
Conv1D(4, kernel\_size=3, padding="same") & (128, 4) \\
\hline
LeakyReLU(0.1) & (128, 4) \\
\hline
MaxPooling1D(pool\_size=2, padding="same") & (64, 4) \\
\hline
\hline
\# Decoder & \\
\hline
\hline
Conv1DTranspose(4, kernel\_size=3, strides=2, padding="same") & (128, 4) \\
\hline
LeakyReLU(0.1) & (128, 4) \\
\hline
Conv1DTranspose(8, kernel\_size=3, strides=2, padding="same") & (256, 8) \\
\hline
LeakyReLU(0.1) & (256, 8) \\
\hline
Conv1DTranspose(16, kernel\_size=5, strides=2, padding="same") & (512, 16) \\
\hline
LeakyReLU(0.1) & (512, 16) \\
\hline
Conv1DTranspose(32, kernel\_size=5, strides=2, padding="same") & (1024, 32) \\
\hline
LeakyReLU(0.1) & (1024, 32) \\
\hline
Conv1D(1, kernel\_size=3, activation="sigmoid", padding="same") & (1024, 1) \\
\hline
\end{tabular}

\caption{
Architecture of the 1D convolutional autoencoder implemented in~\href{https://www.tensorflow.org}{\textcolor{blue}{TensorFlow}}, with an input size of \( \mathrm{N} = 1024 \). The model consists of 1D convolutional layers, 1D max pooling layers, 1D transposed convolutional layers and Leaky ReLU activations with a negative slope of 0.1. 
The final layer uses a sigmoid activation. The model is trained using the Mean Squared Error (MSE) loss function and optimised with the Adam optimizer with an initial learning rate of 0.001, which decays exponentially.
}
\label{tab:CAE_structure}

\end{table}

In our architecture design we choose the latent space to remain tensorial and the number of filters decreases with depth. While this is less common approach, it is technically valid. Our goal is to obtain a more compact tensorial latent space that emphasises the most relevant features. For comparison, we also evaluate an alternative architecture with the number of filters increasing with depth, but we do not find significant differences, see Section~\ref{sec:end}.

\subsubsection{Training}

To train an autoencoder, we generate \(10^6\) normal time series, reserving 20\% of the dataset for validation. Before training, we normalise the dataset using the \texttt{scikit-learn} package, in particular the MinMax scheme (more on other schemes in Section~\ref{sec:end}), while the actual training is performed with the \texttt{TensorFlow} package. 

Specifically, we consider three variations of the training dataset that map the three noise types: (A) pure white Gaussian noise with standard deviation \(\epsilon\); (B) white Gaussian noise with the same standard deviation, supplemented by nuisance effects; (C) an equal-weight combination of white and red Gaussian noise, adjusted to preserve the standard deviation \(\epsilon\), along with nuisance effects. For each noise model we generate two datasets, resulting in a total of six datasets (A1, A2, B1, B2, C1, C2). We then train a separate autoencoder for each dataset, giving six autoencoders in total. This enables an exploration of how nuisance effects and red noise components influence the sensitivity limit beyond white Gaussian noise. Two realisations of each noise model help illustrate the effects of randomness.

\begin{figure}[htbp]
    \centering
    \includegraphics[width=0.48\linewidth]{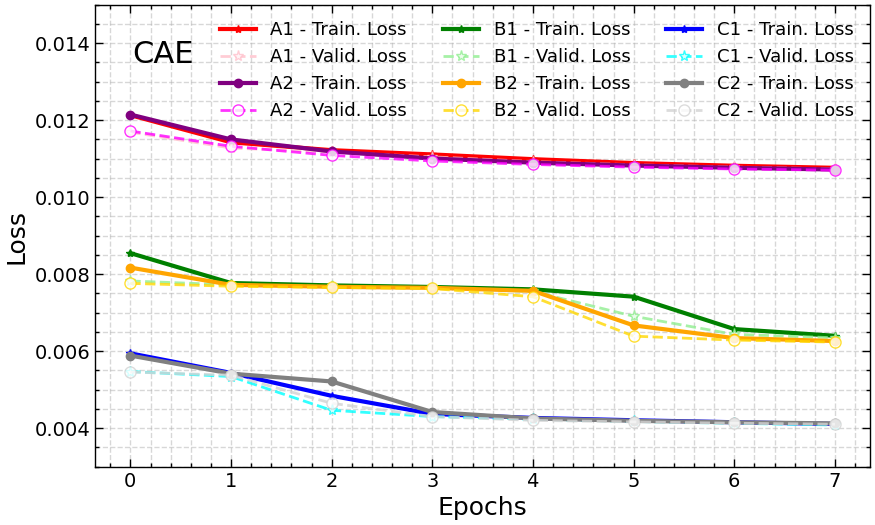}
    \hfill
    \includegraphics[width=0.48\linewidth]{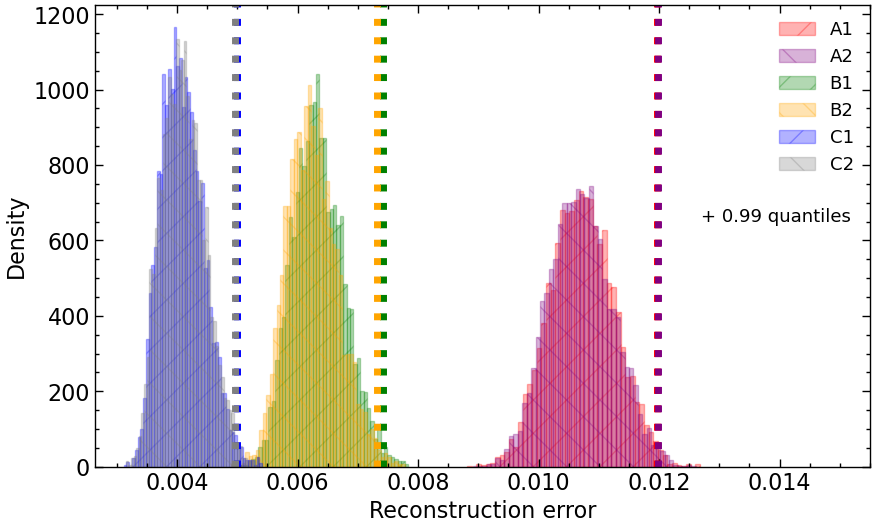} \\
    \vspace{0.5em}
    \includegraphics[width=0.48\linewidth]{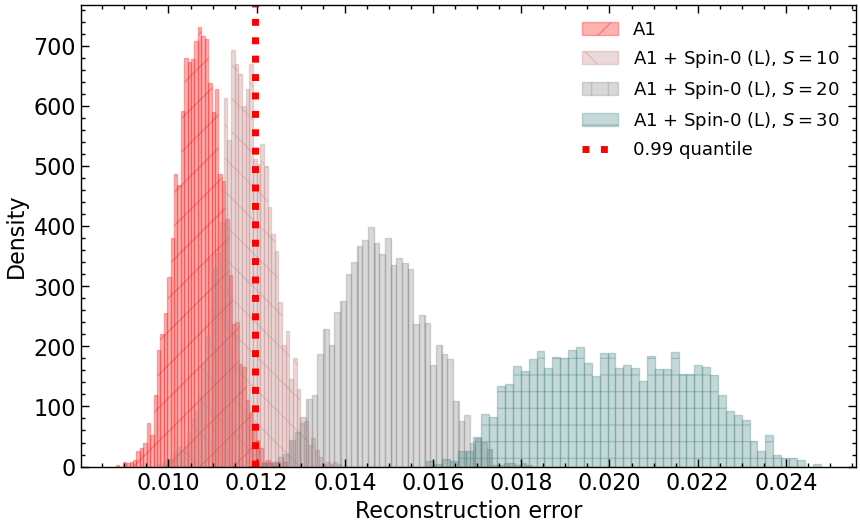}
    \hfill
    \includegraphics[width=0.48\linewidth]{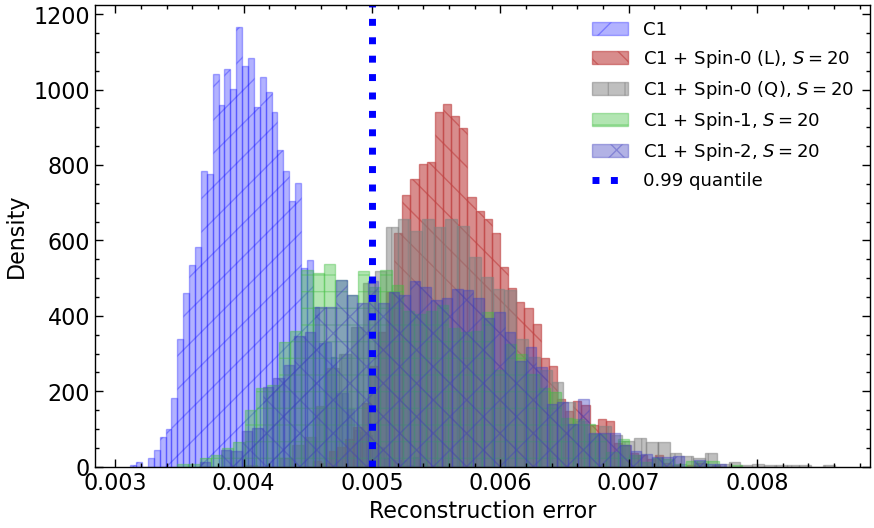}
    \caption{{\itshape Top left:} Loss function versus epochs for both training (solid lines) and validation (dashed lines) datasets. The plot shows results for six different autoencoders corresponding to two realisations for each noise type. {\itshape Top right:} Reconstruction error distributions for test (normal) datasets for the same six autoencoders. Each distribution's 0.99~quantile is shown as a vertical dashed line. {\itshape Bottom left:} Reconstruction error distributions for datasets containing linear spin-0 signals of varying strength $S$ in a dataset with noise of type A; the A1-autoencoder is applied. {\itshape Bottom right:} Reconstruction error distributions for datasets with noise of type C containing different types of ULDM signals with fixed strength $S = 20$; the C1-autoencoder is applied.}
    \label{fig:autoencoder_training_and_reconstruction}
\end{figure}

The training process of the CAE is illustrated in Fig.~\ref{fig:autoencoder_training_and_reconstruction}. In the top-left panel we see that CAE minimises the loss the least with pure white Gaussian noise and the most with a complex signal containing red noise (at the expense of white noise) and nuisance parameters. The autoencoder compresses data by identifying patterns, but white noise lacks structure, making compression difficult. Adding nuisance parameters and red noise introduces correlations, enhancing data structure and leading to a reduction of the loss. However, a lower value for the loss does not necessarily imply better sensitivity, as the model must distinguish the ULDM signal from both white noise and added complexities. We elaborate on this in the next section where we provide the sensitivity limits.

Fig.~\ref{fig:autoencoder_training_and_reconstruction}, top right, shows the reconstruction error distributions for normal data across all three cases after two realisations, resulting in 6 distributions. It also includes 6 threshold values, chosen as the 0.99~quantile (i.e.~the 99th percentile) of each distribution. Fig.~\ref{fig:autoencoder_training_and_reconstruction}, bottom left, illustrates how the reconstruction error distribution of abnormal data shifts to higher values as $S$ increases, which is in line with our expectations. Fig.~\ref{fig:autoencoder_training_and_reconstruction}, bottom right, compares abnormal data distributions across ULDM models with fixed \( S \), showing how the autoencoder sensitivity varies for each model.



\subsubsection{Results}

Our goal is to determine the critical value \( S_c \) such that, given a large time series dataset containing ULDM characterised by \( S_c \), at least a specified fraction of signals with this strength are identified as anomalies. We define the anomaly detection threshold as the 0.99~quantile of the reconstruction error distribution for normal data and require that 99\% of time series with signal strength \( S \) exceed this threshold. We denote the corresponding signal strength as \( S_{99} \).

\begin{figure}[htbp]
    \centering
    \includegraphics[width=0.9\linewidth]{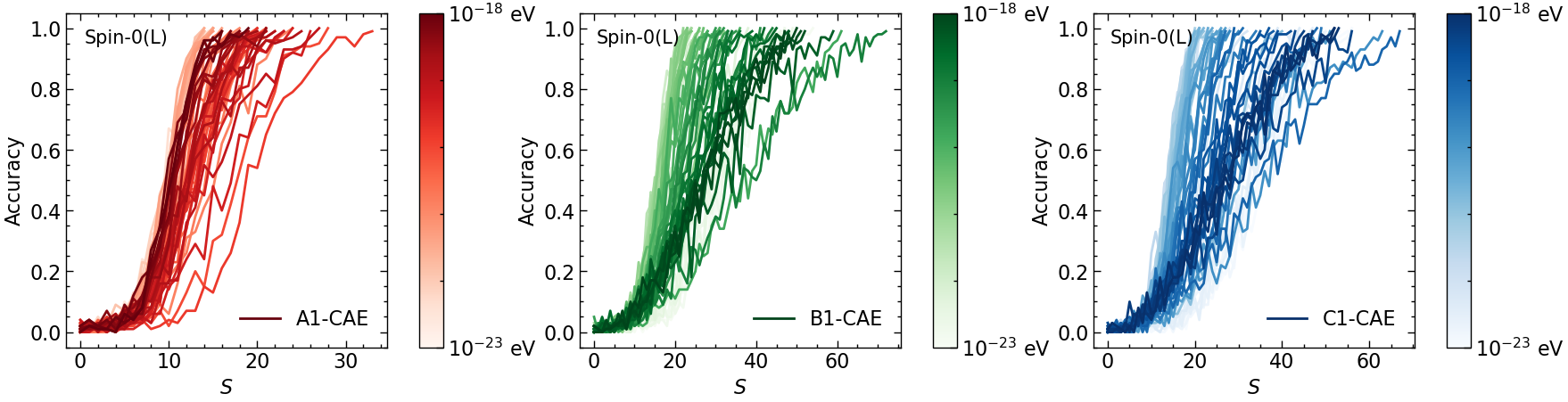} \\
    \vspace{0.2cm}
    \includegraphics[width=0.9\linewidth]{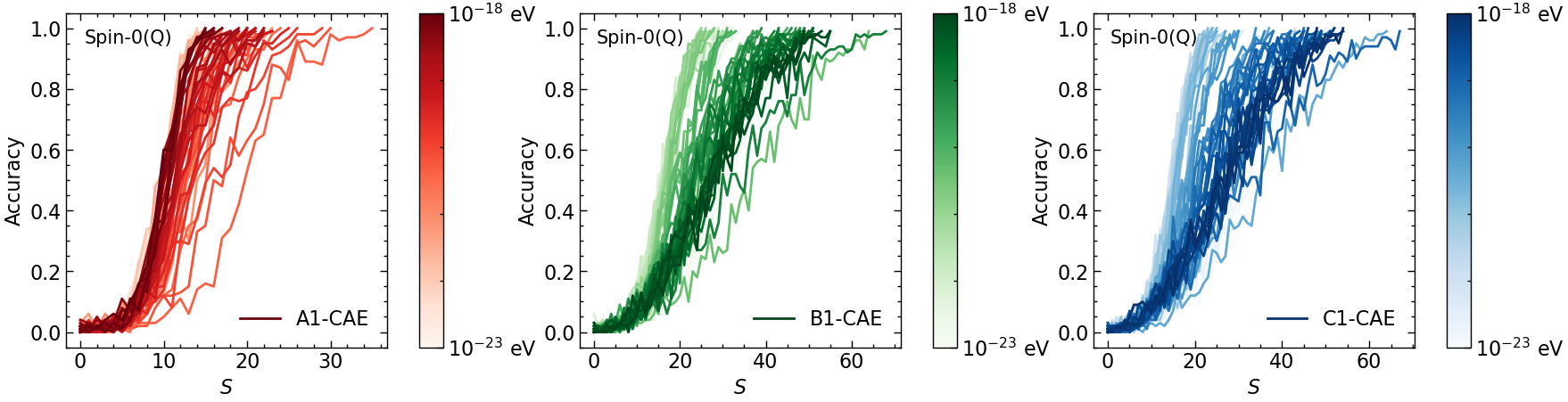} \\
    \vspace{0.2cm}
    \includegraphics[width=0.9\linewidth]{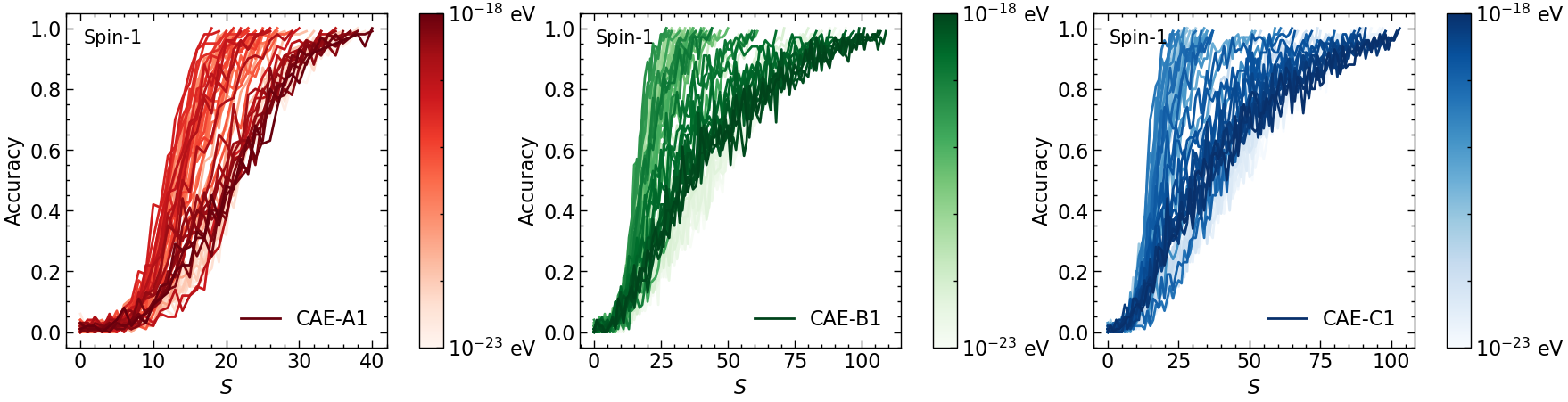} \\
    \vspace{0.2cm}
    \includegraphics[width=0.9\linewidth]{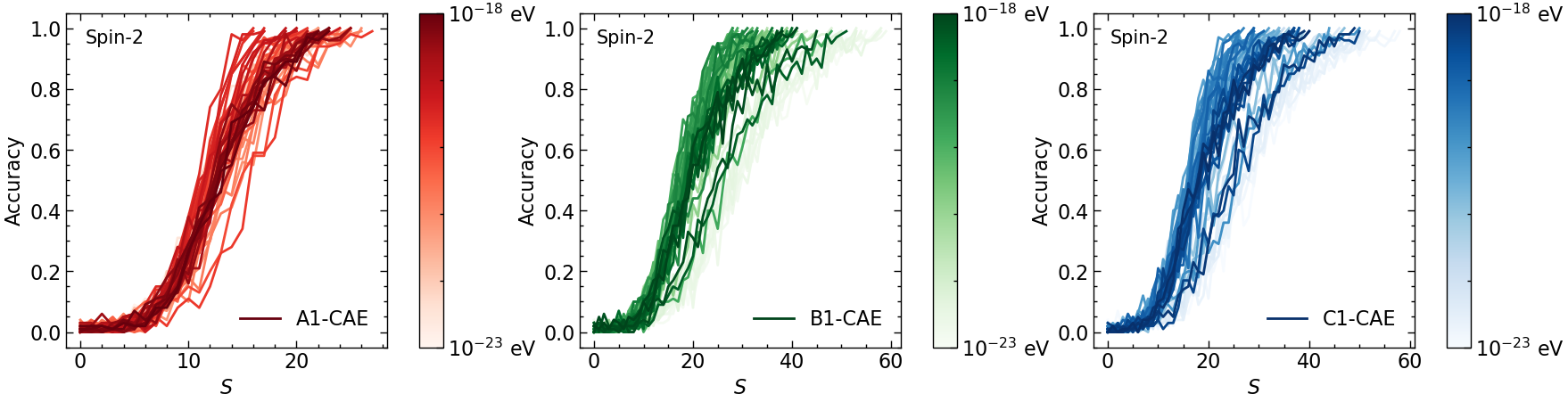}
    \caption{Portion of correctly identified anomalous signals from abnormal data (referred to as accuracy) as a function of signal strength \( S \) for three autoencoders: A1 (left column); B1 (central column) and C1 (right column). The rows correspond to different models: the top row shows spin-0 (L), the second row spin-0 (Q), the third row spin-1 and the bottom row spin-2. Each subplot contains 64 lines, representing 64 selected masses, most of them uniformly sampled on a logarithmic scale from \( 10^{-23}\,\mathrm{eV} \) to \( 10^{-18}\,\mathrm{eV} \), with the lighter colour corresponding to the smallest mass and the progressively darker colours to progressively larger masses. In addition, a few extra points have been added to increase the density around the expected resonances $Q = 1,\ 2,\ 3$.}
    \label{fig:sensitivity_vs_S}
\end{figure}

The fraction of correctly detected anomalous signals of strength \( S \), referred to as the accuracy of the CAE, is shown in Fig.~\ref{fig:sensitivity_vs_S}. Each panel contains 64 lines, representing 64 masses ranging from \(10^{-23}\)\,eV to \(10^{-18}\)\,eV, that is, 52~masses are obtained by scanning the $10^{-23}\,\text{eV}$ to $10^{-18}\,\text{eV}$ in logarithmic steps of $0.1$; then we sample 12~more masses by adding extra points around expected resonances, namely $0.9\,\omega_b,\ 0.95\,\omega_b,\ 1.05\,\omega_b,\ 1.1\,\omega_b$ around $Q = 1$; $1.9\,\omega_b,\ 1.95\,\omega_b,\ 2.05\,\omega_b,\ 2.1\,\omega_b$ around $Q = 2$ and $2.9\,\omega_b,\ 2.95\,\omega_b,\ 3.05\,\omega_b,\ 3.1\,\omega_b$ around $Q = 3$. As both instances of each autoencoder (e.g.~A1 and~A2) yield nearly identical performance, we present the accuracy as a function of \( S \) using only the first instance. We observe that for stronger signals the dataset is more easily distinguished from pure noise, leading to higher accuracy.

\begin{figure}[htbp]
    \centering
    \includegraphics[width=0.48\linewidth]{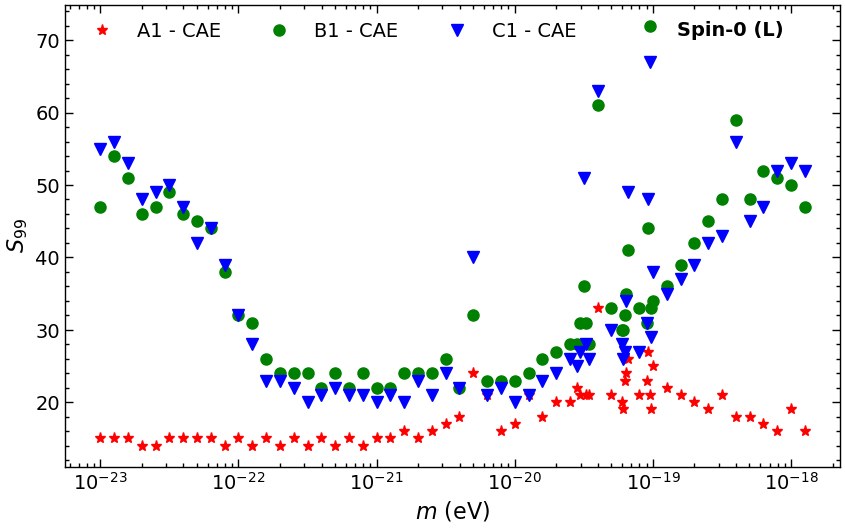}
    \hfill
    \includegraphics[width=0.48\linewidth]{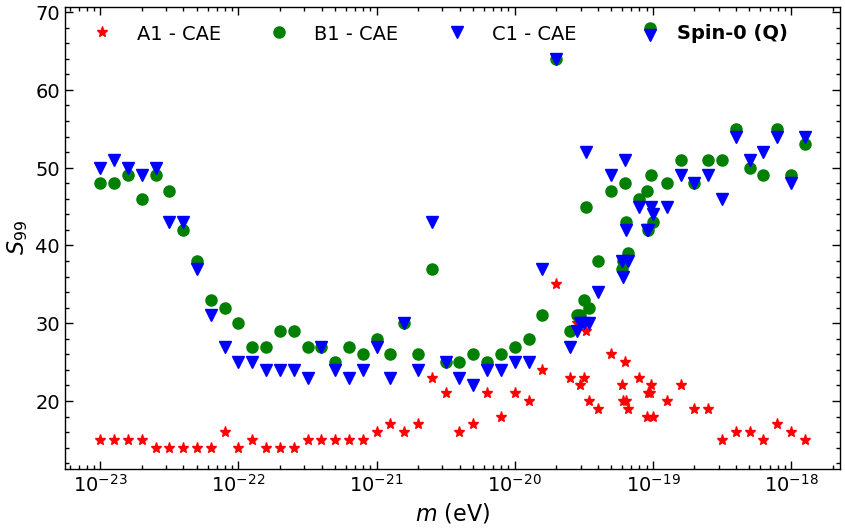} \\
    \vspace{0.5em}
    \includegraphics[width=0.48\linewidth]{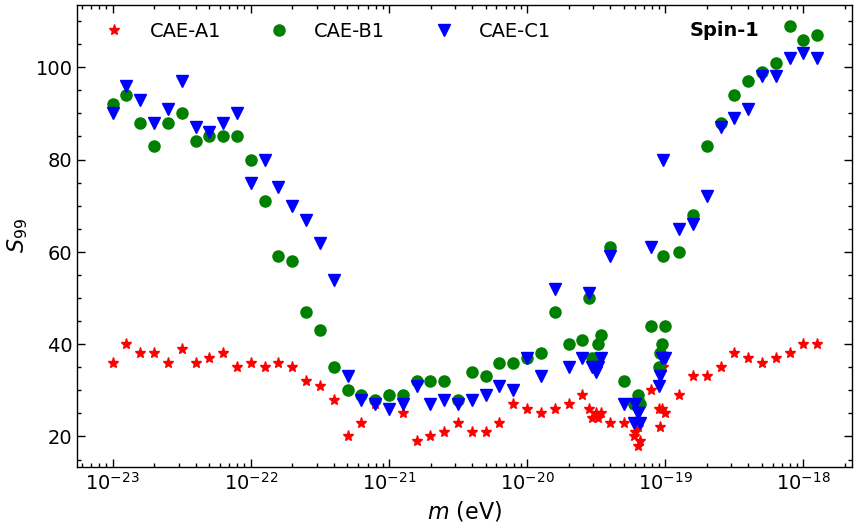}
    \hfill
    \includegraphics[width=0.48\linewidth]{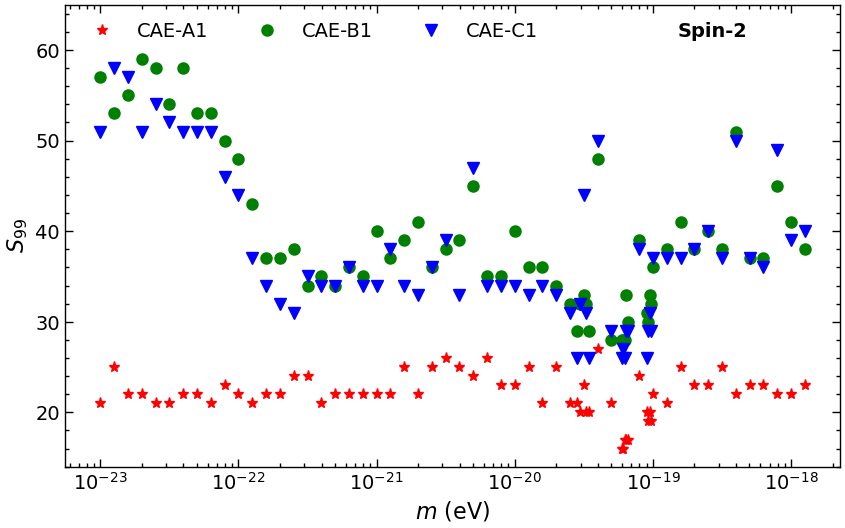}
    \caption{$S_{99}$ versus $m$ for the three autoencoders A1 (red stars), B1 (green circles) and C1 (blue triangles) for all four ULDM models: spin-0 (L) (top left), spin-0 (Q) (top right), spin-1 (bottom left) and spin-2 (bottom right).}
    \label{fig:S_99_vs_m}
\end{figure}

From the plots in Fig.~\ref{fig:sensitivity_vs_S}, we can extract the critical value \( S_{99} \) for each of the 64 masses. In Fig.~\ref{fig:S_99_vs_m} we display the dependence of \(S_{99}\) on the ULDM mass \(m\), in effect a slice along the colour-code of Fig.~\ref{fig:sensitivity_vs_S}. We can observe that \( S_{99} \) is dependent on \( m \), noise and the ULDM model. We will discuss of the strong mass dependence of $S_{99}$ in Section~\ref{sec:discussion}.

Once we obtain \( S_{99} \) for a given \( m \), we can compute \( |\lambda_c (m)|_{\mathrm{ML}} \) using Eq.~(\ref{eq:ML_lambda_estimation_formula}) and subsequently compare it with the Bayesian limit given by Eq.~(\ref{eq:Bayes_lambda_estimation_formula}), valid only for white Gaussian background noise. Notice that, in order to generate sensitivity plots, we must select specific fiducial parameter values that characterise the ULDM configuration at the location of the binary system---this also applies to both the one-step and the two-step Bayesian approaches. Throughout this paper, we adopt the following choices: (a) for spin-0: $\varrho = \frac{1}{\sqrt{2}}$, $\Upsilon = \frac{\pi}{4}$ (b) for spin-1: $\varrho = \frac{1}{\sqrt{2}}$, $\Upsilon = \frac{\pi}{3}$, $\theta = \frac{\pi}{3}$, $\phi = \frac{\pi}{3}$; (c) for spin-2: $\varrho = \frac{1}{\sqrt{2}}$, $\Upsilon = \frac{\pi}{3}$, $\chi = \frac{\pi}{3}$, $\epsilon_S = \epsilon_T = \frac{1}{\sqrt{2}}$.\footnote{Notice that the leading-order contribution to \(\delta x^\text{DM}\), \(\delta \eta^\text{DM}\) and \(\delta \kappa^\text{DM}\) does not depend on the amplitude \(\epsilon_V\) and the angle \(\eta\), so we do not have to choose a value for them.}

\begin{figure}[htbp]
    \centering
    \includegraphics[width=0.48\linewidth]{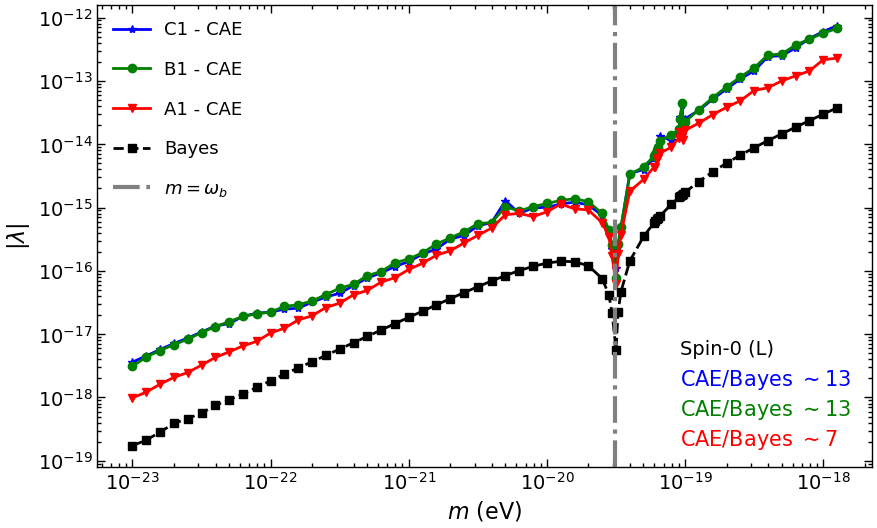}
    \hfill
    \includegraphics[width=0.48\linewidth]{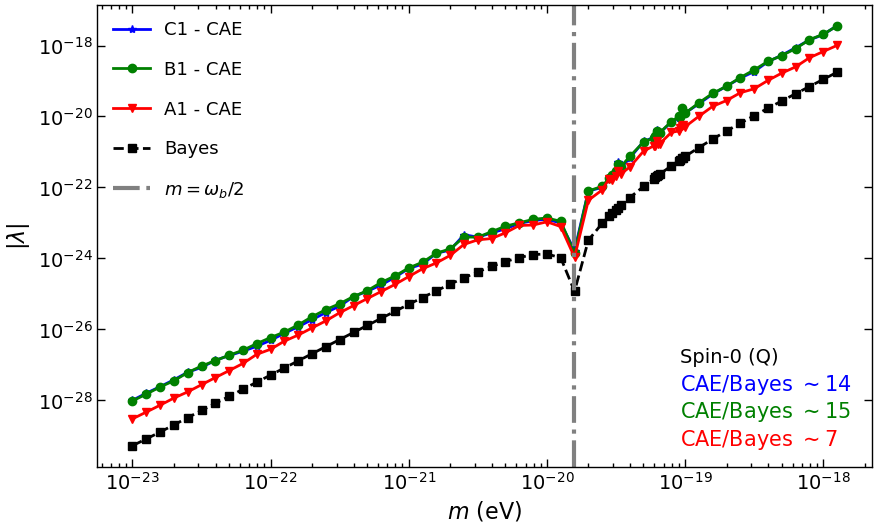} \\
    \vspace{0.5em}
    \includegraphics[width=0.48\linewidth]{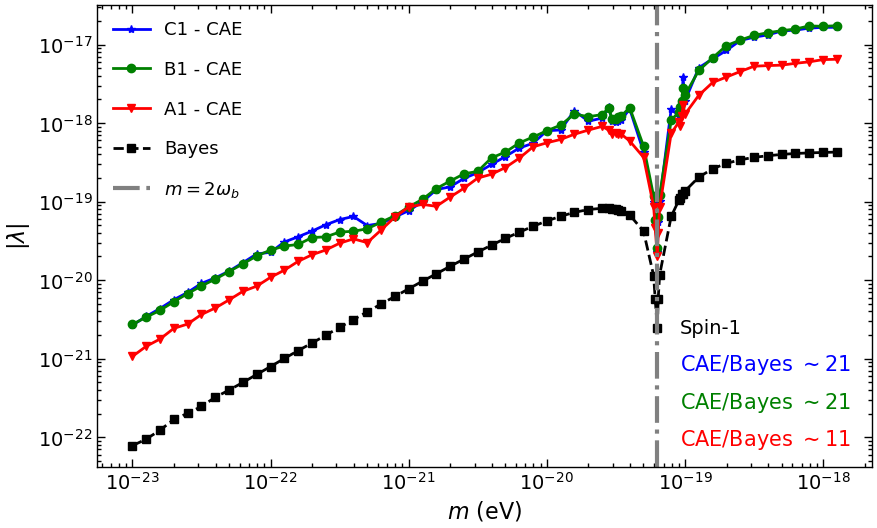}
    \hfill
    \includegraphics[width=0.48\linewidth]{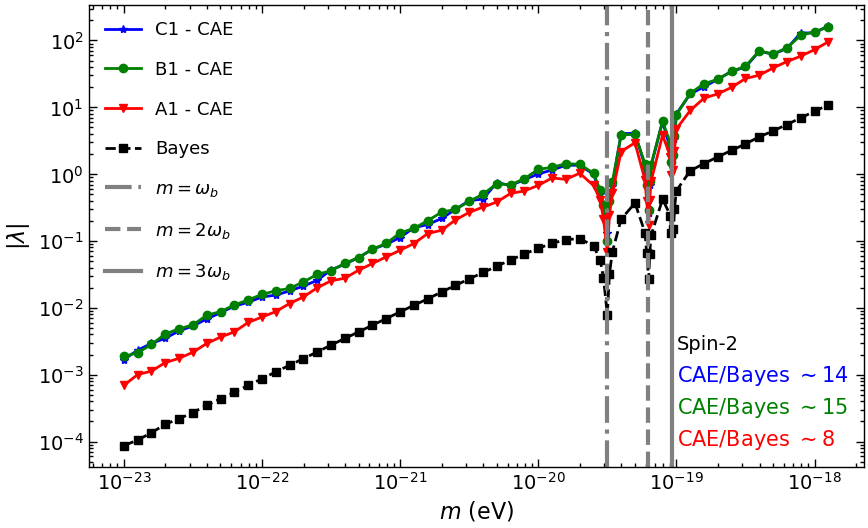}
    \caption{Sensitivity limits from Eq.~(\ref{eq:ML_lambda_estimation_formula}) for 99\% accuracy compared to the Bayesian limit in Eq.~(\ref{eq:Bayes_lambda_estimation_formula}) for the three autoencoders A1 (red triangles), B1 (green squares) and C1 (blue circles) for all four ULDM models: spin-0 (L) (top left), spin-0 (Q) (top right), spin-1 (bottom left) and spin-2 (bottom right). The one-step Bayesian sensitivity is shown in black and the vertical dot-dashed grey line marks the resonant frequency.}
    \label{fig:sensitivity_plots_autoencoder}
\end{figure}

The values of \( |\lambda_c (m)|_{\text{ML}} \) and \( |\lambda_c (m)|_{\text{Bayes}} \) are shown in Fig.~\ref{fig:sensitivity_plots_autoencoder}, along with the ratio between the ML and Bayesian methods, specifically the mean value over all points. We observe the following: (1) the CAE produces sensitivity curves consistently above the Bayesian ones by a factor of \({\cal O}(1)\) to \({\cal O}(10)\); (2) the sensitivity decreases by a factor of two when nuisance effects are considered; (3) adding a red noise component at the expense of white noise leads to either a slight or no improvement in sensitivity.

\begin{figure}[htbp]
    \centering
    \includegraphics[width=0.48\linewidth]{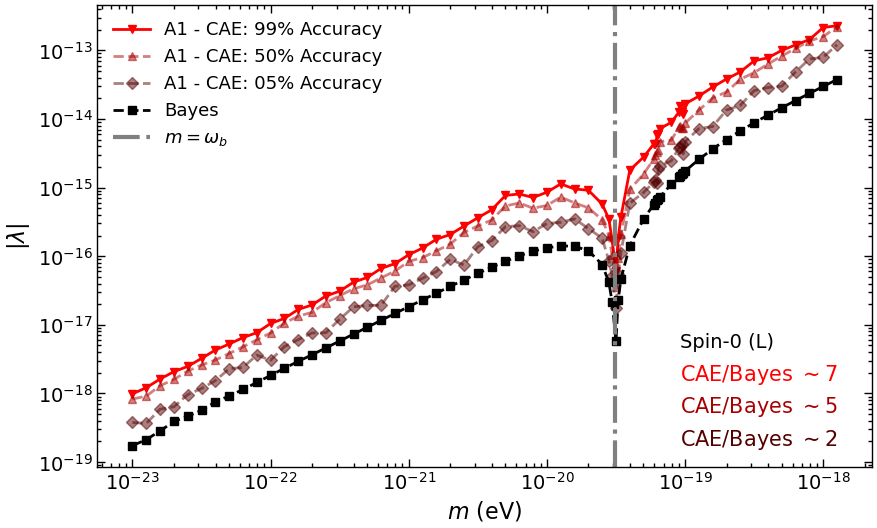}
    \hfill
    \includegraphics[width=0.48\linewidth]{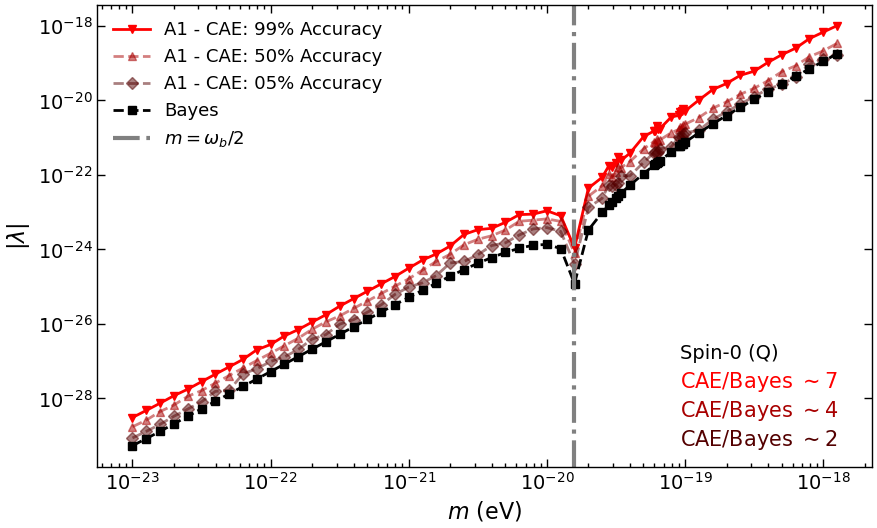} \\
    \vspace{0.5em}    
    \includegraphics[width=0.48\linewidth]{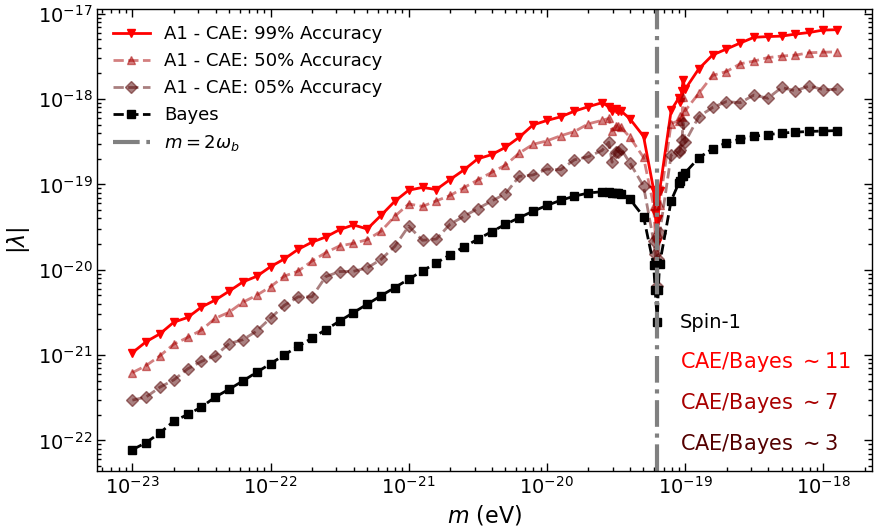}
    \hfill
    \includegraphics[width=0.48\linewidth]{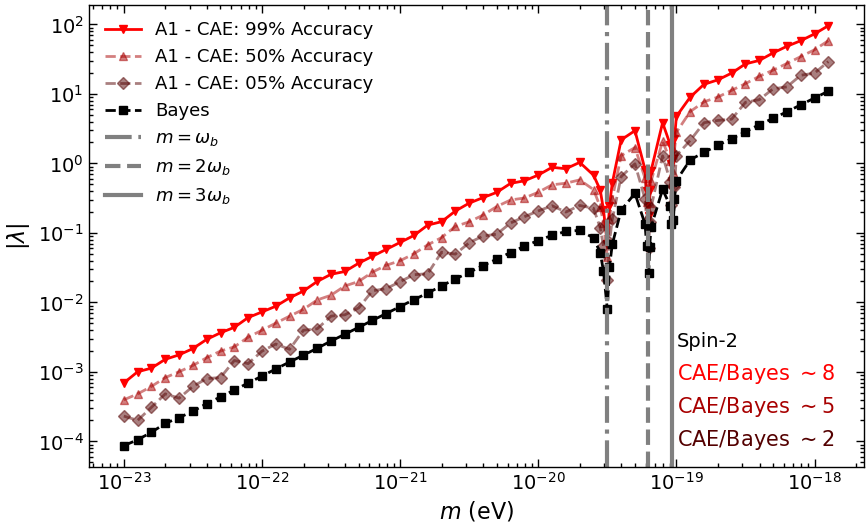}
    \caption{Same af Fig.~\ref{fig:sensitivity_plots_autoencoder} but only for the A1-autoencoder, with required accuracies of 5\% (brown diamonds), 50\% (pink stars) and 99\% (red triangles).}
    \label{fig:sensitivity_plots_autoencoder_various_S}
\end{figure}

We may be able to reduce the gap between the CAE and Bayesian approach by modifying the CAE criteria. For instance, we can keep the 0.99~quantile as a threshold but adjust the criterion to detect 99\% of all anomalous time series. Such tests are shown in Fig.~\ref{fig:sensitivity_plots_autoencoder_various_S} where the autoencoder trained on A1 noise is employed. We observe that less stringent requirements for the detection threshold result in improved sensitivities, as expected.

\subsection{Binary classifier}

An alternative approach involves using a supervised binary classifier trained to distinguish between pure noise and noisy ULDM signals. Here `supervised' means that during training we know which time series represents pure noise (labelled as `0') and which contains  ULDM signals (labelled as `1'). We discuss two cases in which a binary classifier learns to distinguish between pure noise and: (a) noisy spin-0 linear coupling ULDM signals, where only this spin and coupling type are considered for illustration; (b) noisy ULDM signals of any kind, always labelled as `1'.

\subsubsection{Architecture}

To effectively handle noisy time series data, we utilise the CNN architecture detailed in Table~\ref{tab:CNN_structure}.

\begin{table}[htbp]
\centering
\begin{tabular}{|l|c|}
\hline
\textbf{Layers} & \textbf{Output shape} \\
\hline
\hline
InputLayer(input\_shape=(1024, 1)) & (1024, 1) \\
\hline
Conv1D(32, kernel\_size=5, strides=1, activation="relu", padding="same") & (1024, 32) \\
\hline
BatchNormalization() & (1024, 32) \\
\hline
MaxPooling1D(pool\_size=2, strides=2, padding="same") & (512, 32) \\
\hline
Conv1D(64, kernel\_size=3, strides=1, activation="relu", padding="same") & (512, 64) \\
\hline
BatchNormalization() & (512, 64) \\
\hline
MaxPooling1D(pool\_size=2, strides=2, padding="same") & (256, 64) \\
\hline
Flatten() & (flattened) \\
\hline
Dense(64, activation="relu") & (64,) \\
\hline
Dropout(0.3) & (64,) \\
\hline
Dense(1, activation="sigmoid") & (1,) \\
\hline
\end{tabular}

\caption{
Architecture of the 1D convolutional neural network implemented in \href{https://www.tensorflow.org}{\textcolor{blue}{TensorFlow}}, with an input size of \( N = 1024 \). The model consists of 
1D convolutional layers, 
1D max pooling layers, 
batch normalisation layers, dense layers, a flatten layer and 
a dropout layer. The final layer uses a sigmoid activation for binary classification. 
The model is trained using the binary cross-entropy loss 
and optimised with the Adam optimizer with an initial learning rate of 0.001, decaying exponentially.
}
\label{tab:CNN_structure}
\end{table}

\subsubsection{Training}

For the training phase we adopt the \textit{curriculum learning approach}. Initially, the CNN is trained on a dataset where distinguishing between noise and noisy ULDM signals is relatively straightforward. The network is then progressively trained on more challenging datasets that include weaker ULDM signals, making classification increasingly difficult.

Because the CNN training requires the explicit inclusion of ULDM signals (unlike the autoencoder approach), the complexity of the task grows significantly as the parameter space expands. To mitigate this, we initially focus on a specific mass subrange, \( m / \mathrm{eV} \in [10^{-22}, 10^{-21}] \), which simplifies the training of the and binary classifiers while preserving sufficient generality for illustrative purposes. During training, the mass is randomly and uniformly sampled within this range, allowing the classifier to efficiently learn to detect ULDM signals with high accuracy.

\begin{figure}[htbp]
    \centering
    \includegraphics[width=0.48\linewidth]{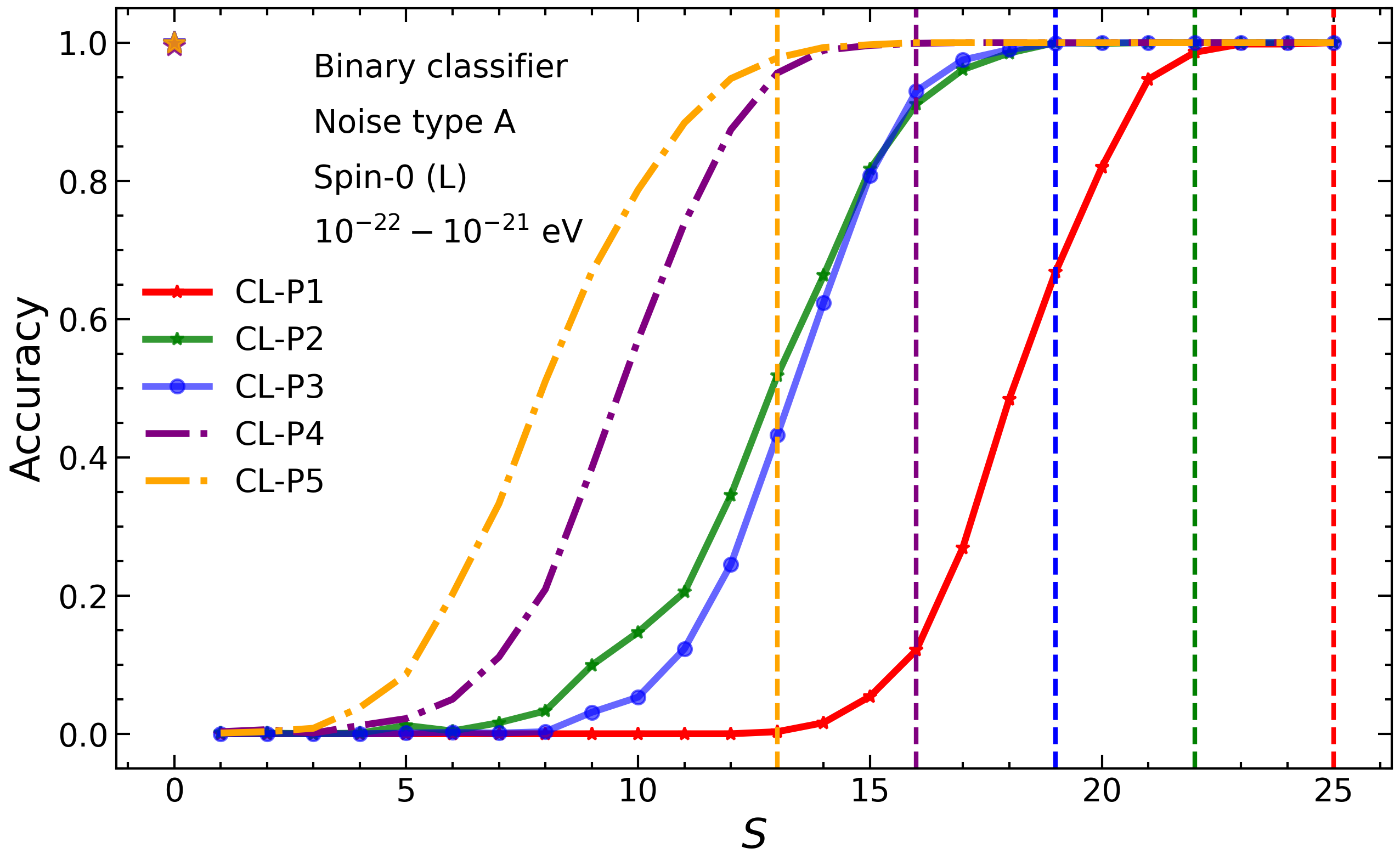}
    \hfill
    \includegraphics[width=0.48\linewidth]{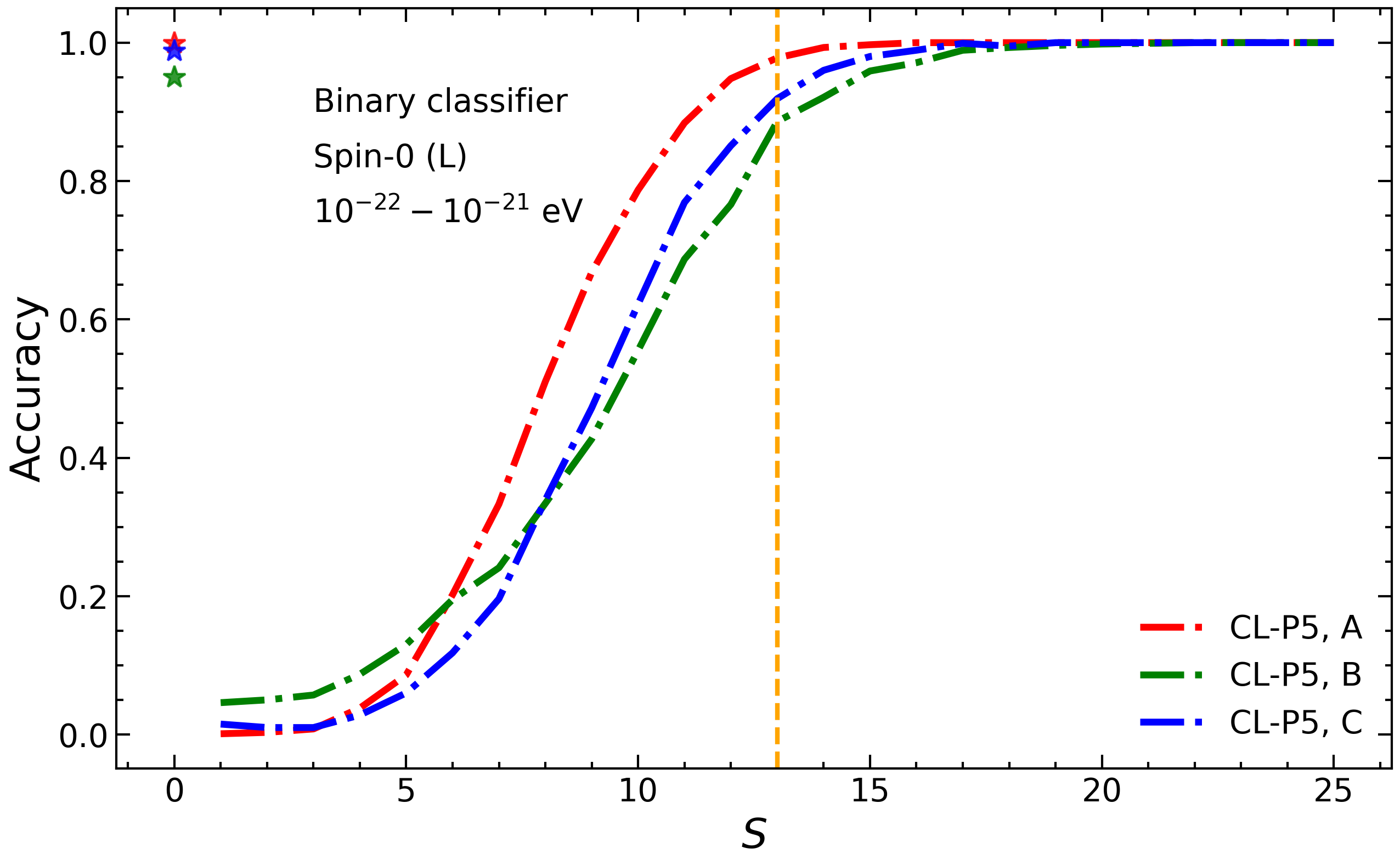}
    \caption{{\itshape Left:} Illustration of the CL approach for spin-0 (L) and noise type A, 5 training phases (P1–P5). The vertical lines indicate the lowest value of \( S \) at which the binary classifier was trained. The colours of the vertical lines correspond to the respective training phase numbers. {\itshape Right:} Accuracy versus $S$ after the training procedure for all three noise types A (red), B (green) and C (blue) for spin-0 (L). The vertical line represents the lowest value of \( S \) at which each binary classifier was trained.}
    \label{fig:bin_classifier_training}
\end{figure}

\paragraph{(a) Linear spin-0 only } The initial training dataset consists of 50\% pure noise and 50\% ULDM signals, with masses \( m / \mathrm{eV} \in [10^{-22}, 10^{-21}] \), mixed with noise at a signal strength of \( S = 25 \). Once the neural network successfully distinguishes between these two classes, we introduce a more challenging dataset that includes weaker signals. Specifically, this new dataset consists of 50\% noise, 25\% ULDM signals at \( S = 25 \) and 25\% ULDM signals at \( S = 22 \). The CNN is then retrained on this dataset. We continue this process iteratively, progressively introducing weaker signals by decreasing \( S \) in steps of 3. All signals with non-zero \( S \) are represented in equal proportions at each stage: in the first stage, \( S = 25 \) (50\%); in the second stage, \( S = 25 \) (25\%), \( S = 22 \) (25\%); in the third stage, \( S = 25 \) (16.7\%), \( S = 22 \) (16.7\%) and \( S = 19 \) (16.7\%), etc. The final dataset includes signals with \( S = 25, 22, 19, 16, 13 \), alongside pure noise; we choose to not train the CNN with signals weaker than \( S = 13 \) because this would lower the accuracy at \( S = 0 \)---notice that the length of the injected noise series is always the same as the overall signal series, however the latter is divided into strong and weak signals. For each non-zero \( S \), we generate 3,000 time series; we found that with this many time series we can train the CNN quickly and efficiently (see Fig.~\ref{fig:bin_classifier_training}, left panel). Additionally, we consider all three noise types (A, B and C), training separate CNNs for each.

\paragraph{(b) All models } Rather than training separate binary classifiers for each spin, coupling and noise type, we train a single classifier capable of distinguishing pure noise of type C (the most complex case) from ULDM signals of any type. As before, 50\% of the data consists of pure noise, while the remaining 50\% comprises ULDM signals, with all signal types equally represented, namely comprising of 25\% (12.5\%) of the total (signal only) data each. Initially, all ULDM signals have a strength of \( S = 35 \). In subsequent training epochs, we introduce weaker signals, decreasing \( S \) in steps of 3. The final dataset includes signals with \( S = 35, 32, 29, 26, 23, 20 ,17, 14, 11 \), alongside pure noise.

\subsubsection{Results}

\paragraph{(a) Linear spin-0 only } The impact of the curriculum learning (CL) approach on the ability of CNNs to detect the ULDM signal is illustrated in Fig.~\ref{fig:bin_classifier_training}, left panel. The CL procedure consists of 5 stages, starting with signals at $S = 25$ (marked by the red vertical line) and concluding at $S = 13$ (marked by the orange vertical line). While the sensitivity could potentially improve further with additional stages, we find this range sufficient for demonstration purposes. Importantly, we ensure that the rate of false positives is minimised, as the accuracy for $S = 0$ is approximately 100\%. In Fig.~\ref{fig:bin_classifier_training}, right panel, we show the accuracy as a function of $S$ after completing the fifth training phase for all three types of noise. These results suggest that there is no substantial difference between the noise types, indicating that increased noise complexity does not noticeably degrade the CNN's performance.

\begin{figure}[htbp]
    \centering
    \includegraphics[width=0.48\linewidth]{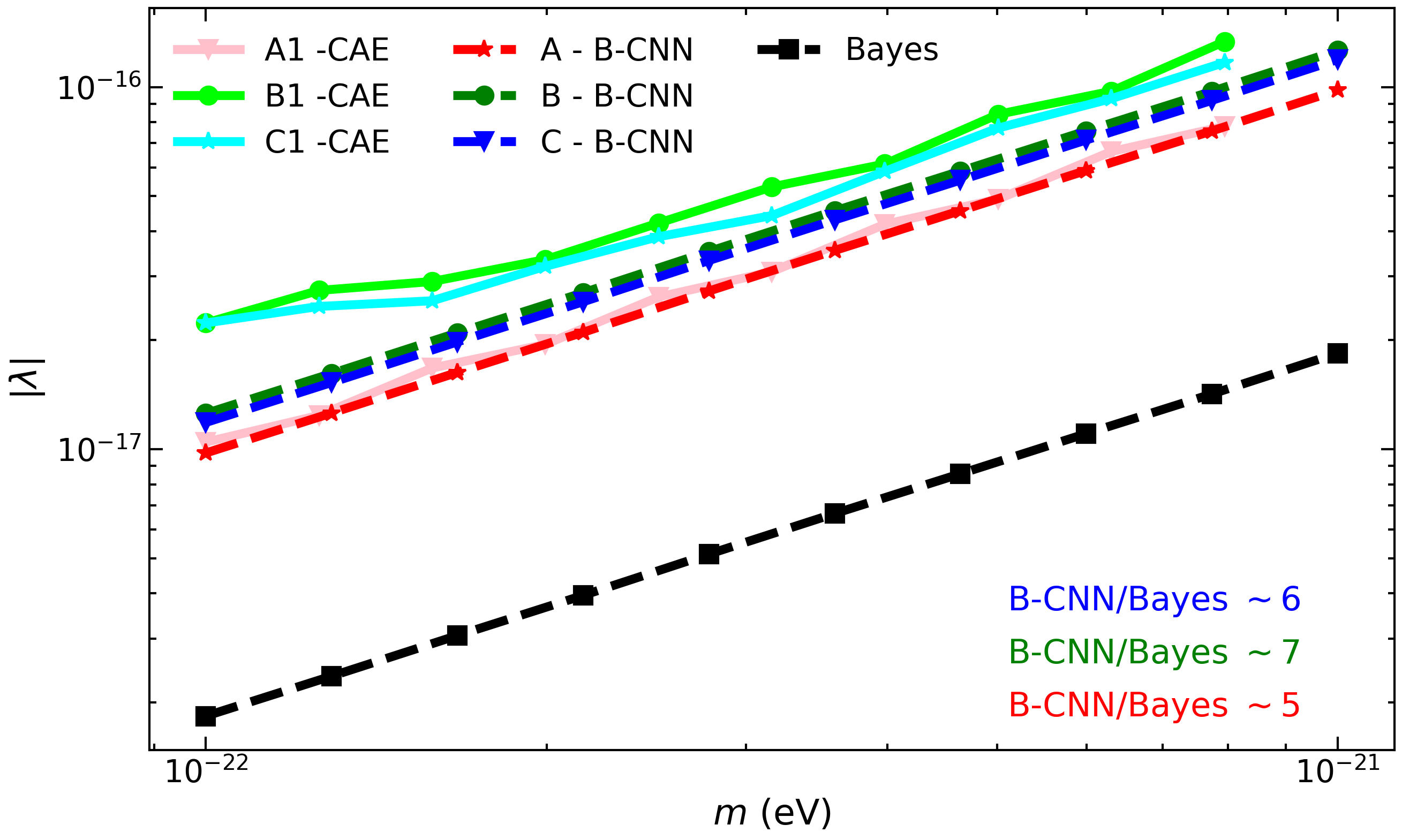}
    \hfill
    \includegraphics[width=0.48\linewidth]{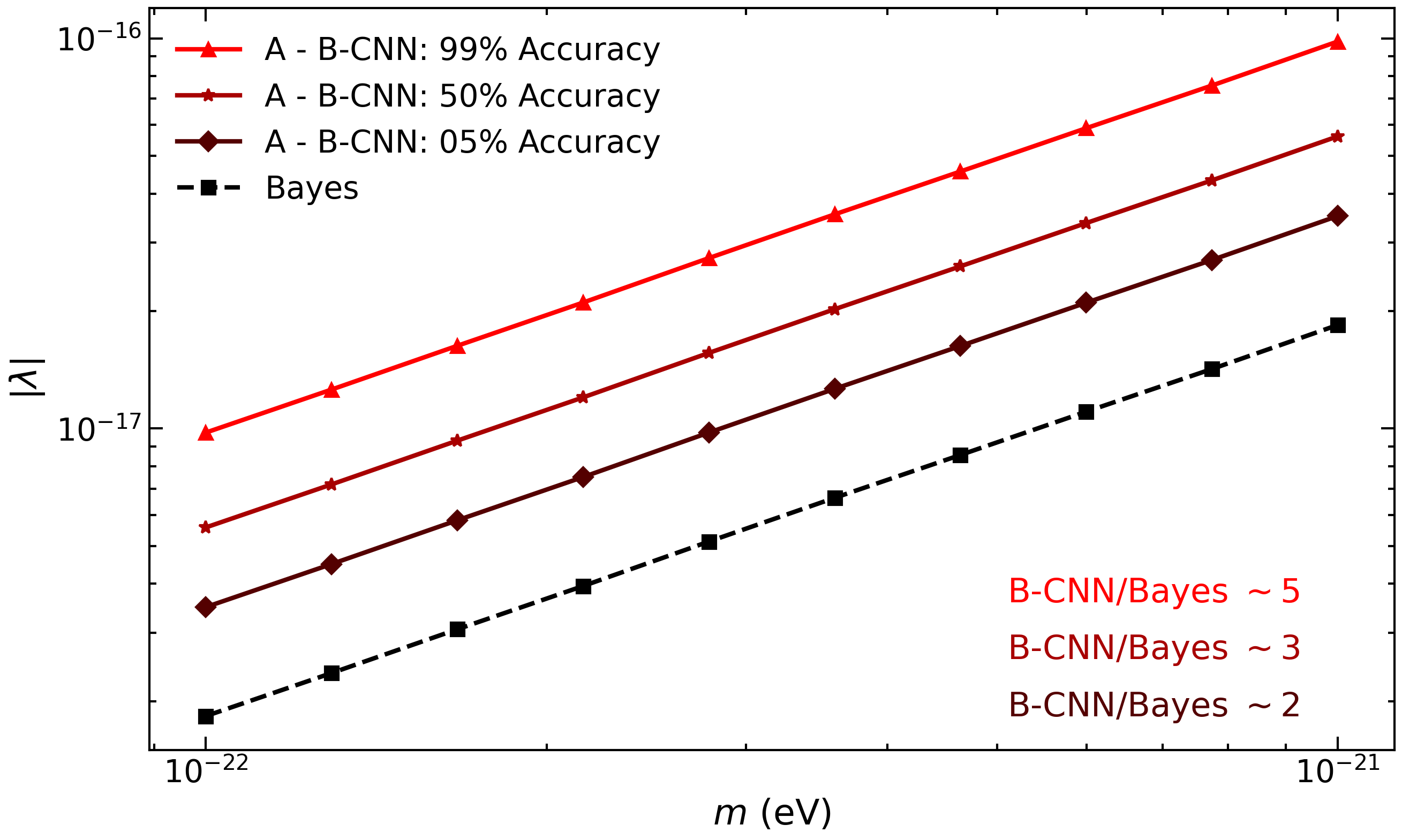}
    \caption{{\itshape Left:} Sensitivity lines from a binary classifier, spin-0 (L), all three noise types, accuracy threshold of 99\% (dashed lines). For comparison, we present autoencoder sensitivity lines (solid lines), which suggest that binary classifiers are mildly more sensitive. {\itshape Right:} Sensitivity lines from a binary classifier, spin-0 (L), noise type A and three different accuracy thresholds (5\%, 50\%, 99\%).}
    \label{fig:bin_classifier_sensitivity_graphs}
\end{figure}

Using Eq.~(\ref{eq:ML_lambda_estimation_formula}), the critical value $S_c$ from Fig.~\ref{fig:bin_classifier_training}, right panel and the agreed fiducial values for the ULDM configuration, we plot the sensitivity lines $|\lambda(m)|$. This is shown in Fig.~\ref{fig:bin_classifier_sensitivity_graphs}, left panel, for the range $m/\mathrm{eV} \in [10^{-22}, 10^{-21}]$ and $S_c = S_{99}$, where 99\% of ULDM signals are correctly distinguished from noise. The figure includes sensitivity curves for all three noise types. Compared to the Bayesian sensitivity line (from the 1-step approach with noise type A), the CNN sensitivity is slightly lower, by a factor of $\sim 6$. This gap might be reduced with continued CL training, the adoption of alternative architectures, or training on larger datasets. However, we consider the $O(1)$ difference acceptable for our purposes. Alternatively, lowering the accuracy threshold could reduce the gap, as shown in Fig.~\ref{fig:bin_classifier_sensitivity_graphs}, right panel.

\begin{figure}[htbp]
	\centering
	\includegraphics[width=0.48\textwidth]{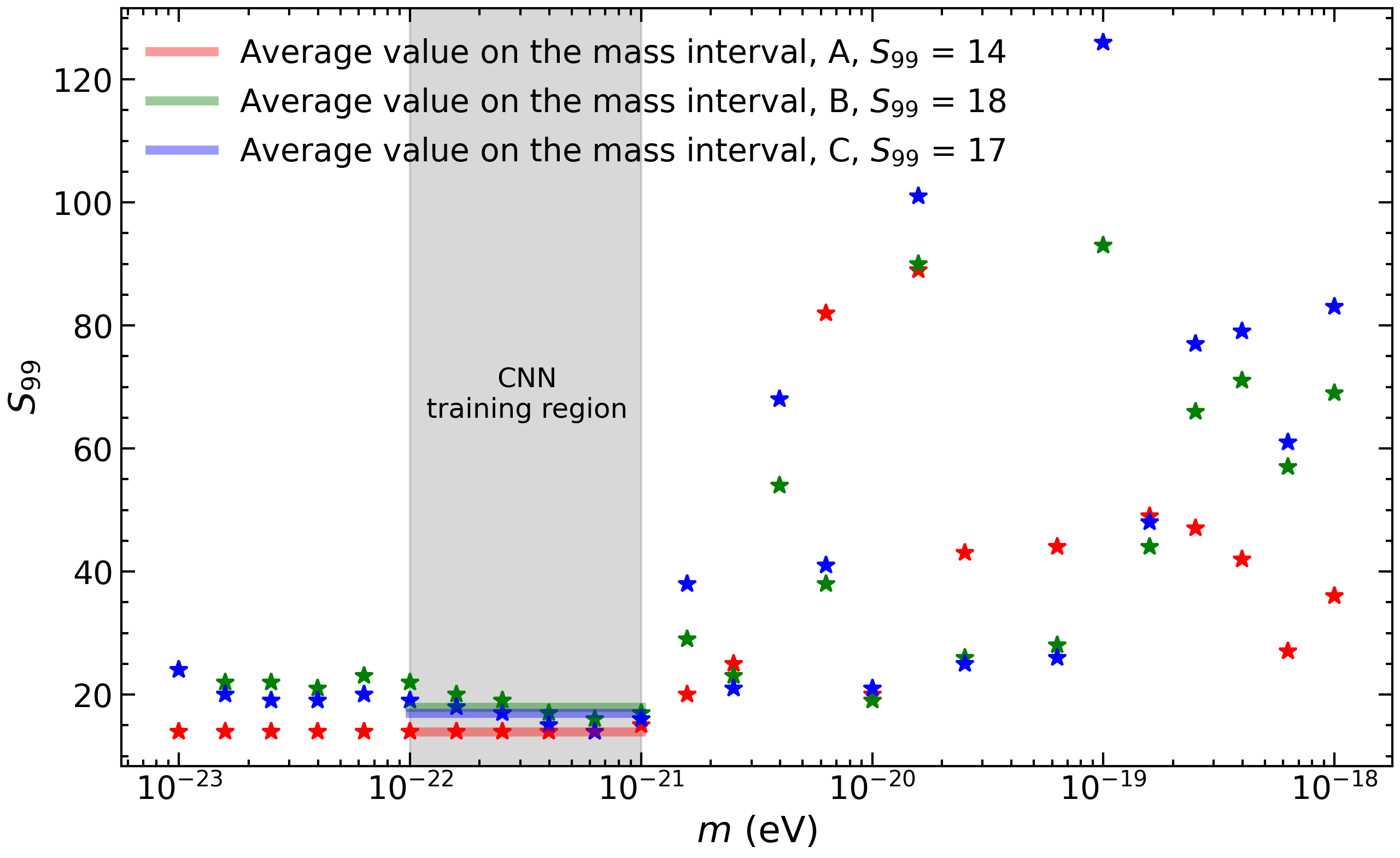}
	\hfill
	\includegraphics[width=0.48\textwidth]{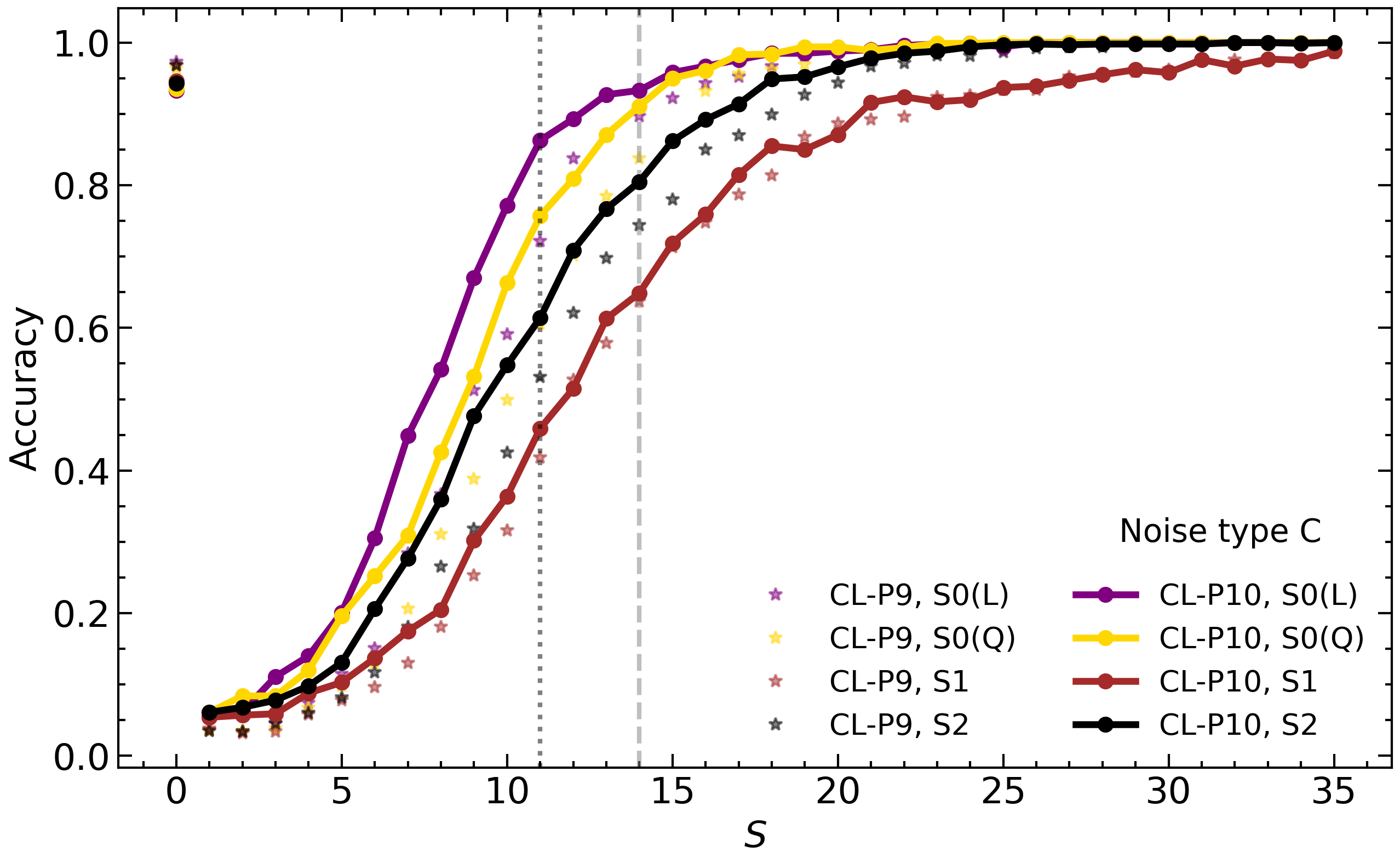}
	\caption{{\itshape Left:} $S$ versus $m$ for all three noise types A, B and C, all ULDM types. {\itshape Right:} Accuracy versus $S$ for a binary classifier, noise type C, all ULDM types, phases 9 and 10. The vertical lines show the lowest value of $S$ at which the binary classifier was trained.}
    \label{fig:S_vs_m_and_accuracy_binary_spin-0_and_binary_all}
\end{figure}

We can also evaluate how well the CNN performs outside the mass interval it was trained on. Fig.~\ref{fig:S_vs_m_and_accuracy_binary_spin-0_and_binary_all}, left panel, shows the value of $S_{99}$ required to detect 99\% of all ULDM signals for the entire mass range. The plot demonstrates that the CNN's performance is optimal and flat within the mass interval it was trained on. However, outside this interval, the performance deteriorates, particularly at higher frequencies. This suggests that, if this CNN were to be applied to an extended mass range, the training mass interval should also be expanded---we will discuss this more in depth in Section~\ref{sec:discussion}.

\paragraph{(b) All models } We extend the previous task to include all spins and couplings. This time, we train a single CNN to distinguish between pure noise (label `0') and any type of ULDM signal (label `1'). The CL approach now has~10 phases (CL-P1 to CL-P10), starting from an initial $S = 35$ and ending at $S = 11$, with a step size of~3. In each phase, the dataset is balanced, with 50\% pure noise and 50\% ULDM, with each ULDM type equally represented. In training phase~1, for each type of ULDM signal with \( S = 35 \), we generate 6,000 time series and obtain initial estimates of the learning parameter values. In the phases 2-7, each non-zero $S$ is represented by 2,000 time series. In phases~8 and~9, ULDM signals with $S \in [26,35]$ in steps of~3 are represented by 1,000 time series, while for weaker $S$, each ULDM signal type is represented by 3,000 time series to enhance sensitivity toward weaker signals.

\begin{figure}[htbp]
	\centering
	\rotatebox{0}{\includegraphics[width=0.7\linewidth]{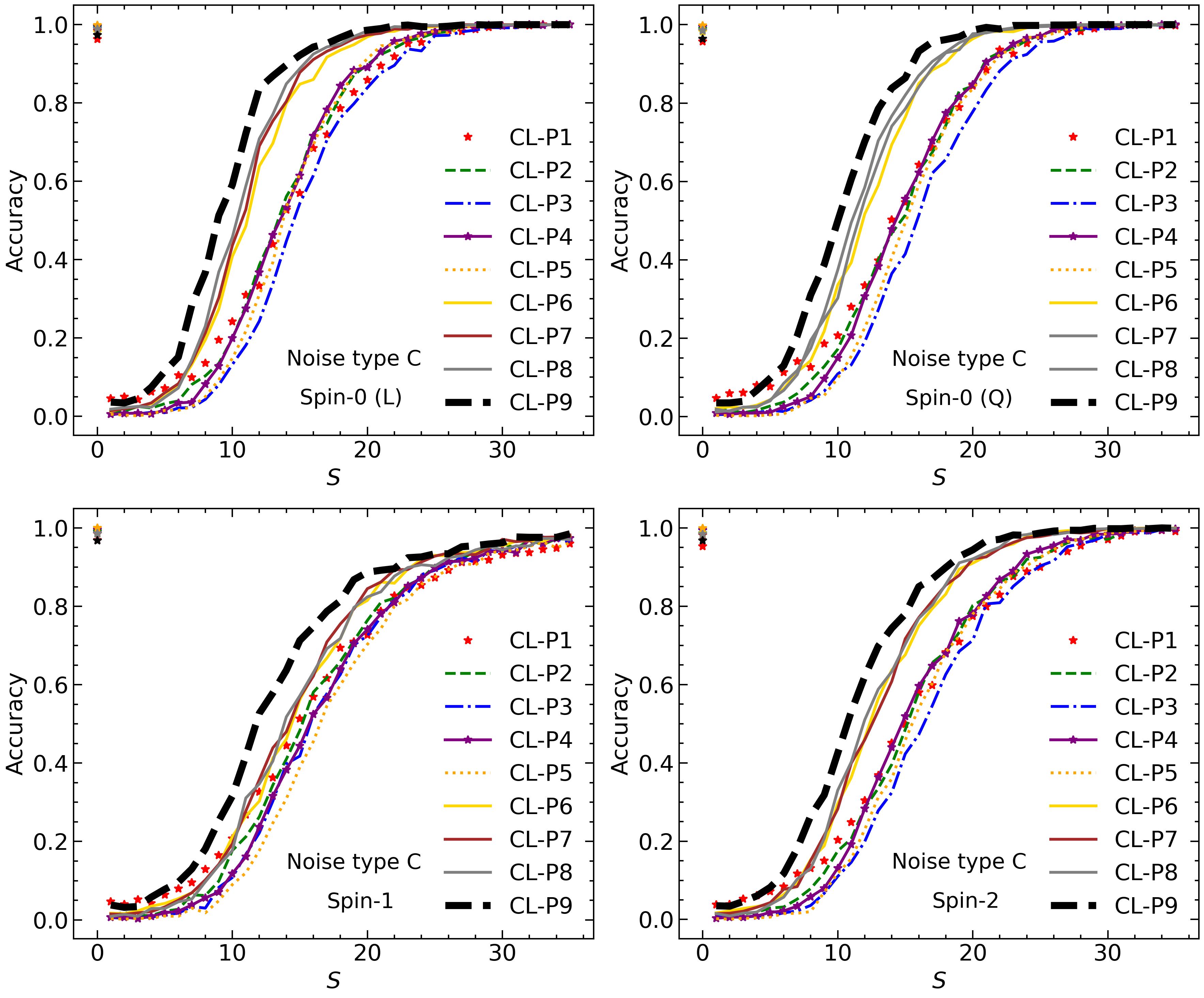}}
	\caption{Accuracy versus $S$ for the (b) binary classifier, noise type C, all ULDM types, phases 1-9 for spin-0 (L) (top left), spin-0 (Q) (top right), spin-1 (bottm left) and spin-2 (bottom right).}
	\label{fig:S_vs_accuracy_binary_all_spins_C_phases1-9}
\end{figure}

The results of this training process are summarised in Fig.~\ref{fig:S_vs_m_and_accuracy_binary_spin-0_and_binary_all}, right panel. In phase~10, we further increase the CNN's sensitivity to weaker signals. Specifically, for $S = 11, 14, 17$, each ULDM signal type has 5,000 time series, for $S = 20, 23$, each has 1,000 series and for larger $S$ only 500 series are provided. The sensitivity improves, as shown in Fig.~\ref{fig:S_vs_m_and_accuracy_binary_spin-0_and_binary_all}, right panel, but the rate of false positives also increases, as the accuracy in pure noise identification drops from about 97\% to about 93\%. We expect that further training will improve sensitivity to weaker ULDM signals but will also increase the false positive rate. How the capability of the binary classifier to detect ULDM signals increased during different phases is shown in Fig.~\ref{fig:S_vs_accuracy_binary_all_spins_C_phases1-9}. 

\begin{figure}[htbp]
	\centering
	\rotatebox{0}{\includegraphics[width=0.95\linewidth]{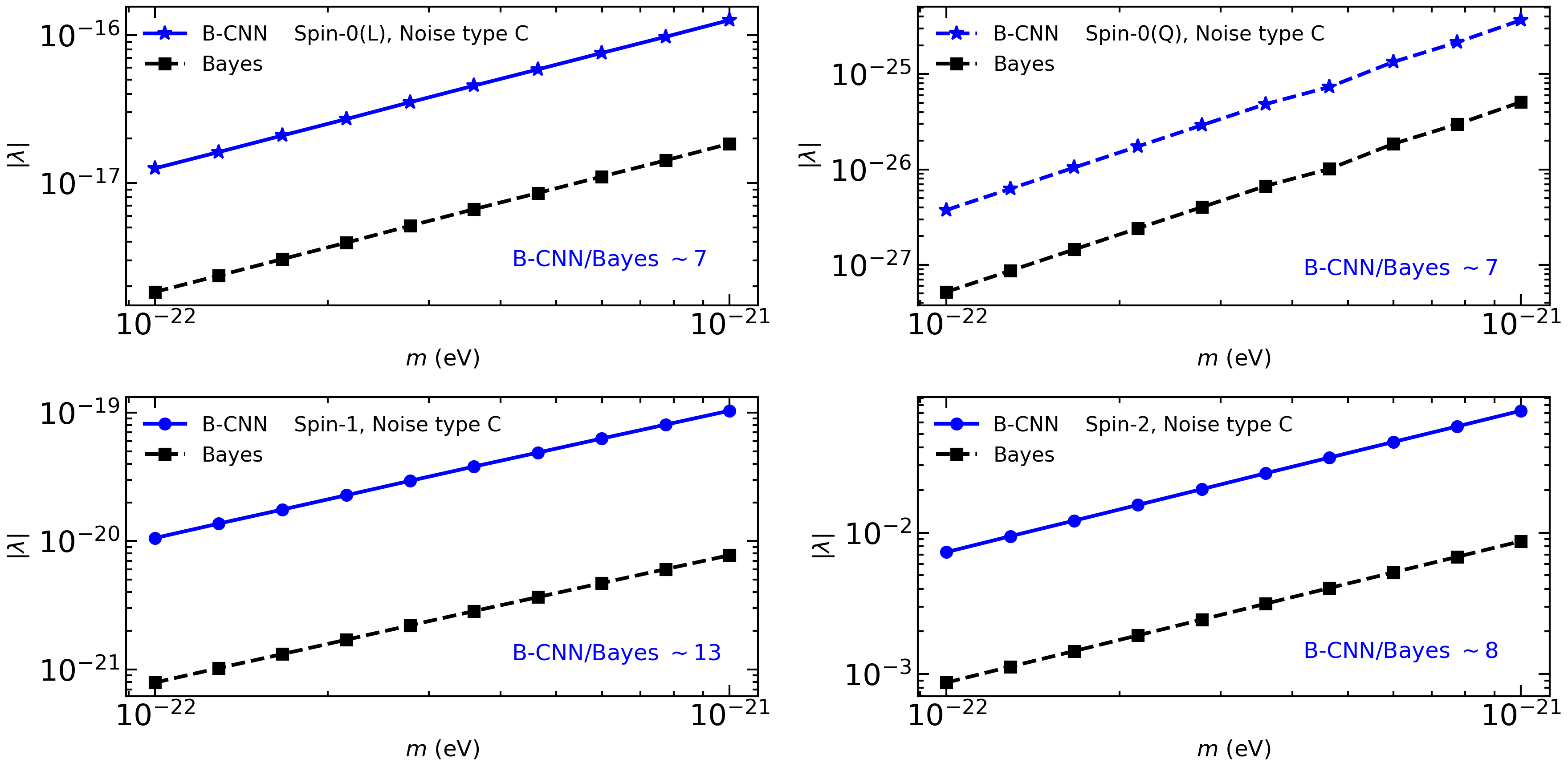}}
	\caption{Sensitivity limits for all four ULDM signal types, derived from $S_{99}$ and the tenth training phase (CL-P10) of the (b) binary classifier for spin-0 (L) (top left), spin-0 (Q) (top right), spin-1 (bottm left) and spin-2 (bottom right).}
	\label{fig:sensitivity_limits_binary_all_spins_C}
\end{figure}

We can again extract $S_{99}$ at the end of the 10th training phase CL-P10 from Fig.~\ref{fig:S_vs_m_and_accuracy_binary_spin-0_and_binary_all}, right panel, from which can we plot the sensitivity lines $|\lambda(m)|$. These are displayed for each ULDM model in Fig.~\ref{fig:sensitivity_limits_binary_all_spins_C} over the mass interval $m/\mathrm{eV} \in [10^{-22}, 10^{-21}]$.

\subsubsection{Comparison with the autoencoder}

In order to compare the sensitivities of the binary classifier and the autoencoder to a linear spin-0 ULDM signal in more details, we train the binary classifier once again, this time on the range \( m / \mathrm{eV} \in [10^{-20}, 10^{-19}] \). We need to do so because the sensitivity derived with the autoencoder has a strong $m$-dependence and we want to check if this dependence is also observed with the binary classifier. Therefore, we build four binary classifiers trained on two different mass ranges and two types of noise (we choose type A and B; the results for type C are very similar to type B).

\begin{figure}[htbp]
	\centering
	\includegraphics[width=0.48\textwidth]{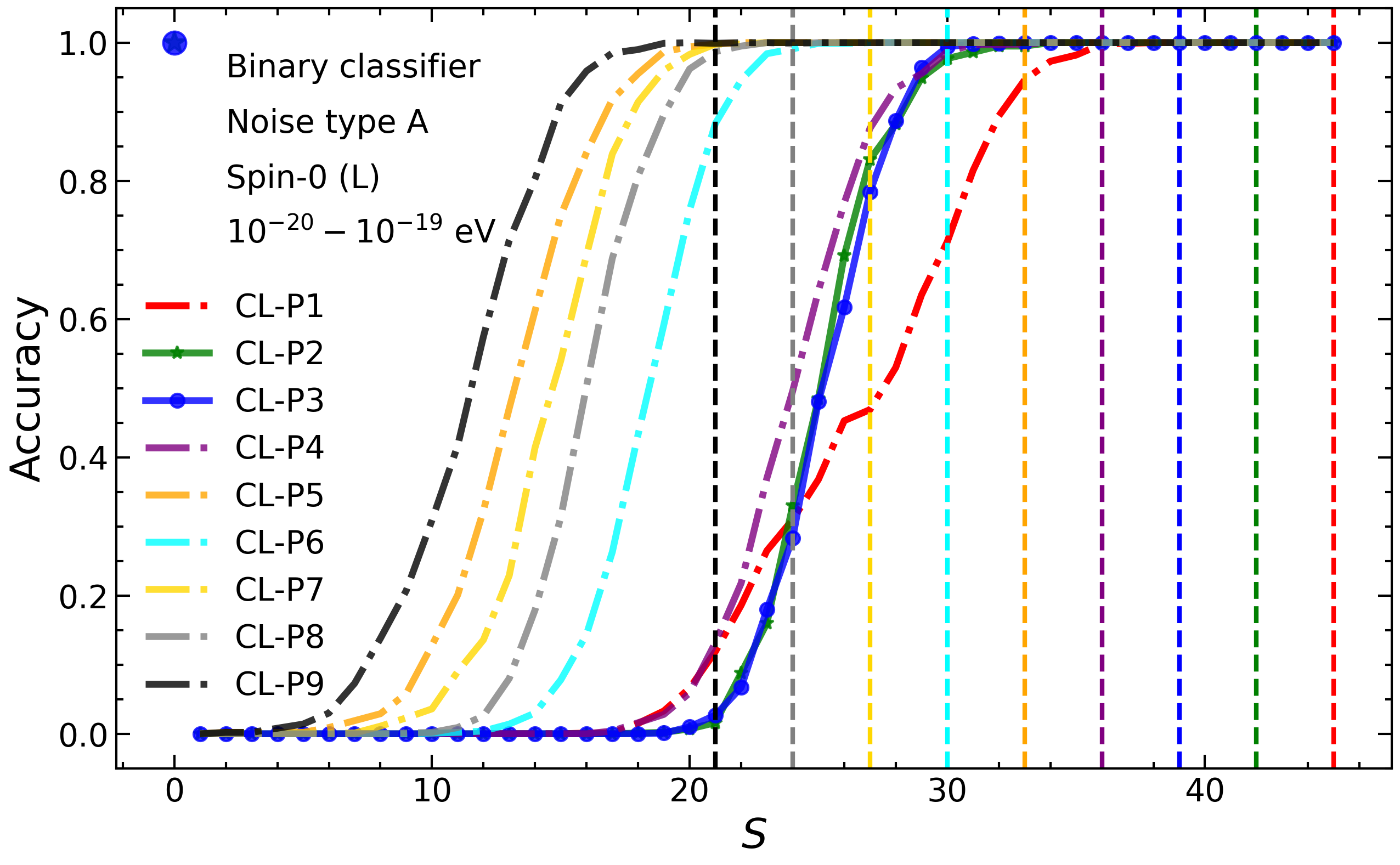}
	\hfill
	\includegraphics[width=0.48\textwidth]{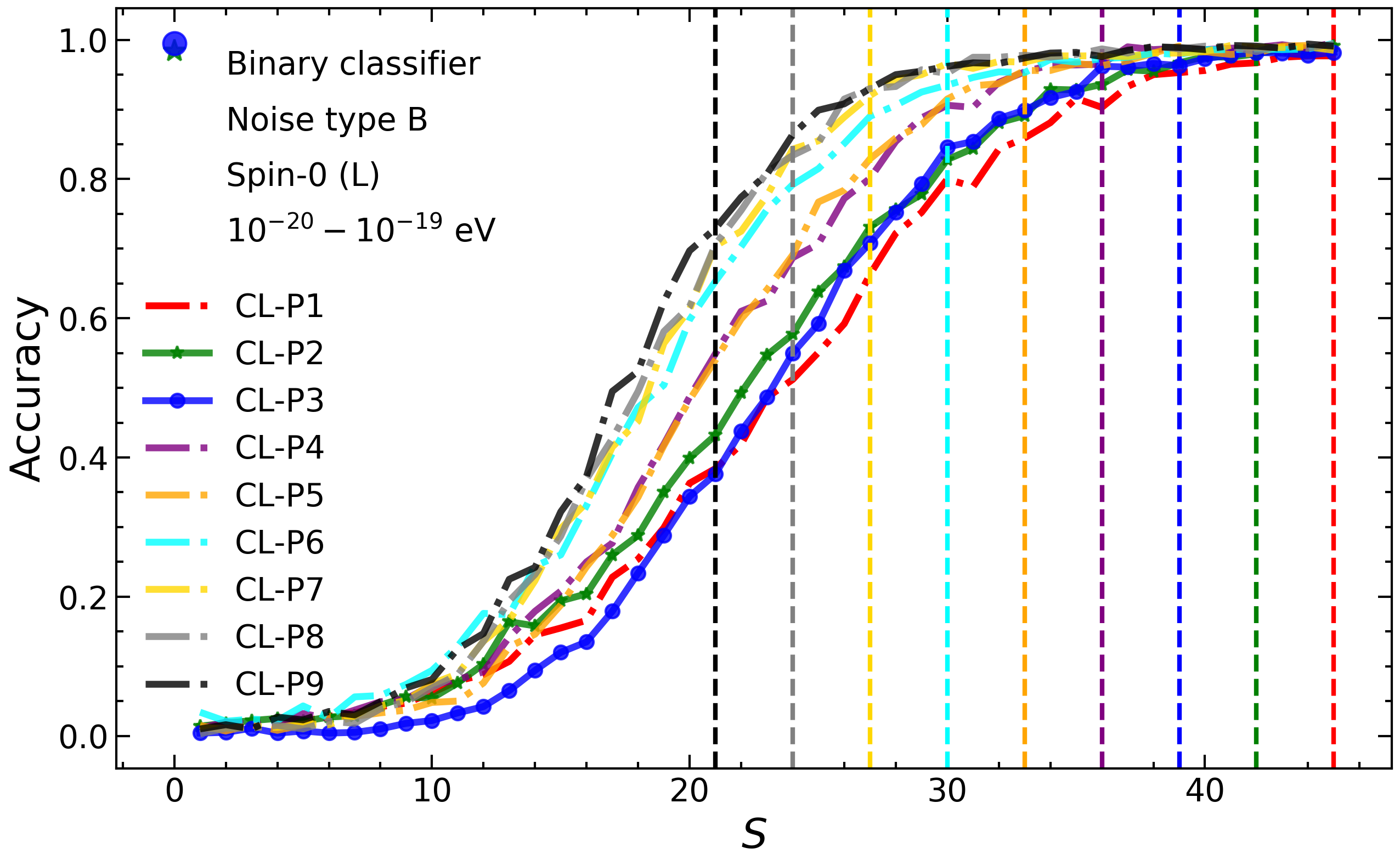}
	\caption{{\itshape Left:} Training a binary CNN classifier on spin-0 (L) ULDM and type A noise. {\itshape Right:} Training a binary CNN classifier on spin-0 (L) ULDM and type B noise.}
	\label{fig:binary_classifier_training_new_mass_interval}
\end{figure}

After some testing it appears that training performs better when starting from larger values of \( S \), a trend also independently indicated by the autoencoder in this mass range. We therefore begin training at \( S = 45 \) and decrease in steps of 3 down to \( S = 21 \), resulting in a total of 9 learning steps. The performance of the trained classifiers is shown in Fig.~\ref{fig:binary_classifier_training_new_mass_interval}. Obviously, the training process is easier for noise A, while in the case of noise B, the training is saturated after stage 5.

\begin{figure}[htbp]
	\centering
	\includegraphics[width=0.48\textwidth]{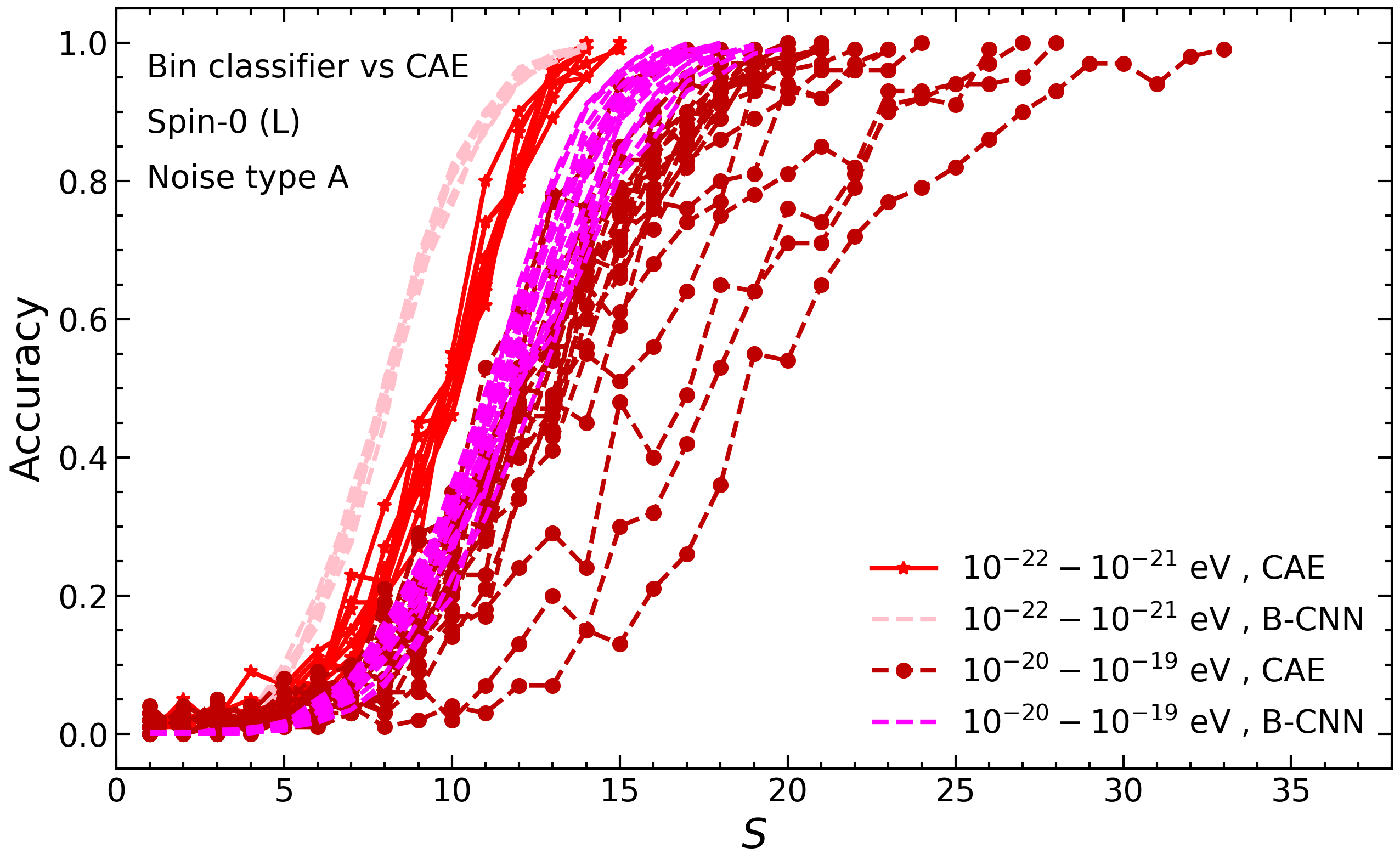}
	\hfill
	\includegraphics[width=0.48\textwidth]{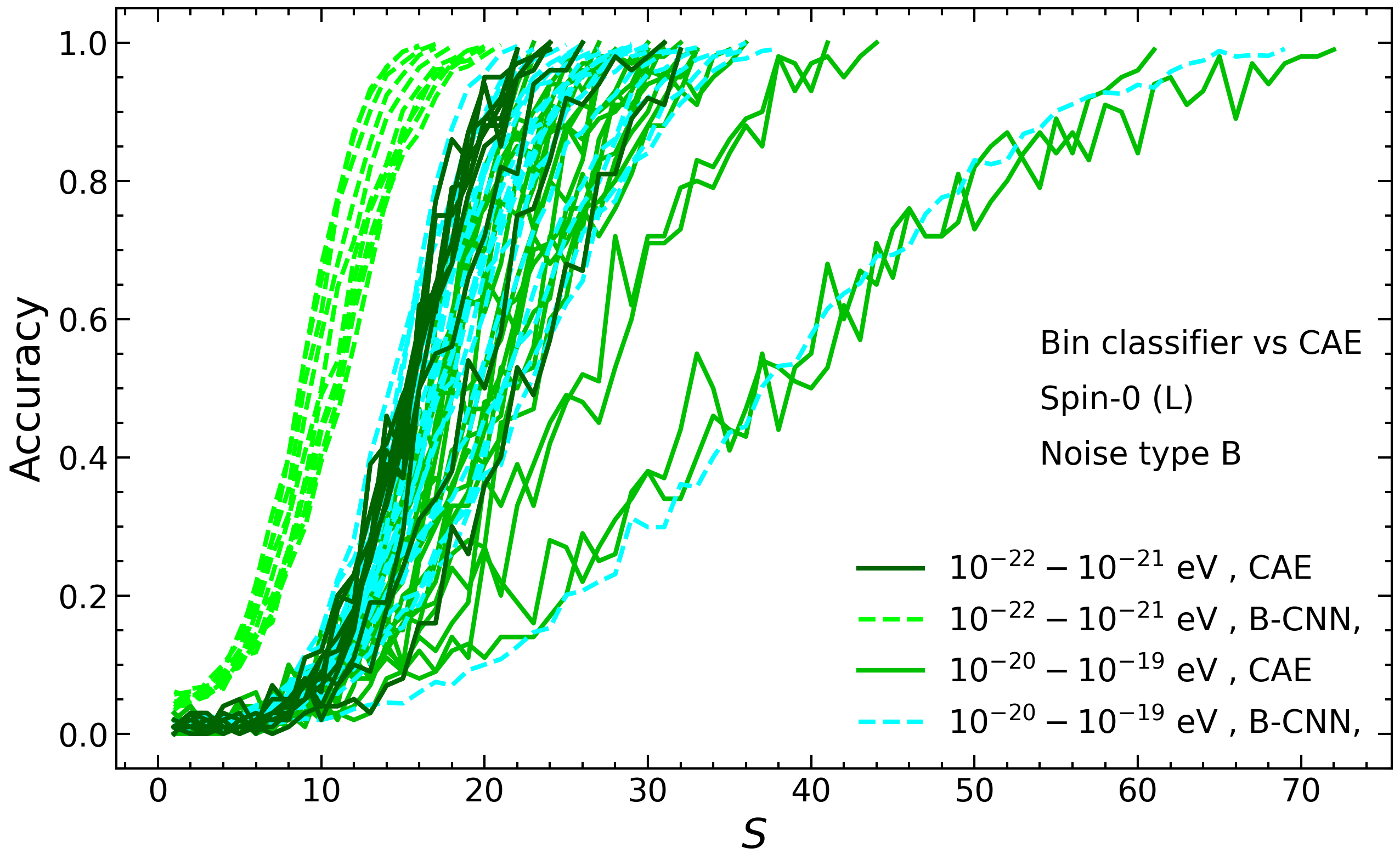}
	\caption{{\itshape Left:} Accuracy versus $S$ for binary classifiers and autoencoder when restricted to noise type A. {\itshape Right:} Accuracy versus $S$ for binary classifiers and autoencoder when restricted to noise type B.}
	\label{fig:accuracy_vs_S_classifiers_and_autoencoder}
\end{figure}

\begin{figure}[htbp]
	\centering
	\rotatebox{0}{\includegraphics[height=7cm]{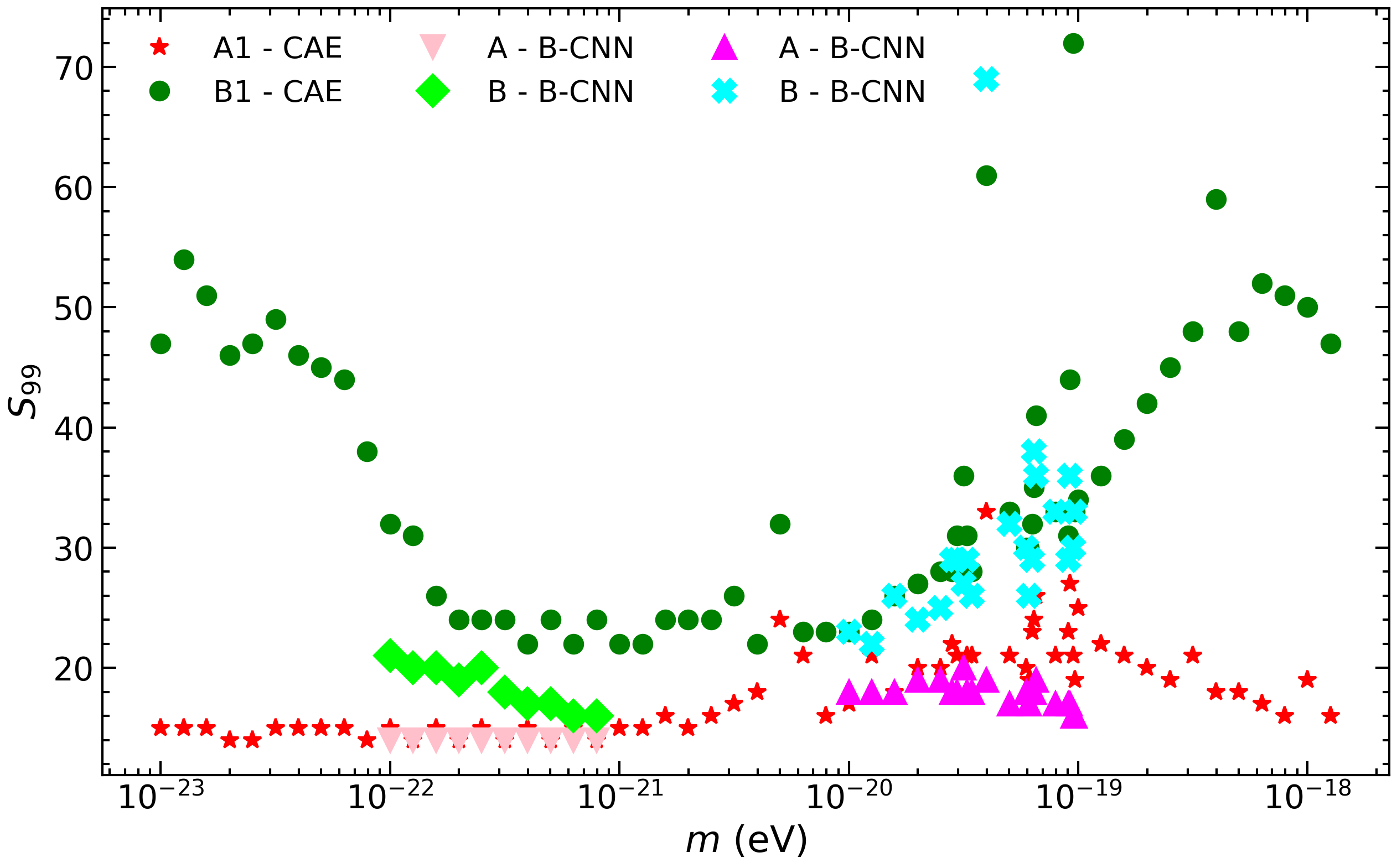}}
	\caption{Values of $S_{99}$ versus mass for 4 classifiers: trained for \(10^{-22}\,\text{eV} <m< 10^{-21}\,\text{eV}\) (type A, pink downward triangles, type B, green diamonds), trained for \(10^{-20}\,\text{eV} <m< 10^{-19}\,\text{eV}\) (type A, magenta upward triangles, type B, teal crosses) and 2 autoencoders (type A, red stars, type B, dark green circles).}
	\label{fig:sensitivity_limits_binary_linear_AB}
\end{figure}

Once all classifiers and autoencoders are trained, we can compare their performance. In Fig.~\ref{fig:accuracy_vs_S_classifiers_and_autoencoder} we show this comparison, whereas the dependence of $S_{99}$ on the mass $m$ is shown in Fig.~\ref{fig:sensitivity_limits_binary_linear_AB}. From these figures we see that, similarly to the autoencoder, the CNNs also lose sensitivity towards low and high frequencies. Moreover, the CNNs produce the same or slightly better results than the autoencoders in the range of masses they were trained on.

\subsection{Multiclass classifier}

We extend the classification task by demonstrating that all five signal types---pure noise and four variations of noisy ULDM signals---can be accurately distinguished from one another, given a sufficiently strong signal magnitude. To achieve this, we design a neural network as a multiclass classifier and enhance its performance using a curriculum learning strategy.

\subsubsection{Architecture}

We employ the CNN architecture designed for robust feature extraction from time-series data. The model structure is detailed in Table~\ref{tab:CNN_structure}.

\begin{table}[htbp]
\centering
\begin{tabular}{|l|c|}
\hline
\textbf{Layers} & \textbf{Output shape} \\
\hline
\hline
InputLayer(input\_shape=(1024, 1)) & (1024, 1) \\
\hline
Conv1D(16, kernel\_size=3, strides=1, activation="relu", padding="same") & (1024, 16) \\
\hline
BatchNormalization() & (1024, 16) \\
\hline
MaxPooling1D(pool\_size=2, strides=2, padding="same") & (512, 16) \\
\hline
Conv1D(32, kernel\_size=3, strides=1, activation="relu", padding="same") & (512, 32) \\
\hline
BatchNormalization() & (512, 32) \\
\hline
MaxPooling1D(pool\_size=2, strides=2, padding="same") & (256, 32) \\
\hline
Conv1D(64, kernel\_size=3, strides=1, activation="relu", padding="same") & (256, 64) \\
\hline
BatchNormalization() & (256, 64) \\
\hline
MaxPooling1D(pool\_size=2, strides=2, padding="same") & (128, 64) \\
\hline
Flatten() & (flattened) \\
\hline
Dense(256, activation="relu") & (256,) \\
\hline
Dropout(0.2) & (256,) \\
\hline
Dense(128, activation="relu") & (128,) \\
\hline
Dropout(0.4) & (128,) \\
\hline
Dense(5, activation="softmax") & (5,) \\
\hline
\end{tabular}

\caption{
Architecture of the 1D convolutional neural network implemented in \href{https://www.tensorflow.org}{\textcolor{blue}{TensorFlow}}, with an input size of \( N = 1024 \). The model consists of 
1D convolutional layers, 1D max pooling layers, batch normalisation layers, dense layers, a flatten layer and dropout layers. 
The final layer uses a softmax activation for multiclass classification. 
The model is trained using the categorical cross-entropy loss
and optimised with the Adam optimiser with an initial learning rate of 0.001, decaying exponentially.
}
\label{tab:CNN_multiclass}
\end{table}

\subsubsection{Training}  

We employ the CL approach once again. We begin by generating 5,000 time series examples for each of the five signal types, with each ULDM signal having \( S = 30 \). Next, we generate a new dataset of the same size for both \( S = 30 \) and \( S = 27 \), effectively doubling the dataset size for the subsequent training phase. This process is repeated for \( S = 30, 27, 24 \), \( S = 30, 27, 24, 21 \) and then for \( S = 30, 27, 24, 21, 18 \). We limit the analysis to noise of type C, which includes both white and red noise as well as nuisance effects. To simplify the training process, we restrict our analysis to a specific mass, \( m = 10^{-21}~\mathrm{eV} \), given the vastness of the parameter space.

\subsubsection{Results}  

\begin{figure}[htbp]
	\centering
	\rotatebox{0}{\includegraphics[width=0.5\linewidth]{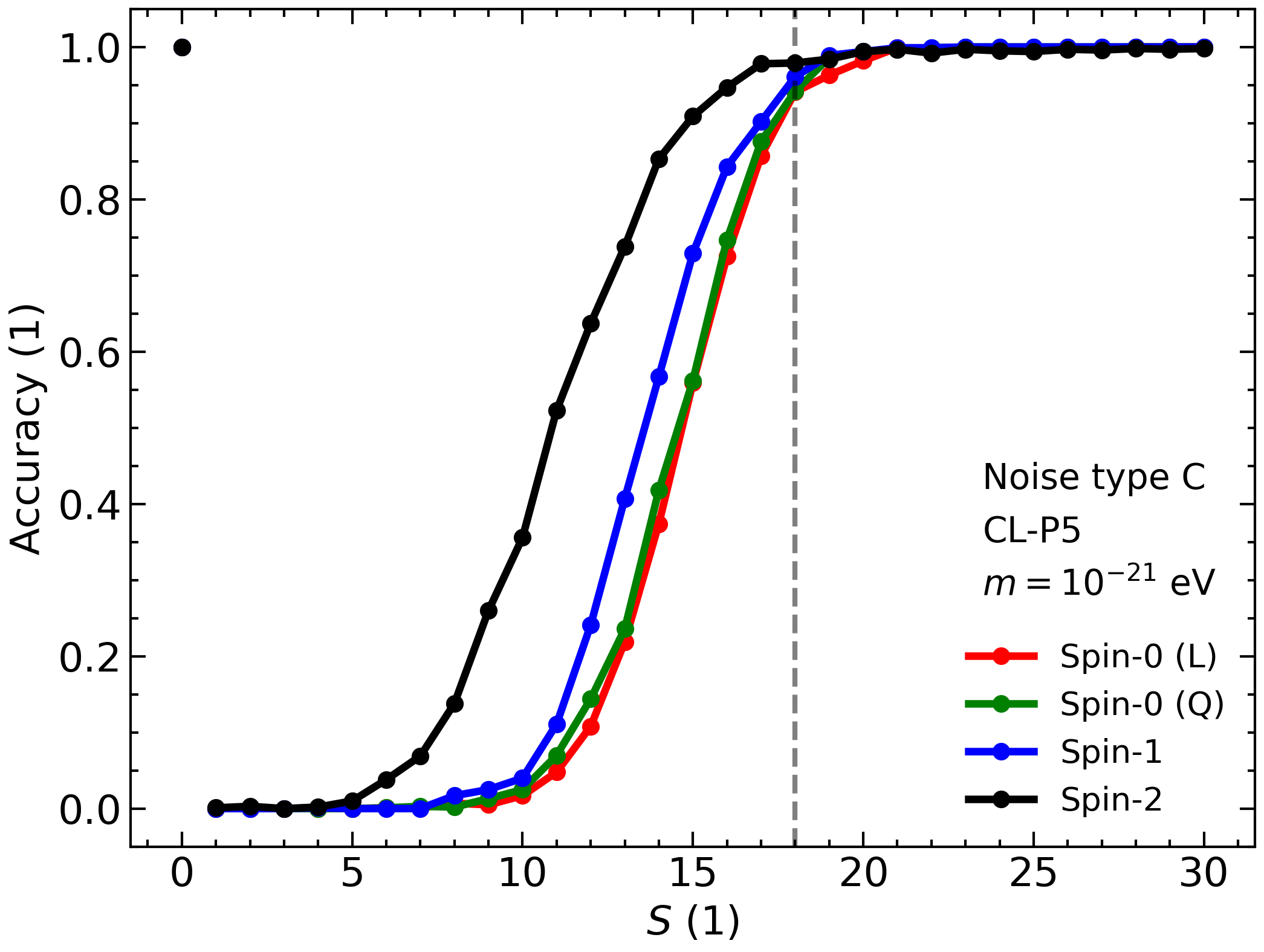}}
	\caption{Accuracy versus $S$ for a multiclass classifier, noise type C, all ULDM types (spin-0 (L) in red, spin-0 (Q) in green, spin-1 in blue and spin-2 in black), phase~5 of the CL.}
	\label{fig:multiclassifier_accuracy_vs_S}
\end{figure}

The ability of the multiclass classifier to detect and correctly classify ULDM signals is shown in Fig.~\ref{fig:multiclassifier_accuracy_vs_S}, which displays the accuracy as a function of \( S \) after the fifth stage of the CL process. The CNN has successfully learnt to recognise ULDM signals of all types up to the last trained value of \( S \) with high accuracy, maintaining an accuracy of approximately 100\% for \( S = 0 \) (i.e.~no false positive detections). 

Further training could enhance sensitivity to weaker signals and improve balance across different ULDM types. This could be achieved through additional training epochs and the use of larger datasets; architectural refinements are also possible. However, we do not pursue these optimisations here, as this section primarily aims to illustrate the technique.

\section{Discussion}
\label{sec:discussion}

\paragraph{Comparison between ML and one-step Bayesian approach } In our analysis, we found that the sensitivity curves produced by both the autoencoder and the binary classifier were broadly consistent with each other across the overlapping mass range, differing by no more than a factor of 2. However, both ML methods consistently demonstrated sensitivity curves up to one order of magnitude weaker than those obtained via the one-step Bayesian approach which we use as a simple benchmark (but applies only under the ideal conditions of white noise and the absence of nuisance signals).

This discrepancy is not surprising. The Bayesian and ML approaches differ in how they incorporate the information stored in the time series. The Bayesian method, being semi-analytic, fully accounts for the noise model and prior knowledge on the signal. In contrast, ML techniques rely exclusively on patterns found in the data provided---they have no access to prior knowledge. Furthermore, and what is perhaps even more relevant, ML techniques involve a degree of compression and decompression of the data which is inherently lossy, in comparison with the semi-analytical Bayesian case for which the final answer only depends on a statistical sampling of the full signal distribution.

\begin{figure}[h!]
    \centering
    \includegraphics[height=7cm]{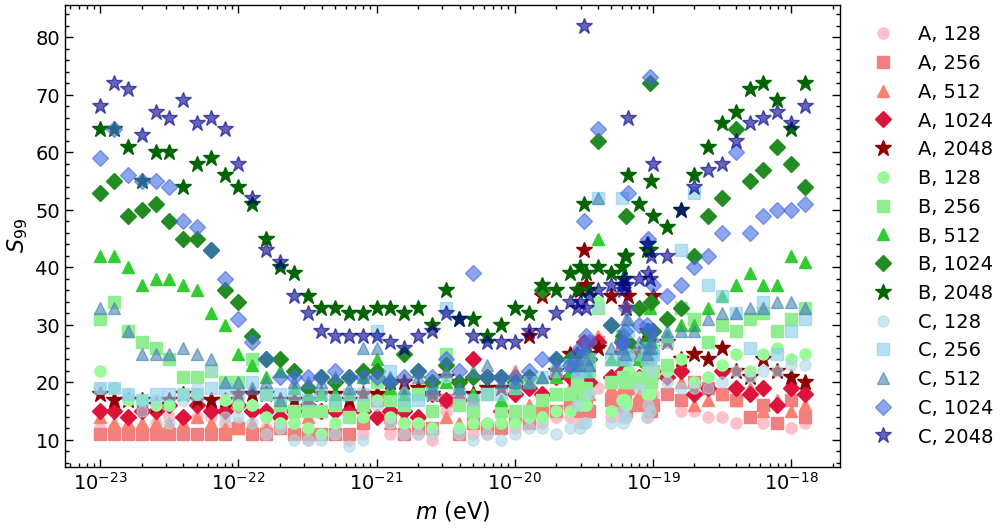}
    \caption{$S_{99}$ versus mass $m$ for the three noise models and five values of the time-series number of points $N$, applied to the spin-0 (L) case; see legend for parameter values for each set.}
    \label{fig:S99_dependence_on_n}
\end{figure}

To explore the difference between Bayesian and ML approaches more quantitatively, we examine how the sensitivity of the autoencoder scales with the number of measurement points, $N$. For simplicity, we restricted the analysis to the spin-0 linear coupling case. We train our autoencoder on new datasets corresponding to the three types of noise (A, B and C), with time series of equal total duration but with $N = 128, 256, 512, 1024, 2048$ points, yielding 15~autoencoders. In Fig.~\ref{fig:S99_dependence_on_n} we display how $S_{99}$ scales with $N$. We observe that, for datasets with denser time series, a higher value of $S_{99}$ is required to detect 99\% of anomalous signals. This is because when $N$ increases while $S$ is held fixed, the amplitude of the signal decreases. Hence, detecting anomalies with small amplitudes thus appears to be more challenging, despite the increased number of data points. 

\begin{figure}[htbp]
    \centering
    \includegraphics[height=7cm]{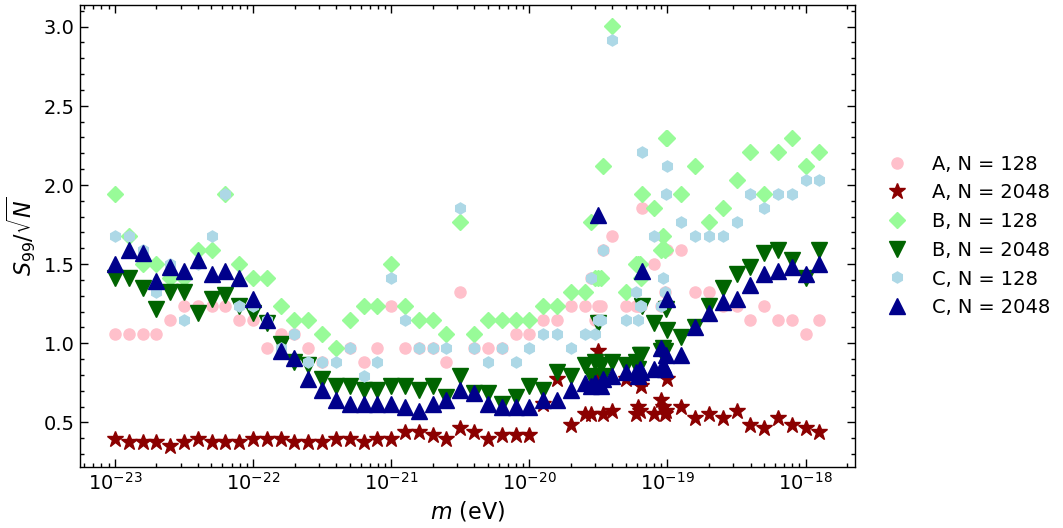}
    \caption{$S_{99}/\sqrt{N}$ versus mass $m$ for all three noise models and time series points $N=128$ and $N=2048$, applied to the spin-0 (L) case; see legend for parameter values for each set.}
    \label{fig:S99divsqrtN_vs_m}
\end{figure}

This seemingly counter-intuitive result can be understood by realising that the sensitivity for varying \(N\) actually scales with \( S_{99} / \sqrt{N} \). We therefore show \( S_{99} / \sqrt{N} \) as a function of mass \( m \) in Fig.~\ref{fig:S99divsqrtN_vs_m} for \( N = 128 \) and \( N = 2048 \), using each type of noise for illustration. In this case we observe that indeed a more dense sampling of the data-taking period leads to a better sensitivity, as expected.


\begin{figure}[h!]
    \centering
    \includegraphics[width=0.48\textwidth]{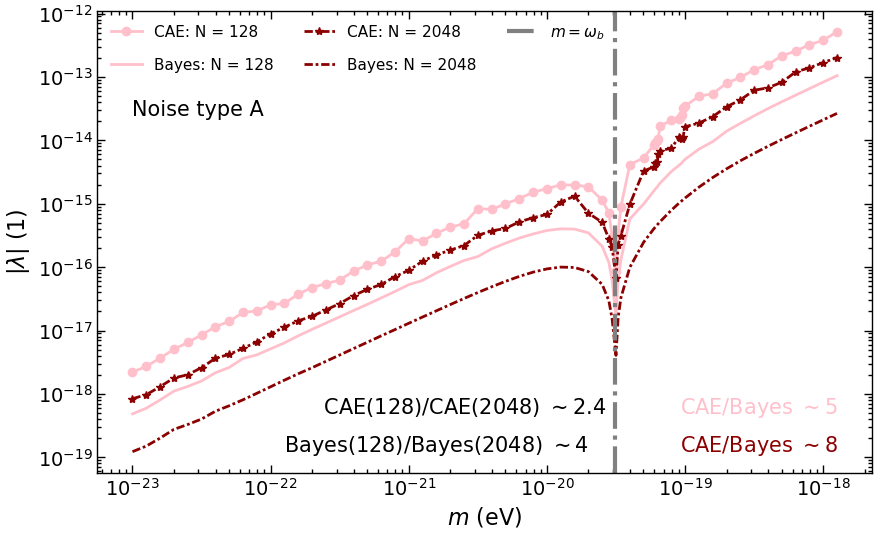}
    \hfill
    \includegraphics[width=0.48\textwidth]{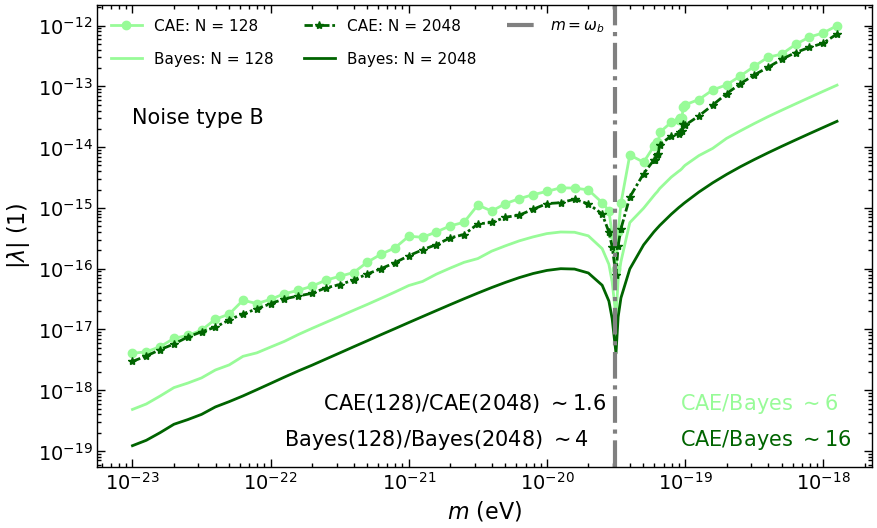}
    \caption{{\itshape Left panel:} Sensitivity limits for $N=128$ (squares, solid line) and $N=2048$ (stars, dot-dashed line) applied to the spin-0 (L) case and with noise of type A. The respective one-step Bayesian limits are shown for comparison. {\itshape Right panel:} Same but for noise type B.}
    \label{fig:autoencoder_dependence_on_n}
\end{figure}

We thus determine the sensitivity limit $|\lambda(m)|$ for each case. For a given noise type and fixed mass $m$ we have five data points corresponding to different values of $N$:
$$\{\lambda(m)_{N=128}^{\mathrm{noise\,type}},\,\lambda(m)_{N=256}^{\mathrm{noise\,type}}, \lambda(m)_{N=512}^{\mathrm{noise\,type}}, \lambda(m)_{N=1024}^{\mathrm{noise\,type}}, \lambda(m)_{N=2048}^{\mathrm{noise\,type}}\}\,,$$
where the noise type can be A, B, or C. We then apply a logarithmic transformation and fit the results using a linear model in log-log space:
$$
\log |\lambda(m)^\mathrm{noise\,type}| = w_0(m)^\mathrm{noise\,type} + w_1(m)^\mathrm{noise\,type} \log N.
$$
After obtaining the slope $w_1(m)^\mathrm{noise\,type}$ for each mass (with the noise type held fixed), we combine the results across all masses by computing the mean and the standard deviation using standard error propagation to obtain the global trend. The observed scaling parameter fits are:
\begin{align*}
w_1^{\text{Bayes}} & = -0.5004 \pm 0.00007\,, \\
w_1^{\text{CAE-A}} & = -0.3155 \pm 0.0030\,, \\
w_1^{\text{CAE-B}} & = -0.1695 \pm 0.0047\,, \\
w_1^{\text{CAE-C}} & = -0.1484 \pm 0.0074\,.
\end{align*}
These results were numerically verified to be robust against variations in the fiducial values of $\varrho$ and $\Upsilon$. Examples of sensitivity curves for $N=128$ and $N=2048$ that illustrate the scaling laws are shown in Fig.~\ref{fig:autoencoder_dependence_on_n}, which confirm that the sensitivity improves with denser sampling.

In summary, although the autoencoder’s sensitivity tends to improve with increasing data, this improvement is weaker compared to the one-step Bayesian method, especially for noise types B and C. In fact, a similar dependence, $|\lambda| \sim 1 / \sqrt{N}$, was observed for the two-step Bayesian approach~\citep{Kus:2024}, which is capable of incorporating the effects of nuisance signals. Therefore, as the number of measurement points increases, the sensitivity with both the ML and Bayesian methods improves, but the Bayesian approach is expected to maintain a clear advantage.

\paragraph{Sensitivity at low and high frequencies } In Fig.~\ref{fig:S_99_vs_m}, we observed that when the autoencoder is trained on pure white Gaussian noise, its sensitivity to the ULDM signal is nearly flat as a function of mass \( m \), i.e.~$S_{99}(m) \simeq \mathrm{const}$. However, when the training data includes more complex noise, the sensitivity decreases at both low and high frequencies, that is, at $k \coloneqq m/\omega_b \ll 1$ and $k\gg1$, respectively. The loss in sensitivity is only by a factor of 3, but it results in a pronounced dependence of \( S_{99} \) on \( m \). This behaviour persists even when the number of samples \( N \) is varied, as shown in Fig.~\ref{fig:S99_dependence_on_n}. To better understand this phenomenon, we focus on the linear spin-0 case for demonstration, as the effect is qualitatively similar for all ULDM models.

\begin{figure}[htbp]
    \centering
    \includegraphics[height=7cm]{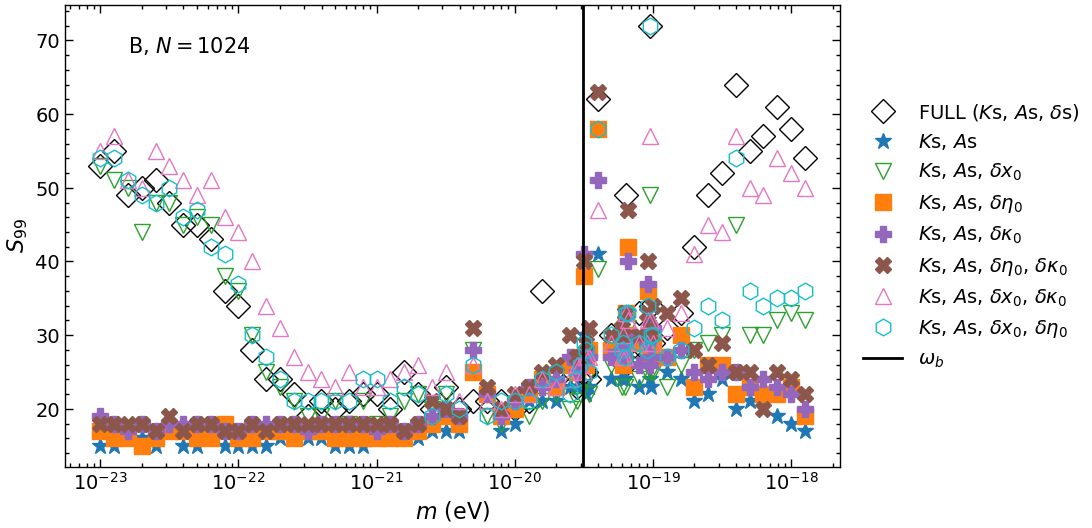}
    \caption{\( S_{99} \) as a function of \( m \) for various training configurations, each involving different combinations of selectively disabled nuisance parameters. All configurations include white Gaussian background noise and at least some nuisance parameters. The `B' in the upper left corner indicates that the noise is derived from noise type B. The results suggest that the dominant factor limiting the autoencoder’s sensitivity to ULDM is the nuisance parameter \( \delta x_0 \); see legend for parameter values for each set.}
    \label{fig:autoencoder_nuisance_signals}
\end{figure}

We begin by investigating how the dependence of \( S_{99} \) on \( m \) changes when individual nuisance parameters are selectively disabled. This helps to identify which nuisance signals are primarily responsible for the observed structure. In Fig.~\ref{fig:autoencoder_nuisance_signals}, we plot \( S_{99}(m) \) for several training configurations. The labels as `Ks' and `As' indicate that the dataset used to train the autoencoder included the nuisance signals induced by \( K_0,\, K_1,\, K_2 \) and \( A_x,\, A_\eta,\, A_\kappa \), respectively; nuisance signals caused by non-zero values \( \delta x_0,\, \delta \eta_0,\, \delta \kappa_0 \) are explicitly indicated when present, with '$\delta$s' meaning that all three terms are included. For instance, the label `Ks, As, $\delta x_0$' means that \( \delta \eta_0,\, \delta \kappa_0 \) are not included in the normal dataset.

Fig.~\ref{fig:autoencoder_nuisance_signals} shows that the overall reduction in sensitivity---especially in the \( k \ll 1 \) regime---is caused by the nuisance parameter \( \delta x_0 \). Notice that the induced nuisance signal from \( \delta x_0 \) may have the largest amplitude among all nuisance signals (as implied by the sampling scheme detailed in Section~\ref{subsec:nuisance_effects}) and is persistent over time---unlike the nuisance signals induced by the `As'. In addition, among the three considered orbital parameter variations, \( \delta x^\mathrm{S0L} \) provides the strongest signal (away from resonance, as found in~\citet{Kus:2024}). However, when the constant term \( \delta x_0 \) is included, it may interfere with the signal from \( \delta x^\mathrm{S0L} \), making it less apparent and thereby reducing the sensitivity.

\begin{figure}[htbp]
    \centering
    \includegraphics[height=7cm]{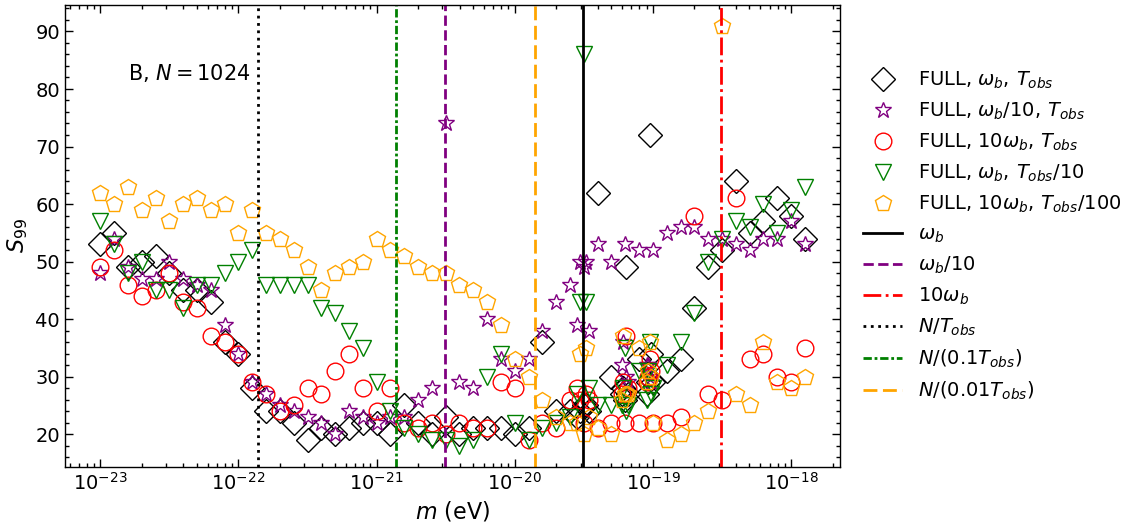}
    \caption{\( S_{99} \) as a function of \( m \) for various combinations of angular velocity and observational times; see legend for parameter values for each set.}
    \label{fig:autoencoder_omega_T}
\end{figure}

In Fig.~\ref{fig:autoencoder_omega_T}, we observe that the autoencoder's sensitivity to the anomalous signal begins to drop, hence \( S_{99} \) rises, when the signal’s characteristic frequency \( m \) becomes smaller than the sampling frequency \( f_s = N / T_{\mathrm{obs}} \), i.e.~when \( m \ll f_s \) which means that \(k \ll f_s/\omega_b \ll 1 \). At this point, the anomalous signal \( \delta x^{\mathrm{S0L}} \) appears as a slowly varying trend---or nearly constant---within the input window, making it harder for the autoencoder to distinguish it from the constant term \( \delta x_0 \).

In the high-frequency regime, \( k \gg 1 \), the peculiar dependence of \( S_{99} \) on \( m \) cannot be attributed to \( \delta x_0 \) alone. The effect arises from a combination of \( \delta x_0 \) with \( \delta \kappa_0 \) and \( \delta \eta_0 \), as none of these individually (even when combined with `Ks' and `As') is sufficient to reproduce the observed structure. This is evident from the comparison of the plots in Fig.~\ref{fig:autoencoder_nuisance_signals}: the set of empty black diamonds (all nuisance parameters included) behaves similarly but has a stronger reduction in sensitivity for $k\gg1$ than the one with empty green downward triangles (\( \delta \eta_0 \) and \( \delta \kappa_0 \) disabled), the pink upward triangles (\( \delta \eta_0 \) disabled) and the cyan hexagons (\( \delta \kappa_0 \) disabled). While \( \delta x_0 \) still appears to play the major role, it cannot by itself produce the observed reduction in sensitivity without the contribution of \( \delta \eta_0 \) and especially \( \delta \kappa_0 \). 

\begin{figure}[htbp]
    \centering
    \includegraphics[height=7cm]{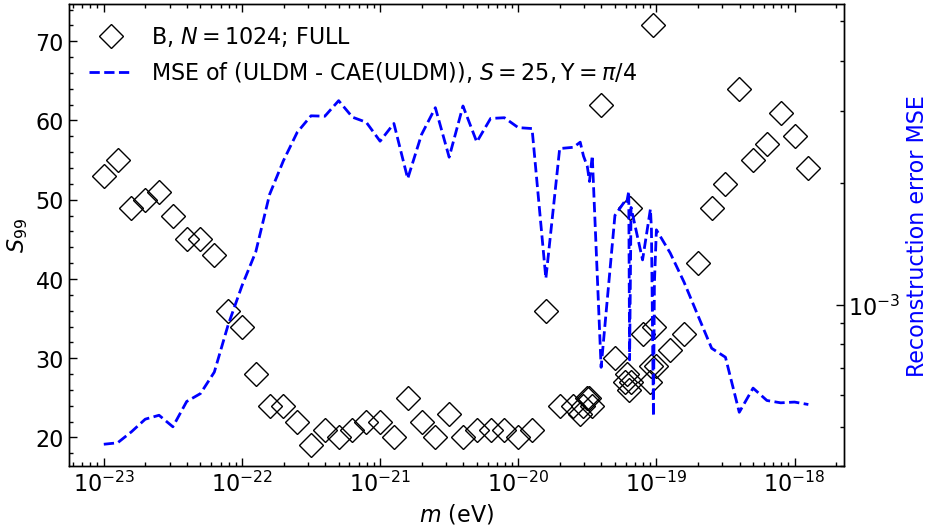}
    \caption{\( S_{99} \) as a function of \( m \), shown alongside the reconstruction error (MSE) of pure ULDM spin-0 (L) signals for various masses.}
    \label{fig:autoencoder_reconstruction_error}
\end{figure}

We empirically find that the deterioration in sensitivity occurs for \( m > \omega_b \), as shown in Fig.~\ref{fig:autoencoder_omega_T}. The apparent drop in sensitivity arises from both the increased complexity of the noise background and the autoencoder’s reduced ability to distinguish the ULDM signal from it, which we interpret as the autoencoder's inability to fully resolve too rapidly oscillating signals. This is illustrated in Fig.~\ref{fig:autoencoder_reconstruction_error}, where the autoencoder is applied directly to a pure ULDM signal and the reconstruction error---measured by the mean squared error (MSE) loss---is evaluated as a function of mass \( m \). The resulting dependence of the MSE on \( m \) complements the observed behaviour of \( S_{99} \) versus \( m \). The results indicate that the reconstruction error is lower for \( m \ll f_s \) and \( m \gg \omega_b \) compared to intermediate masses, suggesting that in these regimes the ULDM signal appears more noise-like to the autoencoder.

In this context, it is worth noting that the shape of the ULDM signal can vary significantly with different values of \( k \). This is most clearly reflected in the behaviour of \( \delta \eta^{\mathrm{S0L}} \) and \( \delta \kappa^{\mathrm{S0L}} \), as shown below:
\begin{align}
    \frac{\delta \eta^{\mathrm{S0L}}}{\alpha \Phi_0 \varrho} &= \cos(\Upsilon) \sin(\Psi') + O(k)\,\,\,\mathrm{for}\,\,\,k\ll1\,, \\
     &= -2  \cos(k\Psi') \sin(\Psi'-\Upsilon)+ O\left(\frac{1}{k}\right)\,\,\,\mathrm{for}\,\,\,k\gg1\,, \\
     \frac{\delta \kappa^{\mathrm{S0L}}}{\alpha \Phi_0 \varrho} &= \cos(\Upsilon) \left( \cos(\Psi') - 1\right)  + O(k)\,\,\,\mathrm{for}\,\,\,k\ll1\,, \\
     &= -2  \cos(k\Psi') \cos(\Psi'-\Upsilon)+  O\left(\frac{1}{k}\right)\,\,\,\mathrm{for}\,\,\,k\gg1\,.
\end{align}


\begin{figure}[htbp]
    \centering
    \includegraphics[height=7cm]{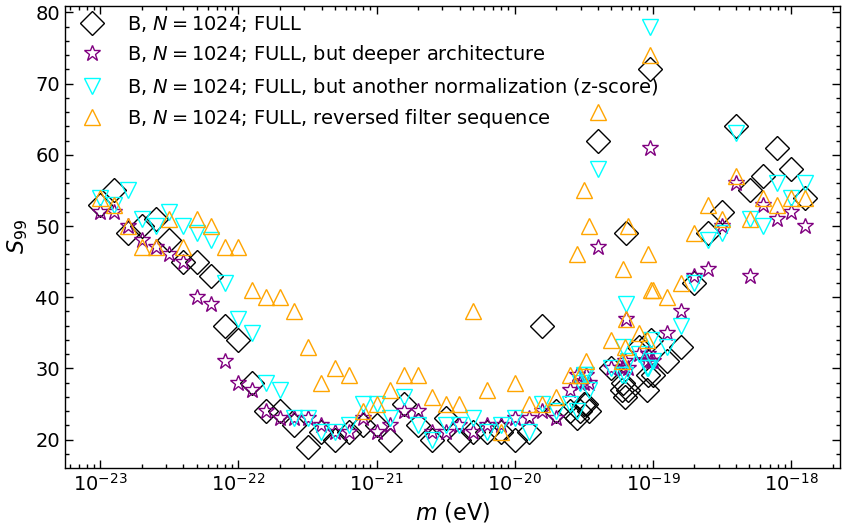}
    \caption{
    \( S_{99} \) as a function of \( m \) for the original setup (black diamonds), compared with results obtained using a deeper autoencoder architecture (purple stars), an alternative normalisation scheme (standardisation instead of MinMax) with a linear (instead of sigmoid) activation function (teal downward triangles) and an architecture with a reversed filter sequence (yellow upward triangles).
    }
    \label{fig:autoencoder_deeperarc_zscore}
\end{figure}

We further demonstrate that the prominent dependence of $S_{99}$ on $m$ cannot be easily mitigated by increasing the depth of the autoencoder. We therefore modify the architecture by extending the encoder: a convolutional layer with~64 filters and kernel size~7 is added at the beginning, followed by LeakyReLU activation and MaxPooling, in accordance with the structure used previously. The decoder is extended symmetrically to reflect these changes. This modification caused the number of trained parameters in the autoencoder to increase by a factor of four, potentially enhancing its capacity to learn finer structural details in the data. While the compression factor of the original architecture was~4, the deeper one has a factor of~8. However, as shown in Fig.~\ref{fig:autoencoder_deeperarc_zscore} (purple stars), this modification does not significantly alter the shape of the \( S_{99}(m) \) dependence.

We also test whether the choice of input normalisation plays a role in the observed behaviour. The original architecture was trained with the MinMax normalisation scheme, which maps all input values to the range \(\left[0, 1\right]\). We repeat the training using standardisation (i.e.~\( z \)-score normalisation), which centres the data and scales it to unit variance. Doing so, we replace the sigmoid activation function in the autoencoder's final layer by a linear function. As Fig.~\ref{fig:autoencoder_deeperarc_zscore} shows, the resulting \( S_{99}(m) \) dependence (green downward triangles) remains largely unchanged, suggesting that neither normalisation scheme has an edge with respect to the other.

Finally, we modify the original autoencoder filter sequence. Our original encoder has a following sequence of filter numbers: \(32 \rightarrow 16 \rightarrow 8 \rightarrow 4\), which makes the input progressively compact and thus compressed---this is an example of undercomplete autoencoder. Now, we compare it with the reversed sequence: \(4 \rightarrow 8 \rightarrow 16 \rightarrow 32\), building an overcomplete autoencoder, which can also be applied as an anomaly detector~\citep{Erdmann:2021}. Fig.~\ref{fig:autoencoder_deeperarc_zscore} shows that, with MinMax normalisation, both architectures achieve comparable anomaly detection sensitivity. We also checked that, when the MinMax normalisation is replaced by the standard \(z\)-score normalisation, the widening architecture fails to perform effectively as an anomaly detector, indicating that more regularisation is needed; we do not pursue this path further. Further work on hyperparameter tuning and alternative architectural designs is planned for future study.



\section{Conclusion and Outlook}
\label{sec:end}

\paragraph{Summary } We have conducted an initial study to investigate the potential of machine learning methods, specifically deep neural networks, to detect ultra-light dark matter signals in pulsar timing residuals.

To this end, we employed three types of classifiers, which we applied to simulated datasets: an anomaly detector (unsupervised ML), a binary classifier and a multiclass classifier (both supervised ML). Using these classifiers, we determined the critical signal strength \( S_c(m) = S_{99}\) required to achieve a true signal detection rate of 99\%. This threshold, in conjunction with the chosen fiducial ULDM parameters, yields an estimate of the sensitivity to the ULDM coupling constant to matter, parametrised by \( \lambda(m) \). Simultaneously, we aimed to minimise the rate of false detections to near zero.

The parameter \( S_c \) plays a role analogous to \( \sqrt{\ln \mathcal{B}} \), the square root of the logarithm of the Bayes factor, as used in the one-step Bayesian approach that assumes delta-function priors on the ULDM parameters. Apart from this distinction, the formulae for sensitivity estimation in the ML and Bayesian approaches, as given by Eqs.~(\ref{eq:ML_lambda_estimation_formula}) and (\ref{eq:Bayes_lambda_estimation_formula}), have the same structure. This similarity makes it straightforward to compare between the two methods.

The one-step Bayesian approach, which we used as a benchmark, yields constraints that outperform the sensitivities of the ML methods by a factor of \( \mathcal{O}(1\text{--}10) \). As demonstrated in~\citet{Kus:2024}, the more comprehensive two-step Bayesian approach---capable of incorporating uncertainties in nuisance parameters---closely matches the one-step method in its sensitivity and exhibits the same scaling law with the number of datapoints \( N \), namely \( 1/\sqrt{N} \). In contrast, the ML approach exhibits a less favorable scaling with data volume, approximately \( \sim 1/N^{0.3} \) at best, thereby granting the two-step Bayesian method a clear sensitivity advantage. However, the semi-analytical two-step method has thus far been implemented only for spin-0 linear coupling scenarios and, due to its complexity, extending it to higher spins presents significant challenges.

The ML methods presented here is the first attempt to place constraints on the ULDM coupling constant across the entire mass range from \( 10^{-23} \, \mathrm{eV} \) to \( 10^{-18} \, \mathrm{eV} \) for ULDM models of spin-0, spin-1 and spin-2, while simultaneously accounting for nuisance signals. Not only did the ML methods enable us to place constraints on the coupling constants of ULDM models, but they also allows us to discriminate between different models.

\paragraph{Relevance } Earlier studies exploring the sensitivity of binary pulsars to all four ULDM models were conducted in~\citet{Blas:2016ddr, Blas:2019hxz, LopezNacir:2018epg, Armaleo:2020yml}. These works share several key features: they focus mostly or exclusively on the secular effects induced by ULDM on orbital parameters---in particular in the orbital period \(P_b\)---that arise near resonant masses; the time evolution of orbital parameters driven by ULDM is not fully taken into account and the constraints are obtained from the requirement that the ULDM effect on the time variation of the period \( \dot{P}_b \) remains below the observational uncertainty; lastly, some of the secular effects are only present in systems with significant eccentricity (\( e \gtrsim 0.1 \)), thereby restricting applicability to a subset of binary pulsars, as high-eccentricity systems are relatively rare compared to those with low eccentricity.

In~\citet{Kus:2024}, a significant advancement was made by developing the two-step Bayesian approach. This method enables sensitivity estimates across the entire mass range from \( 10^{-23} \, \mathrm{eV} \) to \( 10^{-18} \, \mathrm{eV} \), hence not only near resonances. It accounts for the distinctive time-dependent behaviour of ULDM signals beyond secular effects, incorporates the loss of sensitivity due to nuisance signals and allows for the combination of data from all binaries with any eccentricity to produce a global sensitivity curve.

However, this approach has so far been applied only to the spin-0 linear coupling scenario, sensitivity constraints are derived from the isolated variation of individual orbital parameters, that is, only allowing one orbital parameter only to vary at once. This method is challenging to extend to higher spins and it can be computationally expensive.

Given this context, we aimed to explore the applicability of ML methods---already successfully used in GW searches---to the case of ULDM. We employed two distinct ML tools: a convolutional autoencoder used as an anomaly detector and a binary and a multi-class classifier.

A key advantage of this approach is the ability to generate mock data with arbitrary statistical properties and then apply ML algorithms to learn the underlying structure. This allows us to bypass potential technical difficulties associated with analytical integrations, as encountered in the two-step approach.

\paragraph{Outlook } There are several natural directions in which this work can be extended. One important improvement would be to incorporate variations of all orbital parameters, rather than only the three considered in this study. While focusing on three variations allowed for a cleaner investigation of the ML approach, previous work~\citep{Kus:2024} has shown that including other orbital parameter variations can significantly boost sensitivity, at least for the spin-0 linear coupling case. Extending the analysis to include the remaining parameters is a priority for future work.

Another promising avenue is to consider binary systems with higher eccentricities. In this study, we focused on a single example of an ELL1 system, i.e.~a binary with low eccentricity. However, systems with higher eccentricity can exhibit strong resonant behaviour, which has the potential to enhance sensitivity to ULDM. Our ML framework can, in principle, handle such systems without major changes and we plan to investigate this regime in subsequent work.

We also see value in generalising the ML pipeline to incorporate data from multiple pulsars simultaneously. Currently, our neural networks are designed to handle input from a single pulsar, but the architecture could be modified to support multiple sources. This would allow for a more comprehensive framework and would enable a direct comparison with Bayesian approaches, which naturally incorporate multiple pulsars.

Our treatment of timing data assumes equally spaced times-of-arrival samples, with \( N = 1024 \) chosen for illustrative purposes. A more realistic analysis would involve actual times-of-arrival distributions for specific pulsars, including irregular sampling and observational cadences. Adapting our method to these real-world data features will be important to transition toward practical applications.

Incorporating more realistic and complex noise models will bring our analysis closer to real observational conditions. While our current study includes Gaussian white noise or a mix of white and red noise and three nuisance parameters, extending to noise profiles derived from pulsar data will improve the robustness and applicability of the ML approach. Further improvements can also be made on the ML tools themselves by exploring alternative network architectures and training strategies. This may improve upon the generalisation to masses outside of the training range, other ULDM models, and enhance the overall detection performance.

To conclude, the progress of machine learning is relentless: new algorithms are continuously emerging and modern computational and clustering capabilities are increasingly capable to process large datasets. With more sophisticated algorithms, advanced training strategies and detailed data modelling, ML could become a powerful tool to detect ULDM signals in real observational data.

\paragraph{Acknowledgements } DLN acknowledges support from ESIF,  UBA and CONICET. FU acknowledges support from MEYS through the INTER-EXCELLENCE II, INTER-COST grant LUC23115. PK acknowledges support from the European Structural and Investment Funds and the Czech Ministry of Education, Youth and Sports (project No. FORTE---CZ.02.01.01/00/22\_008/0004632). This article is based upon work from the COST Action COSMIC WISPers CA21106, supported by COST (European Cooperation in Science and Technology).

\bibliographystyle{plainnat} 
\bibliography{sample} 

\begin{thebibliography}{39}
\providecommand{\natexlab}[1]{#1}
\providecommand{\url}[1]{\texttt{#1}}
\expandafter\ifx\csname urlstyle\endcsname\relax
  \providecommand{\doi}[1]{doi: #1}\else
  \providecommand{\doi}{doi: \begingroup \urlstyle{rm}\Url}\fi

\bibitem[Agazie et~al.(2023)]{NANOGrav:2023hde}
Gabriella Agazie et~al.
\newblock {The NANOGrav 15 yr Data Set: Observations and Timing of 68 Millisecond Pulsars}.
\newblock \emph{Astrophys. J. Lett.}, 951\penalty0 (1):\penalty0 L9, 2023.
\newblock \doi{10.3847/2041-8213/acda9a}.

\bibitem[Amin et~al.(2022)Amin, Jain, Karur, and Mocz]{Amin:2022pzv}
Mustafa~A. Amin, Mudit Jain, Rohith Karur, and Philip Mocz.
\newblock {Small-scale structure in vector dark matter}.
\newblock \emph{JCAP}, 08\penalty0 (08):\penalty0 014, 2022.
\newblock \doi{10.1088/1475-7516/2022/08/014}.

\bibitem[Aoki and Maeda(2018)]{Aoki:2017cnz}
Katsuki Aoki and Kei-ichi Maeda.
\newblock {Condensate of Massive Graviton and Dark Matter}.
\newblock \emph{Phys. Rev. D}, 97\penalty0 (4):\penalty0 044002, 2018.
\newblock \doi{10.1103/PhysRevD.97.044002}.

\bibitem[Armaleo et~al.(2020)Armaleo, L\'opez~Nacir, and Urban]{Armaleo:2020yml}
Juan~Manuel Armaleo, Diana L\'opez~Nacir, and Federico~R. Urban.
\newblock {Pulsar timing array constraints on Spin-2 ULDM}.
\newblock \emph{JCAP}, 09:\penalty0 031, 2020.
\newblock \doi{10.1088/1475-7516/2020/09/031}.

\bibitem[Baltus et~al.(2021)Baltus, Janquart, Lopez, Reza, Caudill, and Cudell]{Baltus:2021}
Grégory Baltus, Justin Janquart, Melissa Lopez, Amit Reza, Sarah Caudill, and Jean-René Cudell.
\newblock Convolutional neural networks for the detection of the early inspiral of a gravitational-wave signal.
\newblock \emph{Physical Review D}, 103\penalty0 (10), 2021.

\bibitem[Baltus et~al.(2022)Baltus, Janquart, Lopez, Narola, and Cudell]{Baltus:2022}
Grégory Baltus, Justin Janquart, Melissa Lopez, Harsh Narola, and Jean-René Cudell.
\newblock Convolutional neural network for gravitational-wave early alert: Going down in frequency.
\newblock \emph{Physical Review D}, 106\penalty0 (4), 2022.

\bibitem[Blandford and Teukolsky(1976)]{Teukolsky1976}
R.~Blandford and S.~A. Teukolsky.
\newblock {Arrival-time analysis for a pulsar in a binary system}.
\newblock \emph{Astrophysical Journal}, 205:\penalty0 580--591, 1976.
\newblock \doi{10.1086/154315}.

\bibitem[Blas et~al.(2017)Blas, L\'opez~Nacir, and Sibiryakov]{Blas:2016ddr}
Diego Blas, Diana L\'opez~Nacir, and Sergey Sibiryakov.
\newblock {Ultralight Dark Matter Resonates with Binary Pulsars}.
\newblock \emph{Phys. Rev. Lett.}, 118\penalty0 (26):\penalty0 261102, 2017.
\newblock \doi{10.1103/PhysRevLett.118.261102}.

\bibitem[Blas et~al.(2020)Blas, López~Nacir, and Sibiryakov]{Blas:2019hxz}
Diego Blas, Diana López~Nacir, and Sergey Sibiryakov.
\newblock {Secular effects of ultralight dark matter on binary pulsars}.
\newblock \emph{Phys. Rev. D}, 101\penalty0 (6):\penalty0 063016, 2020.
\newblock \doi{10.1103/PhysRevD.101.063016}.

\bibitem[Chase et~al.()Chase, L\'opez~Nacir, and Yunes]{Chase:2025wwj}
Tom\'as~Ferreira Chase, Diana L\'opez~Nacir, and Nicol\'as Yunes.
\newblock {Gravitational waves of quasi-circular, inspiraling black hole binaries in an ultralight vector dark-matter environment}.
\newblock arXiv:2505.21383 [astro-ph.CO].

\bibitem[Collaboration(2023)]{epta2023}
EPTA Collaboration.
\newblock {The second data release from the European Pulsar Timing Array}.
\newblock \emph{Astronomy \& Astrophysics}, 678\penalty0 (A48), 2023.
\newblock \doi{10.1051/0004-6361/202346841}.

\bibitem[Danby(1970)]{Danby:1970}
J.~Danby.
\newblock \emph{{Fundamentals of Celestial Mechanics}}.
\newblock 1970.

\bibitem[Erdmann et~al.(2021)Erdmann, Glombitza, Kasieczka, and Klemradt]{Erdmann:2021}
M.~Erdmann, J.~Glombitza, G.~Kasieczka, and U.~Klemradt.
\newblock \emph{{Deep Learning for Physics Research}}.
\newblock World Scientific Publishing Co Pte Ltd, 2021.
\newblock \doi{10.1142/12294}.

\bibitem[et~al(2021)]{pint2021}
Jing~Luo et~al.
\newblock {PINT: A Modern Software Package for Pulsar Timing}.
\newblock \emph{ApJ}, 911\penalty0 (45), 2021.
\newblock \doi{10.3847/1538-4357/abe62f}.

\bibitem[et~al(2022)]{inpta2022}
Tarafdar~Pratik et~al.
\newblock {The Indian Pulsar Timing Array: First data release}.
\newblock \emph{Publ.Astron.Soc.Austral.}, 39\penalty0 (e053), 2022.
\newblock \doi{10.1017/pasa.2022.46}.

\bibitem[et~al(2023)]{ppta2023}
Zic~Andrew et~al.
\newblock {The Parkes Pulsar Timing Array third data release}.
\newblock \emph{Publ.Astron.Soc.Austral.}, 40\penalty0 (e049), 2023.
\newblock \doi{10.1017/pasa.2023.36}.

\bibitem[Ferreira(2021)]{Ferreira:2021}
Elisa G.~M. Ferreira.
\newblock Ultra-light dark matter.
\newblock \emph{Astron Astrophys Rev}, 29\penalty0 (7), 2021.

\bibitem[Foster et~al.(2018)Foster, Rodd, and Safdi]{Foster:2017hbq}
Joshua~W. Foster, Nicholas~L. Rodd, and Benjamin~R. Safdi.
\newblock {Revealing the Dark Matter Halo with Axion Direct Detection}.
\newblock \emph{Phys. Rev. D}, 97\penalty0 (12):\penalty0 123006, 2018.
\newblock \doi{10.1103/PhysRevD.97.123006}.

\bibitem[Haasteren et~al.(2009)Haasteren, Levin, McDonald, and Lu]{Haasteren:2009}
Rutger~van Haasteren, Yuri Levin, Patrick McDonald, and Tingting Lu.
\newblock {On measuring the gravitational-wave background using Pulsar Timing Arrays}.
\newblock \emph{Monthly Notices of the Royal Astronomical Society}, 395\penalty0 (2):\penalty0 1005–1014, 2009.
\newblock \doi{10.1111/j.1365-2966.2009.14590.x}.

\bibitem[Han and Yang()]{Kim:2024}
Kim.~Jeong Han and Xing-Yu Yang.
\newblock {Gravitational Wave Duet by Resonating Binary Black Holes with Axion-Like Particles}.
\newblock arXiv:2407.14604 [astro-ph.CO].

\bibitem[Hobbs et~al.(2006)Hobbs, Edwards, and Manchester]{tempo2006}
G.~B. Hobbs, R.~T. Edwards, and R.~N. Manchester.
\newblock {TEMPO2, a new pulsar-timing package - I. An overview}.
\newblock \emph{Monthly Notices of the Royal Astronomical Societ}, 369\penalty0 (2), 2006.
\newblock \doi{10.1111/j.1365-2966.2006.10302.x}.

\bibitem[Jacoby et~al.(2003)Jacoby, Bailes, van Kerkwijk, Ord, Hotan, Kulkarni, and Anderson]{Jacoby:2003}
B.~A. Jacoby, M~Bailes, M.~H. van Kerkwijk, S~Ord, A.~Hotan, S.~R. Kulkarni, and S.~B. Anderson.
\newblock Psr j1909-3744, a binary millisecond pulsar with a very small duty cycle.
\newblock \emph{The Astrophysical Journal}, 599\penalty0 (2), 2003.
\newblock \doi{10.1086/381260}.

\bibitem[Knapen et~al.(2017)Knapen, Lin, and Zurek]{Knapen:2017}
Simon Knapen, Tongyan Lin, and Kathryn~M. Zurek.
\newblock {Light Dark Matter: Models and Constraints}.
\newblock \emph{Physical Review D}, 96, 2017.
\newblock \doi{10.1103/PhysRevD.96.115021}.

\bibitem[Kůs et~al.(2024)Kůs, L\'opez~Nacir, and Urban]{Kus:2024}
Pavel Kůs, Diana L\'opez~Nacir, and Federico~R. Urban.
\newblock Bayesian sensitivity of binary pulsars to ultra-light dark matter.
\newblock \emph{Astron Astrophysics}, 690\penalty0 (A51), 2024.

\bibitem[Lange et~al.(2001)Lange, Camilo, Wex, Kramer, Backer, Lyne, and Doroshenko]{Lange:2001rn}
Ch. Lange, F.~Camilo, N.~Wex, M.~Kramer, D.~C. Backer, A.~G. Lyne, and O.~Doroshenko.
\newblock {Precision timing measurements of psr j1012+5307}.
\newblock \emph{Mon. Not. Roy. Astron. Soc.}, 326:\penalty0 274, 2001.
\newblock \doi{10.1046/j.1365-8711.2001.04606.x}.

\bibitem[Liu(2020)]{Liu:2020}
K.~et~al. Liu.
\newblock A revisit of psr j1909-3744 with 15-year high-precision timing.
\newblock \emph{Monthly Notices of the Royal Astronomical Society}, 499\penalty0 (2), 2020.
\newblock \doi{10.1093/mnras/staa2993}.

\bibitem[L\'{o}pez~Nacir and Urban(2018)]{LopezNacir:2018epg}
Diana L\'{o}pez~Nacir and Federico~R. Urban.
\newblock {Vector Fuzzy Dark Matter, Fifth Forces, and Binary Pulsars}.
\newblock \emph{JCAP}, 1810\penalty0 (10):\penalty0 044, 2018.
\newblock \doi{10.1088/1475-7516/2018/10/044}.

\bibitem[L\'opez-S\'anchez et~al.()L\'opez-S\'anchez, Munive-Villa, Skordis, and Urban]{Lopez-Sanchez:2025osk}
Jessica~N. L\'opez-S\'anchez, Erick Munive-Villa, Constantinos Skordis, and Federico~R. Urban.
\newblock {Scaling relations and tidal disruption in spin $s$ ultralight dark matter models}.
\newblock arXiv:2502.03561 [astro-ph.CO].

\bibitem[Lorimer(2008)]{Lorimer:2008se}
D.~R. Lorimer.
\newblock {Binary and Millisecond Pulsars}.
\newblock \emph{Living Rev. Rel.}, 11:\penalty0 8, 2008.
\newblock \doi{10.12942/lrr-2008-8}.

\bibitem[Maggiore(2007)]{Maggiore:2007}
M.~Maggiore.
\newblock \emph{{Gravitational Waves. Vol. 1: Theory and Experiments.}}
\newblock Oxford Master Series in Physics, Oxford University Press, 2007.

\bibitem[Marsh(2016)]{Marsh:2016}
David. J.~E. Marsh.
\newblock {Axion Cosmology}.
\newblock \emph{Phys. Rept.}, 643:\penalty0 1--79, 2016.
\newblock \doi{10.1016/j.physrep.2016.06.005}.

\bibitem[Marzola et~al.(2018)Marzola, Raidal, and Urban]{Marzola:2018}
Luca Marzola, Martti Raidal, and Federico~R. Urban.
\newblock Oscillating spin-2 dark matter.
\newblock \emph{Physical Review D}, 97\penalty0 (024010), 2018.
\newblock \doi{10.1103/PhysRevD.97.024010}.

\bibitem[Moore et~al.(2015)Moore, Taylor, and Gair]{Moore:2014eua}
Christopher~J. Moore, Stephen~R. Taylor, and Jonathan~R. Gair.
\newblock {Estimating the sensitivity of pulsar timing arrays}.
\newblock \emph{Class. Quant. Grav.}, 32\penalty0 (5):\penalty0 055004, 2015.
\newblock \doi{10.1088/0264-9381/32/5/055004}.

\bibitem[Nelson and Scholtz(2011)]{Nelson:2011}
Ann~E. Nelson and Jakub Scholtz.
\newblock {Dark Light, Dark Matter and the Misalignment Mechanism}.
\newblock \emph{Physical Review D}, 84, 2011.
\newblock \doi{10.1103/PhysRevD.84.103501}.

\bibitem[Schive et~al.(2014)Schive, Chiueh, and Broadhurst]{Schive:2014}
Hsi-Yu Schive, Tzihong Chiueh, and Tom Broadhurst.
\newblock {Cosmic Structure as the Quantum Interference of a Coherent Dark Wave}.
\newblock \emph{Nature Physics}, 10\penalty0 (7):\penalty0 496--499, 2014.
\newblock \doi{10.1038/nphys2996}.

\bibitem[Schmidt-May and von Strauss(2016)]{Schmidt:2016}
Angnis Schmidt-May and Mikael von Strauss.
\newblock {Recent developments in bimetric theory}.
\newblock \emph{J. Phys. A: Math. Theor.}, 49:\penalty0 18, 2016.
\newblock \doi{10.1088/1751-8113/49/18/183001}.

\bibitem[Smarra(2024)]{Smarra:2024}
Clemente et~al. Smarra.
\newblock {Constraints on conformal ultralight dark matter couplings from the European Pulsar Timing Array}.
\newblock \emph{Phys. Rev. D}, 110\penalty0 (4):\penalty0 043033, 2024.
\newblock \doi{10.1103/PhysRevD.110.043033}.

\bibitem[Thibault and Donoghue(2010)]{Thibault:2010}
Damour Thibault and John~F. Donoghue.
\newblock {Equivalence Principle Violations and Couplings of a Light Dilaton}.
\newblock \emph{Physical Review D}, 82, 2010.
\newblock \doi{10.1103/PhysRevD.82.084033}.

\bibitem[Weber and de~Boer(2010)]{Weber:2010}
M.~Weber and W.~de~Boer.
\newblock Determination of the local dark matter density in our galaxy.
\newblock \emph{Astron Astrophysics}, 509\penalty0 (A25), 2010.

\end{thebibliography}

\end{document}